\documentclass[review]{elsarticle}

\usepackage{lineno,hyperref}
\modulolinenumbers[5]

\journal{Journal of \LaTeX\ Templates}

\graphicspath{{images/} {images/BNresult/} {images/BNcell/} {images/BN2D/} {images/BNconf/} {images/duct/}}
\usepackage{amsmath}
\usepackage{amssymb}
\usepackage{amsthm}
\usepackage{algorithmicx,algorithm}
\usepackage[noend]{algpseudocode}

\usepackage{mathrsfs}

\theoremstyle{definition}
\newtheorem{example}{Example}
\theoremstyle{remark}
\newtheorem*{rem}{Remark}

\usepackage{multirow}
\usepackage{bm}
\usepackage{geometry}
\usepackage{color}

\begin{document}

\begin{frontmatter}

\title{A staggered-projection Godunov-type method for \\the Baer-Nunziato two-phase  model}
 \tnotetext[mytitlenote]{The second author is supported by NSFC (nos. 11771054, 91852207), the Sino-German Research Group Project, National key project(GJXM92579) and Foundation of LCP.} 

\author[bnu]{Xin Lei}
\ead{leixin@mail.bnu.edu.cn}
\author[iapcm,pku1,pku2]{Jiequan Li\corref{cor1}}
\ead{li\_jiequan@iapcm.ac.cn}

\cortext[cor1]{Corresponding author}

\address[bnu]{School of Science, China University of Geosciences, Beijing, 100083, P. R. China}
\address[iapcm]{Laboratory of Computational Physics,  Institute of Applied Physics and Computational Mathematics, Beijing, 100088, P. R. China}
\address[pku1]{Center for Applied Physics and Technology, Peking University, 100871, P. R. China}
\address[pku2]{State Key Laboratory for Turbulence Research and Complex System, Peking University, 100871, P. R. China}

\begin{abstract}
When describing the deflagration-to-detonation transition in solid granular explosives mixed with gaseous products of combustion, a  well-developed two-phase mixture model is the compressible Baer-Nunziato (BN) model, containing solid and gas phases. If this model is numerically simulated by a conservative Godunov-type scheme, spurious oscillations are likely to generate from porosity interfaces, which may result from  the average process of conservative variables that  violates the continuity  of  Riemann invariants across porosity interfaces.
In order to suppress the oscillations, this paper proposes a staggered-projection Godunov-type scheme over a fixed gas-solid staggered grid, by
enforcing that solid contacts with porosity jumps are always inside gaseous grid cells and other discontinuities appear at gaseous cell interfaces.
This  scheme is based on  a standard Godunov scheme for the Baer-Nunziato model on gaseous cells and   guarantees the continuity of the Riemann invariants associated with the solid contact discontinuities across porosity jumps.
While porosity interfaces are moving, a projection process fully takes into account  the continuity of associated Riemann invariants and ensure that porosity jumps remain inside gaseous cells.
This staggered-projection Godunov-type scheme is well-balanced with good numerical performance not only on  suppressing spurious oscillations near porosity interfaces but also capturing strong discontinuities such as shocks.
\end{abstract}


\end{frontmatter}


\section{Introduction}

In the two-phase flows containing solid and gas phases, dispersed solid particles in the gas can be considered as a continuous fluid that penetrates the gas phase.
Each phase in the gas-solid two-phase flow is usually non-equilibrium and has individual state variables, including the density, velocity, and pressure. 
In 1986, Baer and Nunziato proposed a two-velocity two-pressure model for the compressible two-phase flow \cite{baer_two-phase_1986}, commonly known as the Baer-Nunziato (BN) model.
This model provides a quantitative analysis of the deflagration-to-detonation transition in porous granular explosives.
Neglecting the non-differential source terms due to combustion, drag, heat transfer and chemical reaction in the complete BN model, the governing equations remain balance laws of mass, momentum and energy for each phase, coupled with an evolution equation for the porosity, i.e., the gaseous volume fraction \cite{kapila_two-phase_2001}.
This  simplified model is usually called the {\em homogeneous BN model,} and it consists of  a hyperbolic system of compressible Euler equations plus {\em nozzling terms} of non-conservative products.
The study of the homogeneous BN model plays an important role in simulating the complete BN model; however,  the homogeneous BN model  cannot be written in conervative form so that there is no well-acceptable understanding of the non-conservative products when dealing with strong discontinuities, large deformation of interfaces, mixing of the two phases and interphase exchange
\cite{abgrall_comment_2010}. 
The goal  of this paper is to construct an efficient  numerical scheme to compute  the homogeneous BN model with good fidelity, through a careful analysis of the behavior of solid contacts at porosity interfaces. 

Since shocks may exist in the flow, a numerical scheme should be conservative and thus  Godunov-type schemes \cite{godunov1959difference}  are preferable, for which the associated Riemann problems are solved numerically at cell
interfaces to resolve wave structures properly.
In the Eulerian framework, many Godunov-type methods computing the homogeneous BN two-phase flow model have been proposed, including the operator splitting method \cite{toro_riemann-problem-based_1989,lowe_two-phase_2005}, the unsplit Roe-type wave-propagation method \cite{lowe_two-phase_2005,sainsaulieu_finite_1995,bale_wave_2003, karni_hybrid_2010} and the path-conservative scheme \cite{dumbser_simple_2011,castro_why_2008}.
However, there are some flaws that need to be fixed.  For example, Lowe showed in \cite{lowe_two-phase_2005} the performance of various conservative shock-capturing schemes, exhibiting spurious oscillations (or visible errors) around porosity interfaces.
In \cite{karni_hybrid_2010}, Karni proposed a hybrid algorithm composed of a non-conservative Roe-type scheme across the porosity jump and a conservative Roe-type method away from the porosity jump.
Here, the non-conservative Roe-type scheme utilizes the Riemann invariants across the porosity jump to avoid spurious oscillations.
However, the hybrid scheme is not fully conservative and needs to track the porosity interface.
 In addition to Godunov-type methods, a non-conservative finite-volume approach that uses residuals instead of fluxes was designed \cite{abgrall_high_order_2018}, and it is compatible with local conservation.


As far as  a Godunov-type method is applied to a non-conservative hyperbolic system, a key point is
to compute numerical fluxes by solving the associated  Riemann problem at  each cell interface.
If the Riemann problem contains a porosity jump, the  solution  consists of  complicated wave structures on account of the nozzlling terms, especially because of the resonance phenomenon meaning waves of different families coincide \cite{isaacson_nonlinear_1992,goatin_riemann_2004}.
In \cite{andrianov_riemann_2004}, Andrianov and Warnecke analyzed various wave structures of the homogeneous BN model and designed an exact Riemann solver for given wave patterns, which is called the {\em inverse Riemann solver}.
Schwendeman {\em et al.}~\cite{schwendeman_riemann_2006} proposed an exact Riemann solver for solving the Riemann problem directly.
In the subsequent paper \cite{deledicque_exact_2007},  a different exact Riemann solver, also dealing with  resonance,  was developed.
Tokareva and Toro \cite{tokareva_hllc-type_2010} designed a HLLC-type approximate Riemann solver  to treat the solid contact.
However, for some initial data, the Riemann solution may  be not unique or even do not exist \cite{andrianov_riemann_2004,deledicque_exact_2007}.
This paper is mainly restricted to the case that the Riemann solution exists  uniquely.

Another key point  is about the numerical approximation of non-conservative products.
As introduced in \cite{saurel_multiphase_1999,andrianov_simple_2003,andrianov_riemann_2004,schwendeman_riemann_2006} and summarized in \cite{deledicque_exact_2007}, there are various numerical methods for discretizing nozzlling terms and corresponding numerical results satisfy the Abgrall criterion (or free-streaming condition) generalized from \cite{abgrall_how_1996}, which requires   uniform velocity and pressure  be preserved for two-phase flows.
Nevertheless, the improper  numerical integration of non-conservative products across  porosity jumps could cause spurious oscillations, which  will be analyzed in this paper.
Across the porosity interface, the Riemann invariants for a $0$-characteristic field keep constant across the associated solid contact. However, this is not the case numerically.   It is observed that these Riemann invariants cannot always keep constant with visible errors and spurious oscillations when the conservative scheme is adopted for an isolated moving porosity interface.  Hence, in order to eliminate the oscillations, the conservation requirement of numerical solutions across  porosity interfaces should be appropriately liberated.
Re and Abgrall built a non-conservative pressure-based method for weakly compressible BN-type model \cite{re_non-equilibrium_2019,abgrall_simulation_2020}. This method utilizes a staggered description of the flow variables to avoid spurious pressure oscillations. 
Furthermore, in this paper, we concentrate on getting oscillation-free solutions of general compressible BN flows that contain shock waves, contact discontinuities, and strong rarefactions.
The robust approximation of non-conservative products also depends on the positivity and well-balanced property.
Recent work in this aspect includes a entropy-satisfying scheme in \cite{coquel_positive_2017} and a well-balanced scheme in \cite{thanh_well_balanced_2019} proven to preserve positive phase volume fractions, densities and internal energies.

With the above thinkings in  mind, a staggered-projection Godunov-type scheme is designed to suppress spurious oscillations near porosity interfaces.
For such a  scheme  with piecewise constant initial data, solid contacts, which are associated with  the $0$-characteristic field, are assumed to always exist inside  fixed grid cells, named as {\em ``gaseous cells''}, whereas other waves in the Riemann solution are generated from  gaseous cell interfaces.
We take these solid contacts as interfaces of  {\em ``solid cells''},  which are staggered with the gaseous cells. In each gaseous cell, the standard Godunov scheme is used to update all conservative variables, and the states on both sides of the solid contact are computed in terms of  the same family of Riemann invariants associated with the $0$-characteristic field.
The numerical integration of nozzling terms are computed precisely so that  this scheme is well-balanced.
Moreover, since there is no porosity jump at any gaseous cell interface, it becomes routine to  solve the Riemann problem at  each cell interface for decoupled two phases.
Technically, in order to ensure that moving porosity jumps are always located inside fixed gaseous cells, the porosity discontinuities  are projected back to their original locations at each time step. 
The principle of this projection process is to keep constant  Riemann invariants for the $0$-characteristic field, and thus the total energy conservation has to be given up  across porosity jumps for avoiding the   contradiction  with the former.
Afterwards, the staggered-projection Godunov-type scheme is extended to space-time second-order accuracy using  the generalized Riemann problem (GRP) method for hyperbolic balance laws \cite{Li-2}, and extended to a multi-dimensional regular grid by the dimensional splitting method \cite{strang_construction_1968} accordingly.

This paper is organized as follows. In Section \ref{sec:RP}, we summarize the characteristic analysis of the homogeneous BN model in \cite{baer_two-phase_1986,embid_mathematical_1992} and the properties of the solid contact in the exact Riemann solution are reviewed.
A brief introduction of standard Godunov methods for the homogeneous BN model is described in Section \ref{sec:Godunov-1D} with detailed analysis of   visible numerical errors across porosity interfaces.
The staggered-projection Godunov-type scheme with the second-order GRP version and two-dimensional extension are presented in Section \ref{sec:Stag-Godunov}.
In Section \ref{sec:ex-BN}, several examples are presented to demonstrate the performance. 
Finally, conclusions are made in Section \ref{sec:conclu}. The technical representation of BN model and  approximate solutions of the nonlinear equations related to Riemann invariants is put in Appendix.

\section{The Riemann problem for homogeneous BN model} \label{sec:RP}

In this section, we provide some properties of the homogeneous BN model and discuss the Riemann problem for the later design of Godunov-type schemes. 
In particular, we investigate the property of solid contacts  associated with porosity jumps. 

\subsection{Properties of homogeneous BN model}\label{sec:char-analy}

A detailed characteristic analysis of the compressible BN two-phase model was presented in \cite{embid_mathematical_1992}, and those to be used  in the following sections are summarized below.
For this model, the gaseous products and the dispersed solid particles  are considered as two different phases,  referring to the gas phase and the solid phase, respectively.
In the one-dimensional (1-D) case, the state of each phase is determined by a volume fraction $\alpha_k$, density $\rho_k$, velocity $u_k$, pressure $p_k$, and a total specific energy $E_k$, where $k=s$ and $g$ represent the solid phase and the gas phase, respectively.
The gaseous volume fraction $\alpha_g$ measures the porosity of the solid phase and satisfies the saturation constraint $\alpha_g=1-\alpha_s$.
Ignoring the non-differential source terms, the system of governing equations for the homogeneous BN model is written as
\begin{equation}\label{eq:BN-Euler}
\bm{u}_t+\bm{f}(\bm{u})_x=\bm{h}(\bm{u})(\alpha_s)_x,
\end{equation}
with
\begin{equation*}
\bm{u}=
\begin{bmatrix}
\alpha_s\\
\alpha_s \rho_s\\
\alpha_s \rho_s u_s\\
\alpha_s \rho_s E_s\\
\alpha_g \rho_g\\
\alpha_g \rho_g u_g\\
\alpha_g \rho_g E_g
\end{bmatrix},\quad
\bm{f}=
\begin{bmatrix}
0\\
\alpha_s \rho_s u_s\\
\alpha_s \rho_s u_s^2+\alpha_s p_s\\
\alpha_s u_s (\rho_s E_s + p_s)\\
\alpha_g \rho_g u_g\\
\alpha_g \rho_g u_g^2+\alpha_g p_g\\
\alpha_g u_g (\rho_g E_g  + p_g)
\end{bmatrix},\quad
\bm{h}=
\begin{bmatrix}
-u_s\\
0\\
p_g\\
p_g u_s\\
0\\
-p_g\\
-p_g u_s
\end{bmatrix},
\end{equation*}
where $E_k=e_k+\frac{1}{2}u_k^2$, $e_k=e_k(\rho_k,p_k)$ is the specific internal energy specified by the equations of state (EOS) for each phase. 
The non-conservative product $\bm{h}(\bm{u})(\alpha_s)_x$ is called the nozzling term since it resembles  the non-conservative term of the quasi-1-D compressible duct flow model \eqref{eq:duct-Euler} in a converging-diverging nozzle that we will discuss briefly later on.
The advection equation for the volume fraction,
$(\alpha_s)_t+u_s(\alpha_s)_x=0$,
represents that the porosity is passively advected with the local solid velocity $u_s$.
We assume that each isolated phase satisfies the thermodynamic (Gibbs) relation,
\begin{equation}\label{eq:second-law}
\mathrm{d}e_k=T_k \mathrm{d}\eta_k+\frac{p_k}{\rho_k^2}\mathrm{d}\rho_k,
\end{equation}
where $\eta_k$ and $T_k$ are the entropy and temperature for the phase $k$, respectively.

For smooth solutions, system \eqref{eq:BN-Euler} can be expressed in terms of  primitive variables
\begin{equation*}
\textbf{v}_t+\bm{A}(\textbf{v}) \textbf{v}_x=\bm{0},
\end{equation*}
with
\begin{equation*}
\textbf{v}=\begin{bmatrix}
\alpha_s\\
\rho_s\\
u_s\\
p_s\\
\rho_g\\
u_g\\
p_g
\end{bmatrix},\quad
\bm{A}=
\begin{bmatrix}
u_s & 0 & 0 & 0 & 0 & 0 & 0\\
0 & u_s & \rho_s & 0 & 0 & 0 & 0\\
(p_s-p_g)/(\alpha_s\rho_s) & 0 & u_s & 1/\rho_s & 0 & 0 & 0\\
0 & 0 & \rho_s c_s^2 & u_s & 0 & 0 & 0\\
\rho_g(u_s-u_g)/\alpha_g & 0 & 0 & 0 & u_g & \rho_g & 0\\
0 & 0 & 0 & 0 & 0 & u_g & 1/\rho_g\\
\rho_g c_g (u_s-u_g)/\alpha_g & 0 & 0 & 0 & 0 & \rho_g c_g & u_g
\end{bmatrix},
\end{equation*}
where $c_k$ is the sound speed related to the entropy $\eta_k$ through the thermodynamic relation  \eqref{eq:second-law},
\begin{equation*}
c_k=\sqrt{\frac{\partial p_k(\rho_k,\eta_k)}{\partial \rho_k}}, 
\end{equation*}
the matrix $\bm{A}$ has seven real eigenvalues
\begin{equation*}
\begin{aligned}
&\lambda_0=u_s,\\
&\lambda_{1,s}=u_s-c_s,\quad \lambda_{2,s}=u_s,\quad \lambda_{3,s}=u_s+c_s,\\
&\lambda_{1,g}=u_g-c_g,\quad \lambda_{2,g}=u_g,\quad \lambda_{3,g}=u_g+c_g.
\end{aligned}
\end{equation*}
According to the characteristic analysis in \cite{embid_mathematical_1992}, the $\lambda_{i,s}$-fields, $i=1,2,3$,  are the same as the three eigen-fields of the one-dimensional  Euler equations for the solid phase, and so are the $\lambda_{i,g}$-characteristic fields for the gas phase. Linearly degenerate fields $\lambda_0$, $\lambda_{2,s}$ and $\lambda_{2,g}$ define contact discontinuities (solid contacts and gas contacts); other genuinely nonlinear fields define nonlinear waves (rarefaction waves and shocks). 
Note that $\lambda_{1,s}<\lambda_{2,s}<\lambda_{3,s}$ and $\lambda_{1,g}<\lambda_{2,g}<\lambda_{3,g}$. However, there are no orderings among $\lambda_{i,s}$ and $\lambda_{i,g}$, $i=1,2,3$, which leads to the occurrence of resonance.  Typically, 
under the sonic condition $u_s=u_g\pm c_g$,  a solid contact is located inside a rarefaction  wave or coincides with a shock of  the gas phase and thus the resonance occurs. Only as the following conditions hold, 
\begin{equation*}
(u_g-u_s)^2 \neq c_g^2,\quad \alpha_s\neq 0, \quad \alpha_g\neq 0,
\end{equation*}
the coefficient matrix $\bm{A}$ has a complete set of eigenvectors and the system \eqref{eq:BN-Euler} is hyperbolic. Note that  $\lambda_0=\lambda_{2,s}=u_s$. The resonance may occur, which is analogous to the resonant case of the duct flow that we will discuss later on.

\subsection{Properties of the Riemann solution}\label{sec:BN-RP}

The Riemann problem is the building block of  Godunov-type schemes. Assume,  for the 1-D homogeneous BN model \eqref{eq:BN-Euler},    that piecewise constant data at initial time $t=t_n$ that can be shifted to $t=0$, 
\begin{equation*}
\bm{u}(x,t=0)=\left\{
\begin{aligned}
&\bm{u}_L, & x<x_{0},\\
&\bm{u}_R, & x>x_{0},
\end{aligned}
\right.
\end{equation*}
where $\bm{u}_L$ and $\bm{u}_R$ are constant states on both sides of  certain position $x=x_{0}$. The solution of this  Riemann problem   has self-similarity,
\begin{equation*}
\bm{u}(x,t)=\bm{v}(\xi),\quad\  \xi=\frac{x-x_{0}}{t},\ t>0, 
\end{equation*}
 denoted as $\textbf{RP}(\xi; \bm{u}_L,\bm{u}_R)$ and
this solution consists of shock waves $\mathcal{S}_k$, rarefaction waves $\mathcal{R}_k$, and contact discontinuities $\mathcal{C}_k$ ($k=s,g$) for each phase.
For the time being, we leave aside the discussion on the coalescence of these waves for different phases to the next section, and pay attention to solid contacts $\lambda_0=\lambda_{2,s}$.  A  solid contact is associated with $\lambda_0$-field, and the corresponding  five Riemann invariants are
\begin{equation*}
\begin{aligned}
&u_s,\quad \eta_g,\\
&Q:=\alpha_g \rho_g (u_g - u_s),\\
&P:=\alpha_s p_s + \alpha_g p_g + \alpha_g \rho_g (u_g - u_s)^2,\\
&H:=h_g+\frac{(u_g - u_s)^2}{2},
\end{aligned}
\end{equation*}
where $h_g = e_g + p_g/\rho_g$ is the enthalpy of the gas phase. Note that the state variables of the gas phase and the solid pressure $p_s$ do not remain constant across the solid contact, except the situation that the porosity on both sides of the contact is identical, i.e., $(\alpha_s)_L=(\alpha_s)_R$. 
Away from the solid contact, the porosity is constant and the nozzling term is zero, thus the governing equations \eqref{eq:BN-Euler}  are decoupled and reduce to the Euler equations for the two individual phases.
The state variables of one phase are constant across the waves of the other phase.

\begin{figure}[htb]
\begin{minipage}{0.5\linewidth}
\begin{center}
\includegraphics[width=0.95\textwidth]{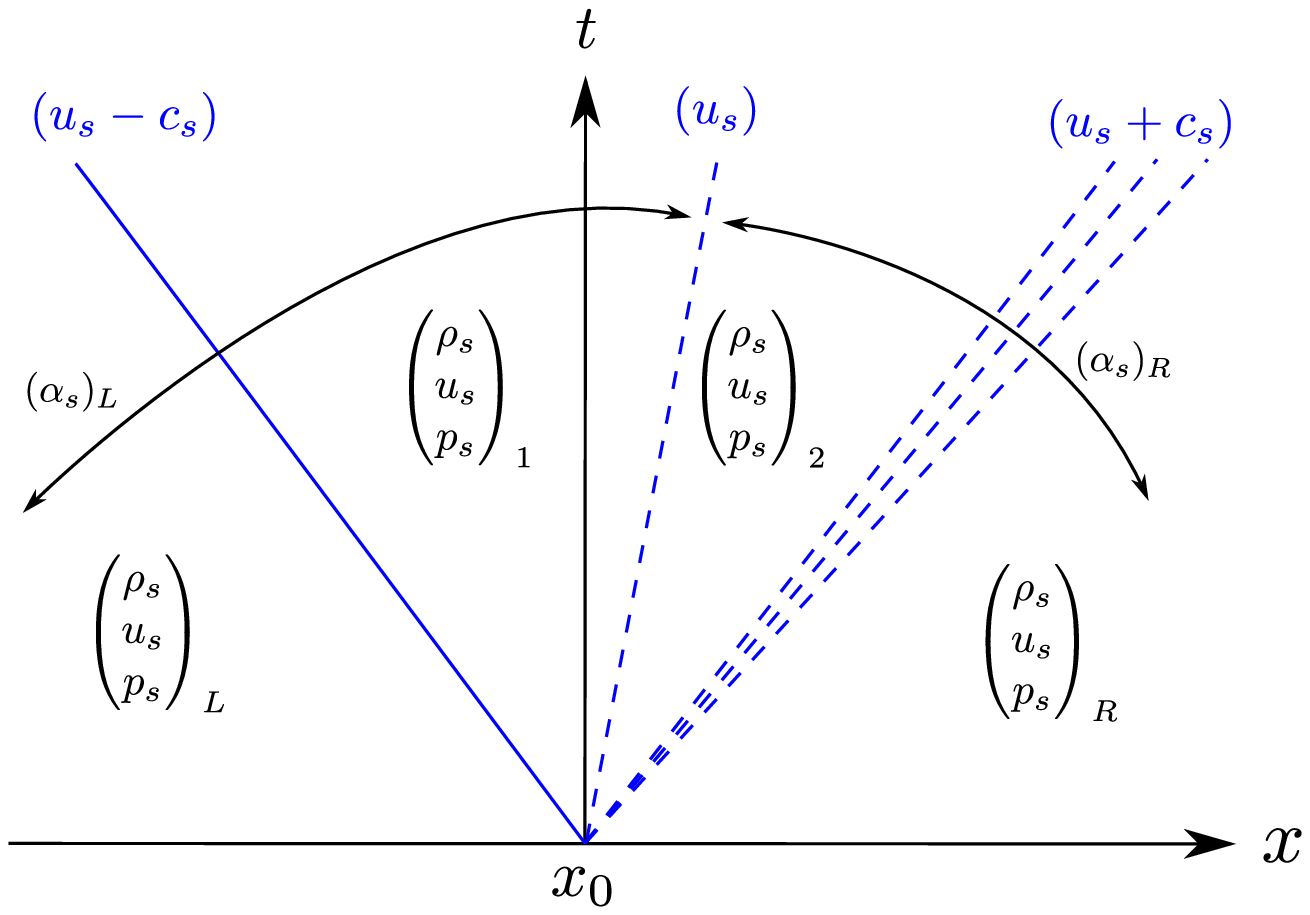}

Intermediate states of the solid phase
\end{center}
\end{minipage}
\hfill
\begin{minipage}{0.5\linewidth}
\begin{center}
\includegraphics[width=0.95\textwidth]{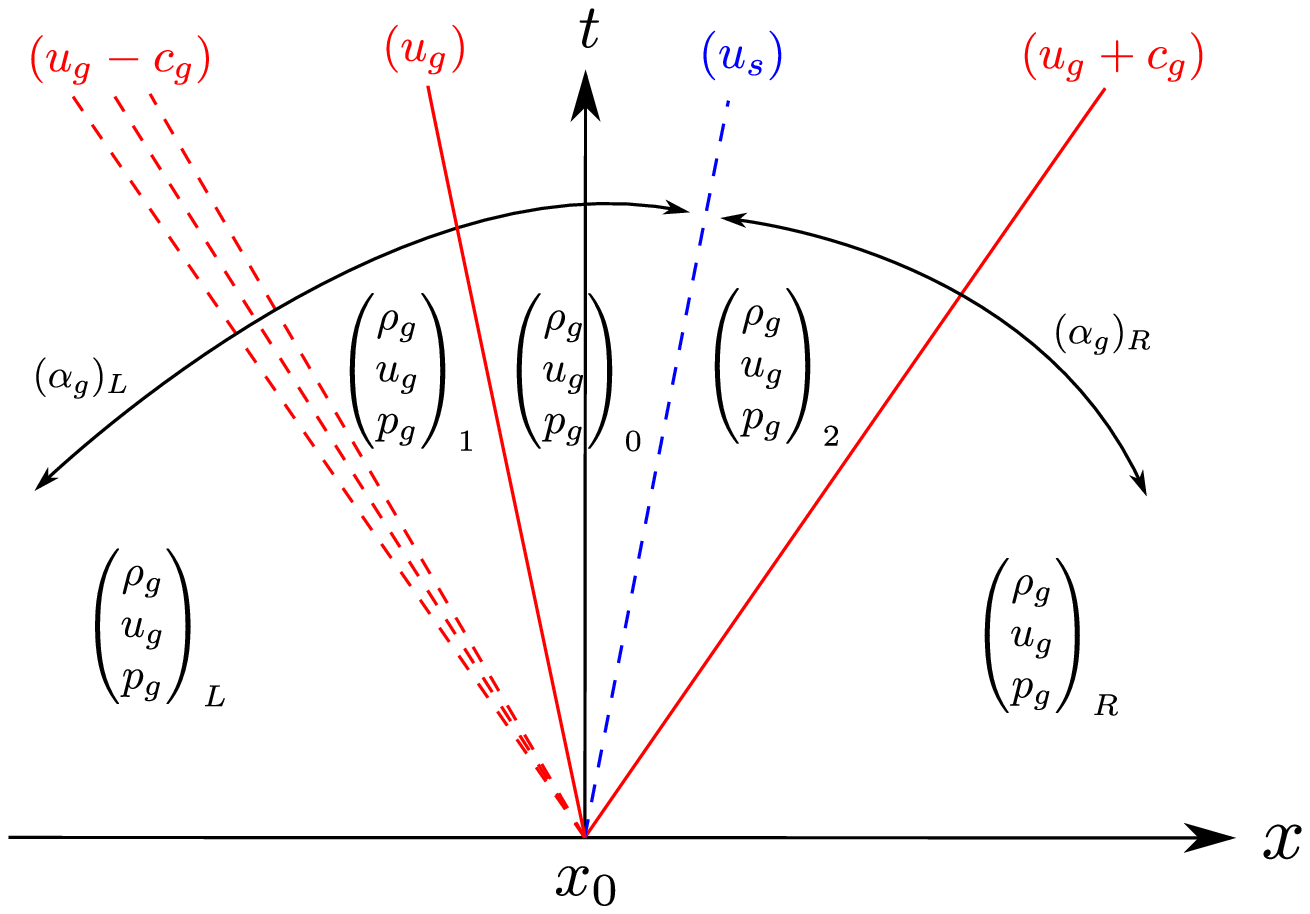}

Intermediate states of the gas phase
\end{center}
\end{minipage}
\caption{The wave configuration to the Riemann problem for a subsonic configuration $\lambda_{2,g}<\lambda_0<\lambda_{3,g}$.}\label{fig:BN-conf}
\end{figure}

Upon the position of the solid contact relative to the gas phase, wave configurations of the Riemann problem are classified into two categories:  (i) {\em the solid contact is outside of gas waves},  $\lambda_0<\lambda_{1,g}$, or $\lambda_0>\lambda_{3,g}$;  (ii) {\em the solid contact is inside the gas waves},  $\lambda_{1,g}<\lambda_0<\lambda_{2,g}$ or  $\lambda_{2,g}<\lambda_0<\lambda_{3.g}$.
For certain initial data,  there may be more than one Riemann solution from different configurations; and  for certain other initial data, the Riemann solution does not exist \cite{andrianov_riemann_2004}.
This ill-posedness issue may reflect the invalidity of the Riemann problem for the homogeneous BN model, and it has not been effectively resolved \cite{deledicque_exact_2007}.
Therefore, in this paper, we are chiefly concerned with the situation that the Riemann solution exists uniquely,
and try to get around this issue by solving more simplified Riemann problems.
For simplicity, we take the configuration (ii) as an example for analysis, as shown in Figure \ref{fig:BN-conf}. In this configuration, the gas phase is subsonic relative to the solid phase, i.e., $(u_g - u_s)^2 < c_g^2$. It holds in the context of granular explosives. 
For the solid phase in this configuration, the intermediate states on the left and right sides of the solid contact are labelled by subscripts $1$ and $2$, respectively. The gas phase is labelled similarly, in addition to indicating the intermediate state between the gas contact and the solid contact by the subscript $0$.

\begin{figure}[htb]
\begin{center}
\includegraphics[width=0.6\textwidth]{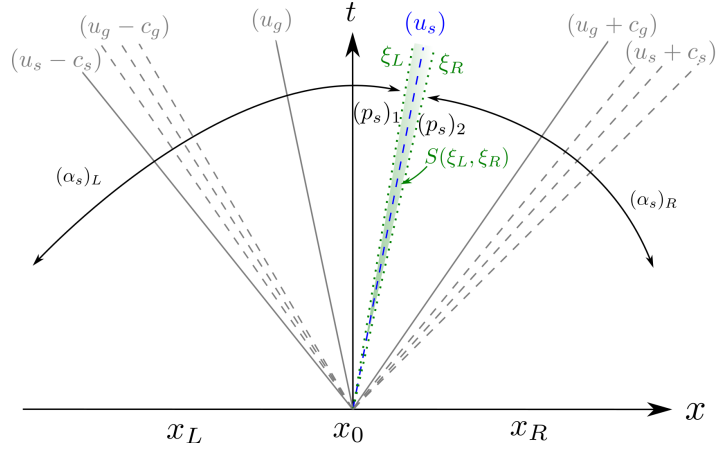}
\end{center}
\caption{The solution in a small sector covering
a solid contact propagating to the right.}\label{fig:solid-contact}
\end{figure}

A schematic diagram for the {\color{red} complete} wave configuration above  $\lambda_{2,g}<\lambda_s<\lambda_{3,g}$ is shown in Figure \ref{fig:solid-contact}.
There are two rays $\xi=\xi_L$ and $\xi=\xi_R$,  $\xi_L=(u_s)^*-\epsilon$, $\xi_R=(u_s)^*+\epsilon$ and $\epsilon>0$, bounding a sectorial region $S(\xi_L,\xi_R)$  by recalling that the similarity variable $\xi=(x-x_{0})/t$  and the solid velocity $(u_s)_1=(u_s)_2=:u_s^*$. 
The solution can only be discontinuous across the solid contact $\xi=(u_s)^*$.
The porosity is constant outside the region $S(\xi_L,\xi_R)$, namely $(\alpha_s)_x\equiv 0$ and hence the nozzling terms take effect only along the solid contact.
Since the Riemann solution $\bm{u}(x,t)$ is self-similar locally, we make the change of variables $(x,t)\rightarrow (\xi,t)$ to recast system \eqref{eq:BN-Euler} in the form
\begin{equation}\label{eq:BN-eta}
-\xi \bm{v}_\xi+\bm{f}(\bm{v})_\xi=\bm{h}(\bm{v})(\alpha_s)_\xi.
\end{equation}
Integrating  system \eqref{eq:BN-eta} from $\xi_L$ to $\xi_R$, we obtain
\begin{equation*}
\int_{\xi_L}^{\xi_R}\left[-\xi \bm{v}_\xi+\bm{f}(\bm{v})_\xi\right] \mathrm{d}\xi
=\int_{\xi_L}^{\xi_R}\bm{h}(\bm{v})(\alpha_s)_\xi \mathrm{d}\xi.
\end{equation*}
Specified to the solid phase, we have
\begin{subequations}\label{eq:integrated-eq}
\begin{align}
& \int_{\xi_L}^{\xi_R}\left[-\xi(\alpha_s \rho_s)_\xi+(\alpha_s \rho_s u_s)_\xi\right] \mathrm{d}\xi =0,\\
& \int_{\xi_L}^{\xi_R}\left[-\xi(\alpha_s \rho_s u_s)_\xi+(\alpha_s \rho_s u_s^2+\alpha_s p_s)_\xi\right] \mathrm{d}\xi =\int_{\xi_L}^{\xi_R}p_g (\alpha_s)_\xi \mathrm{d}\xi,\\
& \int_{\xi_L}^{\xi_R}\left[-\xi(\alpha_s \rho_s E_s)_\xi+(\alpha_s \rho_s u_s E_s+\alpha_s u_s p_s)_\xi\right] \mathrm{d}\xi =\int_{\xi_L}^{\xi_R}p_g u_s (\alpha_s)_\xi \mathrm{d}\xi,
\end{align}
\end{subequations}
and
\begin{equation*}
\int_{\xi_L}^{\xi_R} -u_s(\alpha_s)_\xi \mathrm{d}\xi =-u_s^*\left[(\alpha_s)_R -(\alpha_s)_L\right].\\
\end{equation*}
Since the Riemann invariant $u_s$ is a constant across the solid contact,   (\ref{eq:integrated-eq}a) always holds, and the equations (\ref{eq:integrated-eq}b) and  (\ref{eq:integrated-eq}c) hold under the condition
\begin{equation}\label{eq:BN3-eta}
\int_{\xi_L}^{\xi_R}p_g (\alpha_s)_\xi \mathrm{d}\xi=\int_{\xi_L}^{\xi_R}(\alpha_s p_s)_\xi \mathrm{d}\xi =(\alpha_s)_R (p_s)_2-(\alpha_s)_L (p_s)_1.
\end{equation}
This provides a way to discretize the nozzling term.

\section{Analysis of spurious oscillations by standard Godunov schemes}\label{sec:Godunov-1D}

In this section we review the standard Godunov method for the homogeneous BN model and analyze spurious oscillations in the solution.
For the 1-D model, the computational domain $[0, L]$ are divided into $M$ fixed grid cells $I_i=[x_{i-\frac12},x_{i+\frac12}], i=1,2,\ldots,M$,   $\Delta x=x_{i+\frac12}-x_{i-\frac12}=L/M$,  the cell interface $x_{i+\frac12}=i\Delta x$,  and the cell center $x_i=(i-1/2)\Delta x$, respectively.
The Godunov scheme assumes the initial data to be  piece-wise constant at time $t=t_n$, 
\begin{equation*}
\bm{u}_i^{n} =\frac{1}{\Delta x}\int_{x_{i-\frac{1}{2}}}^{x_{i+\frac{1}{2}}} \bm{u}(x,t_{n}) \mathrm{d}x.
\end{equation*}
The next step is to compute numerical fluxes along all cell interface $x=x_{i+\frac{1}{2}}$ by solving the local Riemann problem $\textbf{RP}\left(\xi; \bm{u}_i^n,\bm{u}_{i+1}^n\right)$, $\xi=\frac{x-x_{i+\frac 12}}{t-t_n}$.
For a  time increment $\Delta t$  restricted by the CFL condition, we can evolve the solution to the next time $t_{n+1}=t_n+\Delta t$ by the finite-volume discretization
\begin{equation}\label{eq:BN-Godunov}
\bm{u}_i^{n+1}-\bm{u}_i^n+\frac{\Delta t}{\Delta x}\left(\bm{f}_{i+\frac{1}{2}}^n-\bm{f}_{i-\frac{1}{2}}^n\right)=\frac{\Delta t}{\Delta x}\bm{S}_i^n,
\end{equation}
where the numerical flux along the cell interface $x=x_{i+\frac 12}$ is
\begin{equation*}
\bm{f}_{i+\frac{1}{2}}^n=\frac{1}{\Delta t}\int_{t_n}^{t_{n+1}}\bm{f}\left(\bm{u}(x_{i+\frac{1}{2}},t)\right)\mathrm{d}t=\bm{f}\left(\bm{u}_{i+\frac{1}{2}}^n\right),
\end{equation*}
with the Riemann solution  $\bm{u}_{i+\frac{1}{2}}^n=\textbf{RP}(0; \bm{u}_i^n, \bm{u}_{i+1}^{n})$ and the time average of the numerical integral for the nozzling term
\begin{equation*}
\bm{S}_i^n=\frac{1}{\Delta t}\int_{t_n}^{t_{n+1}}\int_{x_{i-\frac{1}{2}}}^{x_{i+\frac{1}{2}}}\bm{h}(\bm{u})(\alpha_s)_x \mathrm{d}x \mathrm{d}t.
\end{equation*}
The numerical fluxes can be evaluated by exact Riemann solvers \cite{schwendeman_riemann_2006,deledicque_exact_2007} or approximate Riemann solvers, such as a Roe solver \cite{karni_hybrid_2010}, {\color{red} a  Rusanov‐type solver \cite{menshov_generalized_2018}} or a more precise HLLC solver \cite{tokareva_hllc-type_2010}.
In addition, the non-conservative product in  \eqref{eq:BN-Euler} entails extra difficulties for the numerical simulation. As summarized in \cite{deledicque_exact_2007}, there are several approaches to approximate the integral of the nozzling term $\bm{S}_i^n$, two of which are listed below.  

The first follows the finite-volume method  \cite{andrianov_simple_2003} and the integral of the nozzling term is approximated as
\begin{equation}\label{eq:nozzl-discr}
\bm{S}_i^n\simeq \bm{h}(\bm{u}_i^n)\left((\alpha_s)_{i+\frac{1}{2}}^n-(\alpha_s)_{i-\frac{1}{2}}^n\right),
\end{equation}
where $(\alpha_s)_{i+\frac{1}{2}}^n$ is the exact value of $\alpha$ along the cell interface.
This approach assures that the numerical solutions satisfy the free-streaming condition proposed by Abgrall \cite{abgrall_how_1996}. That is, the   uniformity of  pressure and velocity  remains  during their time evolution.

\begin{figure}[htb]
\begin{center}
\includegraphics[width=0.65\textwidth]{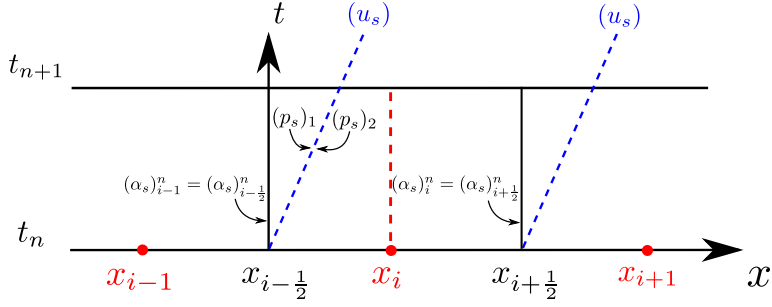}
\caption{Solid contacts in the local Riemannian problems at cell interfaces.}\label{fig:nozzling-cell}
\end{center}
\end{figure}

The other approach was first proposed in  \cite{schwendeman_riemann_2006}, which precisely evaluates the integral of the nozzling term and seems more reasonable.
According to the  interpretation of the Riemann solution in Section \ref{sec:BN-RP}, $(\alpha_s)_x\equiv 0$ in the region of the control volume away from the solid contact.
Suppose $u_s>0$ in the computational domain under consideration. Then the solid contact propagates into a cell from its left interface.
As shown in Figure \ref{fig:nozzling-cell}, the integral of the nozzling term over the control volume $[x_{i-\frac{1}{2}},x_{i+\frac{1}{2}}]\times[t_n, t_{n+1}]$ can be divided into left and right parts,
\begin{equation*}
\bm{S}_i^n= \bm{S}_{i-\frac12,i}^{n} + \bm{S}_{i,i+\frac12}^{n} :=
\frac{1}{\Delta t}\int_{t_n}^{t_{n+1}}\int_{x_{i-\frac{1}{2}}}^{x_i}\bm{h}(\bm{u})(\alpha_s)_x\mathrm{d}x\mathrm{d}t +
\frac{1}{\Delta t}\int_{t_n}^{t_{n+1}}\int_{x_i}^{x_{i+\frac{1}{2}}}\bm{h}(\bm{u})(\alpha_s)_x\mathrm{d}x\mathrm{d}t.
\end{equation*}
The CFL restriction of the time step $\Delta t<\Delta x/(2u_s)$ ensures that the solid contact does not enter into the interval $[x_{i},x_{i+\frac{1}{2}}]$ during  the time interval $[t_n,t_{n+1}]$. 
Then $(\alpha_s)_x\equiv 0$ in this interval implies
\begin{equation*}
\bm{S}_{i,i+\frac12}^{n}=\frac{1}{\Delta t}\int_{t_n}^{t_{n+1}}\int_{x_i}^{x_{i+\frac{1}{2}}}\bm{h}(\bm{u})(\alpha_s)_x\mathrm{d}x\mathrm{d}t=0.
\end{equation*}
According to the essential condition \eqref{eq:BN3-eta}, the integral of the non-conservative product $p_g(\alpha_s)_x$ in $[x_{i-\frac{1}{2}},x_i]$ satisfies 
\begin{equation}\label{eq:non-con-int1}
\begin{aligned}
&\frac{1}{\Delta t}\int_{t_n}^{t_{n+1}}\int_{x_{i-\frac{1}{2}}}^{x_i}p_g(\alpha_s)_x\mathrm{d}x\mathrm{d}t
=\frac{1}{\Delta t}\int_{t_n}^{t_{n+1}}\int_{\xi_L}^{\xi_R} p_g(\alpha_s)_\xi \mathrm{d}\xi \mathrm{d}t\\
=&\frac{1}{\Delta t}\int_{t_n}^{t_{n+1}}\int_{\xi_L}^{\xi_R} (\alpha_s p_s)_\xi \mathrm{d}\xi \mathrm{d}t
=(\alpha_s)_i (p_s)_2-(\alpha_s)_{i-1} (p_s)_1.
\end{aligned}
\end{equation}
Because  $(u_s)_1 = (u_s) _2$, the integrals of other non-conservative products in the nozzling terms in $[x_{i-\frac{1}{2}},x_i]$ are obtained as
\begin{align*}
&\frac{1}{\Delta t}\int_{t_n}^{t_{n+1}}\int_{x_{i-\frac{1}{2}}}^{x_i}p_g u_s(\alpha_s)_x\mathrm{d}x\mathrm{d}t = (u_s)_1\left[ (\alpha_s)_i (p_s)_2-(\alpha_s)_{i-1} (p_s)_1\right],\\
&\frac{1}{\Delta t}\int_{t_n}^{t_{n+1}}\int_{x_{i-\frac{1}{2}}}^{x_i}-u_s(\alpha_s)_x\mathrm{d}x\mathrm{d}t = -(u_s)_1\left[(\alpha_s)_i -(\alpha_s)_{i-1} \right].
\end{align*}
From the formula \eqref{eq:non-con-int1}, it is observed  that the non-conservative product $p_g(\alpha_s)_x$ is approximated by $(p_s\alpha_s)_x$ across the solid contact. This numerical approach is not consistent with the governing equations \eqref{eq:BN-Euler}.
In fact, neither of these approaches is robust enough. Spurious oscillations may arise in the vicinity of porosity jumps (cf. numerical results of \textit{Test 3} in \cite{deledicque_exact_2007}), especially when the solid contact approaches another wave.


\subsection{Well-balancing in capturing  almost  stationary solid contacts}\label{subsec:stati-contact-trouble}

A well-balanced scheme refers to a scheme that accurately preserves specific steady-state solutions. For \eqref{eq:BN-Euler} 
a steady-state solution satisfies,  
\begin{equation}\label{eq:well-balan-eq}
\bm{f}(\bm{u})_x=\bm{h}(\bm{u}) (\alpha_s)_x.
\end{equation}
An individual stationary solid contact, with the solid velocity $u_s\equiv 0$ and a porosity jump, is a typical steady state solution.
The Godunov schemes based on the above two numerical integration approaches for the nozzling term are not necessarily well-balanced since they do not involve  such a solution. For a non-well-balanced scheme, numerical errors are likely to arise from the porosity jump.
Hence  an approach for constructing a well-balanced scheme is necessarily introduced.

\begin{figure}[htb]
\begin{center}
\includegraphics[width=0.6\textwidth]{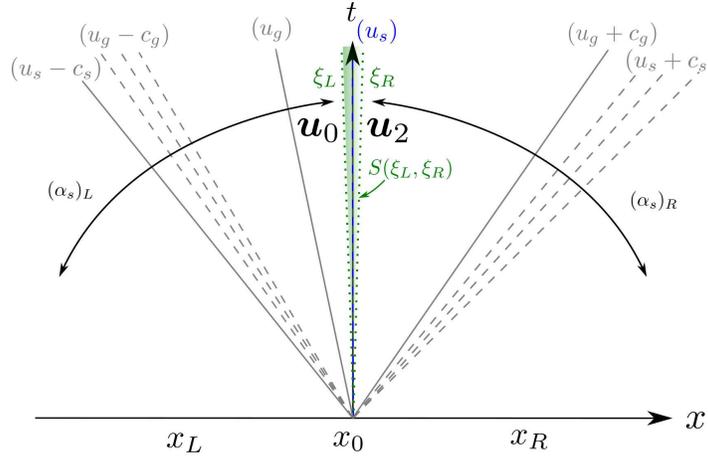}
\end{center}
\caption{The Riemann solution in a small sector covering a stationary solid contact.}\label{fig:0-contact}
\end{figure}
As shown in Figure \ref{fig:0-contact}, we consider a stationary solid contact. In the sectorial region $S(\xi_L,\xi_R)$ bounded by these two rays close to the stationary solid contact, we integrate equation \eqref{eq:well-balan-eq} to get
\begin{equation}\label{eq:duct-eta}
\int_{x_L}^{x_R} \bm{h}(\bm{u})(\alpha_s)_x \mathrm{d}x =\bm{f}(\bm{u}_2)-\bm{f}(\bm{u}_0),
\end{equation}
where $\bm{u}_0$ and $\bm{u}_2$ are the states on the left and right sides of the solid contact, respectively. In general, $\bm{f}(\bm{u}_2)\neq\bm{f}(\bm{u}_0)$ across the porosity jump coinciding with the cell interface.
The inequality of the numerical flux is a difficulty that needs to be settled down in  the numerical simulation of the stationary solid contact.
By integrating system \eqref{eq:BN-Euler} over the interval $(x_{i-\frac{1}{2}},x_{i+\frac{1}{2}})$ from time $t_n$ to $t_{n+1}$, the Godunov scheme for the stationary solid contact becomes
\begin{equation}
\bm{u}_i^{n+1}-\bm{u}_i^n+\frac{\Delta t}{\Delta x}\left(\bm{f}_{i+\frac{1}{2}}^{n,-}-\bm{f}_{i-\frac{1}{2}}^{n,+}\right)=\frac{\Delta t}{\Delta x}\bm{S}_i^n = 0,
\label{eq:solid-fv}
\end{equation}
where the numerical fluxes takes limiting values  along the  inner sides of the interface of cell $I_i$,
\begin{equation*}
\bm{f}_{i-\frac{1}{2}}^{n,+}=\bm{f}\left(\bm{u}(x_{i-\frac{1}{2}}+0,t_n)\right),\quad
\bm{f}_{i+\frac{1}{2}}^{n,-}=\bm{f}\left(\bm{u}(x_{i+\frac{1}{2}}-0,t_n)\right),
\end{equation*}
and $\bm{S}_i^n=0$ due to the fact that  $(\alpha_s)_x=0$ in $(x_{i-\frac{1}{2}},x_{i+\frac{1}{2}})$. 
This method is essentially equivalent to the unsplit  Roe-type wave propagation method in \cite{lowe_two-phase_2005}, which performs a wave decomposition based on the flux difference plus the non-conservative product, inspired by \eqref{eq:duct-eta}.  That is,
\begin{equation*}
\bm{f}_{i-\frac{1}{2}}^{n,+}=\bm{f}_{i-\frac{1}{2}}^{n,-}+\int_{x_{i-1}}^{x_i} \bm{h}(\bm{u})(\alpha_s)_x \mathrm{d}x:=\bm{f}_{i-\frac{1}{2}}^{n,-}+\bm{S}_{i-\frac12}^n,\quad
\bm{f}_{i+\frac{1}{2}}^{n,-}=\bm{f}_{i+\frac{1}{2}}^{n,+}-\bm{S}_{i+\frac12}^n.
\end{equation*}
Hence, \eqref{eq:solid-fv} can be symbolically written in the form   the Godunov scheme  \eqref{eq:BN-Godunov},
\begin{equation*}
\bm{u}_i^{n+1}-\bm{u}_i^n+\frac{\Delta t}{\Delta x}\left(\bm{f}_{i+\frac{1}{2}}^{n,\pm}-\bm{f}_{i-\frac{1}{2}}^{n,\pm}\right)=\frac{\Delta t}{\Delta x}\bm{S}_{i\pm\frac12}^n.
\end{equation*}
This scheme is not robust when the solid velocity $u_s$ is extremely close  but not equal to $0$, since it is hard to take out which numerical fluxes $\bm{f}_{i+\frac{1}{2}}^{n,\pm}$ and integrals of the nozzling term $\bm{S}_{i\pm\frac12}^n$.
Later on, we design a staggered method to enhance robustness. The principle of this method is to set porosity jumps inside cells rather than at cell interfaces. Then, numerical fluxes must be invariant across cell interfaces, and the integral of the nozzling term can be evaluated precisely.
Alternatively, we may use the residual framework to resolve this issue, as done in \cite{Ricchiuto-2015}, which is left for the future work. 

\begin{rem}
The 1-D homogeneous BN two-phase model is readily compared with the quasi-1-D compressible duct flow model \cite{andrianov_riemann_2004,warnecke_solution_2004}.
If the solid phase is assumed incompressible $(\rho_s)_t = 0$ and stationary $u_s\equiv 0$,  which means that  the granular bed does not move,
the porosity becomes a function only associated with $x$.
The solid phase fits the property of a duct of variable cross-section, and the gas phase formally satisfies a duct flow model. 
Here, the porosity $\alpha_g(x)$ acts as the variable cross-section $A(x)$ in the duct flow.
Using a one-to-one correspondence
$(\alpha_g,\rho_g,u_g,p_g,E_g) \longleftrightarrow (A,\rho,u,p,E)$ by removing the subscripts $g$ for specification compliance, system \eqref{eq:BN-Euler} reduces to  the system of compressible  Euler equations in a duct
\begin{equation}\label{eq:duct-Euler}
\bm{u}_t+\bm{f}(\bm{u})_x=\bm{h}(\bm{u})A_x,
\end{equation}
where notations $\bm{u}$, $\bm{f(u)}$ and $\bm{h(u)}$   represent 
\begin{equation*}
\bm{u}=
\begin{bmatrix}
A\\
A \rho\\
A \rho u\\
A \rho E
\end{bmatrix},\quad
\bm{f}(\bm{u})=
\begin{bmatrix}
0\\
A \rho u\\
A(\rho u^2+p)\\
Au(\rho E+p)
\end{bmatrix},\quad
\bm{h}(\bm{u})=
\begin{bmatrix}
0\\
0\\
p\\
0
\end{bmatrix}.
\end{equation*}
Eigenvalues of this reduced system are
\begin{equation*}
\lambda_0=0,\ \lambda_1=u-c,\ \lambda_2=u,\ \lambda_3=u+c.
\end{equation*}
In the solution to the Riemann problem for the  duct flow model, the $0$-characteristic field corresponds to the stationary solid contact, called $0$-contact.
The Riemann invariants across the $0$-contact are
\begin{equation*}
A\rho u, \quad \eta, \quad h+\frac{u^2}{2},
\end{equation*}
which are actually some Riemann invariants across the stationary solid contact.
In practice, it is essentially the same that solving the duct flow model as doing the homogeneous BN model with the stationary solid phase.
\end{rem}


\subsection{Dilemma in the computation of moving solid contacts}\label{subsec:solid-contact-oscillation}


Some numerical experiments in \cite{andrianov_riemann_2004,lowe_two-phase_2005,karni_hybrid_2010} revealed the fact that spurious oscillations are likely to arise near porosity jumps by using conservative shock-capturing schemes and cannot eliminated completely. In this subsection, we discuss the mechanism of these spurious oscillations.
For  a single solid contact, the Riemann invariants $\psi$, $\psi=u_s, \eta_g, Q, P, H$ are constant across porosity jumps.
The Godunov average of conservative variables may yield wrongly the computation of  all of these Riemann invariants. Therefore, visible errors inevitably generate from porosity interfaces. This observation is verified  in the following.

\begin{figure}[htb]
\centering
\includegraphics[width=0.45\textwidth]{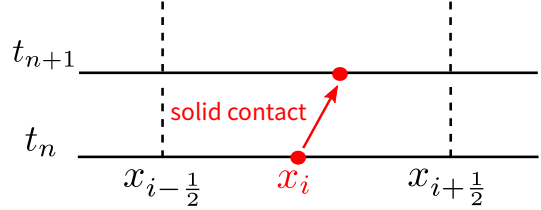}
\caption{Propagation of a solid contact in a time step.}\label{fig:single-solid-time}
\end{figure}

As shown in Figure \ref{fig:single-solid-time}, we consider a single  solid contact  within the cell $I_i$ propagating to the right from time $t_n$ to $t_{n+1}$. Assume the solid velocity remains constant $(u_s)_0>0$ throughout the computational domain.
Using  the conservative Godunov scheme \eqref{eq:BN-Godunov}, the conservation of mass, total momentum and total energy of the two phases should be maintained. Thus we have no choice but to evolve four conservative variables with  inalterable computational formulae.
These conservative variables are the density of each phase $\alpha_s \rho_s$ and $\alpha_g \rho_g$, total momentum $\mathcal{M}:=\alpha_s \rho_s u_s + \alpha_g \rho_g u_g$ and total energy $\mathcal{E}:=\alpha_s \rho_s E_s + \alpha_g \rho_g E_g$. Specifically, we have
\begin{align*}
&(\alpha_s \rho_s)_{i}^{n+1}=(\alpha_s \rho_s)_{i}^n - \frac{\Delta t}{\Delta x} \delta_x(\alpha_s \rho_s u_s)_i^n,\\
&(\alpha_g \rho_g)_{i}^{n+1}=(\alpha_g \rho_g)_{i}^n - \frac{\Delta t}{\Delta x} \delta_x(\alpha_g \rho_g u_g)_i^n,\\
&\mathcal{M}_{i}^{n+1}=\mathcal{M}_{i}^n - \frac{\Delta t}{\Delta x} \delta_x (P+Q u_s+\mathcal{M}u_s)_i^n,
\end{align*}
where the notation of the central difference $\delta_x (\bullet)_i^n=(\bullet)_{i+\frac{1}{2}}^n-(\bullet)_{i-\frac{1}{2}}^n$ is used.
Since the Riemann invariants $u_s$, $P$ and $Q$ are constant across the solid contact, $\delta_x (P + Qu_s)_i^n=0$ holds. 
Then
\begin{equation}\label{eq:Q-inv-cal}
\begin{aligned}
Q_{i}^{n+1}=&[\alpha_g \rho_g (u_g-u_s)]_{i}^{n+1} = \mathcal{M}_{i}^{n+1} - \left[(\alpha_s \rho_s)_{i}^{n+1} + (\alpha_g \rho_g)_{i}^{n+1}\right] (u_s)_0\\
=&\mathcal{M}_{i}^n - \left[(\alpha_s \rho_s)_{i}^{n} + (\alpha_g \rho_g)_{i}^{n}\right](u_s)_0=[\alpha_g \rho_g (u_g-u_s)]_{i}^{n}=Q_{i}^n,
\end{aligned}
\end{equation}
Therefore the Godunov scheme maintains the Riemann invariant $Q$ constant across the solid contact.


In addition to $u_s$ and $Q$,  other three Riemann invariants should remain constant across the solid contact too,  
\begin{equation}\label{eq:deter-var1}
(\eta_g)_i^{n+1}=(\eta_g)_i^{n},\quad
P_i^{n+1}=P_i^{n},\quad
H_i^{n+1}=H_i^{n}.
\end{equation}
By the conservation property of the Godunov scheme, we determine four conservative variables at time $t_{n+1}$,
\begin{equation}\label{eq:deter-var2}
(\alpha_s \rho_s)_i^{n+1},\quad
(\alpha_g \rho_g)_i^{n+1},\quad
\mathcal{M}_i^{n+1},\quad
\mathcal{E}_i^{n+1}.
\end{equation}
These seven quantities in \eqref{eq:deter-var1} and \eqref{eq:deter-var2} are independent, which will be illustrated later.
However, except for the aforementioned solid velocity $u_s$, there are only six undetermined primitive variables  in \eqref{eq:BN-Euler} with the closure of the EOS for each phase and the saturation constraint, 
\begin{equation}\label{eq:prim-var}
(\alpha_s)_i^{n+1},\quad (\rho_s)_i^{n+1},\quad (p_s)_i^{n+1},\quad (\rho_g)_i^{n+1}\quad (u_g)_i^{n+1},\quad (p_g)_i^{n+1}.
\end{equation}
The number of undetermined variables, six,  is less than the number of independent  determined quantities, seven, which results in an overdetermined problem.
This dilemma shows that it is impossible to eliminate numerical errors by an  conservative scheme in the vicinity of porosity jumps completely.

Take Example \ref{ex:test-1} in Section \ref{sec:ex-BN}  as an example.  Assume the  initial data to generate the solution of  a single solid contact propagating to the right and initially resting on the left boundary of the cell $I_i, i=151$. We set the spatial step $\Delta x=1/300$ and time step $\Delta t=1/2400
$. At time $t_1=\Delta t$, we can work out primitive variables in \eqref{eq:prim-var} based on all constant Riemann invariants in \eqref{eq:deter-var1} and conservative variables $(\alpha_s \rho_s)_i^{n+1}, (\alpha_g \rho_g)_i^{n+1}, \mathcal{M}_i^{n+1}$, and  the total energy $\mathcal{E}_i^{n+1}=(\alpha_s \rho_s E_s + \alpha_g \rho_g E_g)_i^{n+1}=10.4106$.
Besides, we can evaluate the total energy using  the conservative Godunov scheme \eqref{eq:BN-Godunov},
\begin{equation*}
\mathcal{E}_{i}^{n+1}=\mathcal{E}_{i}^n - \frac{\Delta t}{\Delta x} \delta_x [\alpha_s u_s(\rho_s E_s+p_s) + \alpha_g u_g(\rho_g E_g+p_g) ]_i^n,
\end{equation*}
which updates the total energy  to be $\mathcal{E}_i^{n+1}=10.4015$, not equal to the above value.
Therefore, we conclude that  the complete conservation in the Godunov scheme, including the conservation of total energy, yields a contradiction to the 
 fact  that all Riemann invariants for the $\lambda_0$-field remain unchanged across the solid contact.
Spurious oscillations near porosity jumps cannot be fully suppressed by adjusting the numerical integration approaches for the nozzling term  \cite{deledicque_exact_2007}.
The only possible way to eliminate oscillations  is to release the conservation,  most  reasonably  the conservation of total energy.
In  Section \ref{sec:Stag-Godunov}, a projection method based on Riemann invariants is developed to implement the release of conservation.

\subsection{Coalescence of gaseous shock and solid contact}\label{subsec:coalescence}


It is observed that  spurious oscillations become more violent as the solid contact approaches other strong discontinuities, which 
corresponds to a resonance phenomenon.
Example \ref{ex:test-3} in Section \ref{sec:ex-BN}  shows an instance of contiguous gaseous shock and solid contact with porosity jump.
The numerical results in \cite{andrianov_riemann_2004} displayed oscillations in the vicinity of the porosity interface.
The proximity of the solid contact and gaseous shock is likely to be the cause.

\begin{figure}[htb]
\begin{center}
\includegraphics[width=0.65\textwidth]{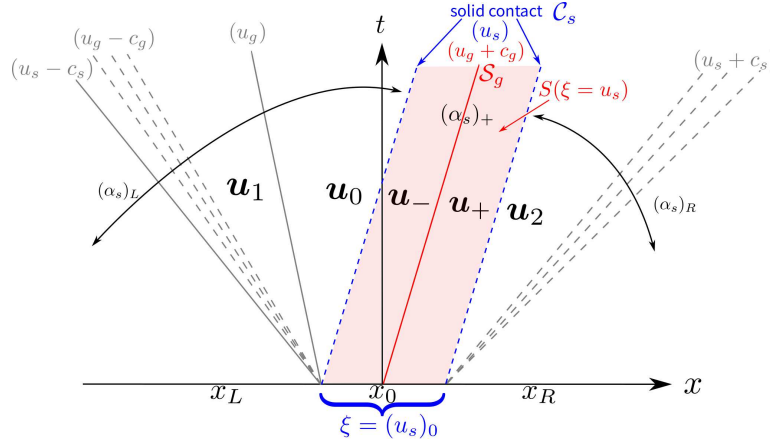}
\end{center}
\caption{The Riemann solution with the coalescence of gaseous shock and solid contact.}\label{fig:shock-contact}
\end{figure}

Let us consider a Riemann problem related to the  coalescent gaseous shock and solid contact.
As shown in Figure \ref{fig:shock-contact}, the Riemann solution includes a transonic solid contact coinciding with a gaseous shock, i.e., a resonant wave associated with  $\lambda_{3,g} = \lambda_0$.
Inspired by the research on duct flows in \cite{han_exact_2012}, the resonant wave consists of three connected waves: a subsonic solid contact $\mathcal{C}_s(\bm{u}_0,\bm{u}_-)$ with the left state $\bm{u}_0$ and right state $\bm{u}_-$, a gaseous shock $\mathcal{S}_g(\bm{u}_-,\bm{u}_+)$ and a supersonic solid contact $\mathcal{C}_s(\bm{u}_+,\bm{u}_2)$.
These three waves coalesce on the ray $\xi=(u_s)_0$, viewed as a region $S(\xi=u_s)$, where the intermediate solid volume fraction $(\alpha_s)_-=(\alpha_s)_+$ represents the porous location of the gaseous shock.
Recall in  Section \ref{sec:BN-RP} that the expression of the integral condition \eqref{eq:BN3-eta} is the same.
However, we have to grasp the resonant wave with great accuracy;  otherwise probably visible errors arise in the intermediate solid pressure $(p_s)_1$ or $(p_s)_2$. As a matter of fact, this difficulty is hard to be avoided because of round-off errors. It is almost impossible to judge whether the wave speeds are idential. In other words, the solid contact is transonic.

As indicated  in \cite{liu_nonlinear_1987}, the resonance phenomenon induces instability of the flow field. Such a Riemann problems are extremely difficult to solve. A straightforward idea is to divide these resonant waves through grid cells and resolve them separately. In the next section, we implement this idea via the strategy of   staggered gas-solid grids, as illustrated in  Figure \ref{fig:BN-stag-cell1}.

\section{Staggered-projection Godunov-type schemes}\label{sec:Stag-Godunov}


In 2007, Saurel {\em et al.} developed a Lagrange-projection method in \cite{saurel_relaxation-projection_2007}. This method is utilized to suppress spurious pressure oscillations that appear at contact discontinuities for compressible flows governed by complex gas EOS.
Inspired by the similarity between this problem and the one we studied, we explore the application of the Lagrange-projection method in order to suppress spurious oscillations that appear at solid contacts.
The first step of the Lagrange-projection method rests upon the   Lagrangian formulation.
The second step is the projection of the solution on a fixed (Eulerian) grid.
The homogeneous BN model contains two phases, which undoubtedly make it difficult to solve in Lagrangian coordinates.
To make the numerical method as simple as possible, at the first step, we compute the solid phase in Lagrangian coordinates and the gas phase in Eulerian coordinates.
Then, the question is how to design a gas-solid grid as the basis for the projection on the grid.

For the coalescence of solid contacts and other gaseous waves discussed in the previous section, the difficulty of the resonance can be circumvented by separating solid contacts from other gaseous waves.
By adjusting the Godunov scheme, we assume solid contacts with porosity jumps only appear at cell centers of fixed grid cells, while other waves generate from the boundaries of these cells. 
Such fixed cells are identified as gaseous grid cells.
The cell centers of these  gaseous cells can be regarded as boundaries of a column of solid grid cells.
Then this hypothesis happens to construct a staggered gas-solid grid.
The Riemann problems at gaseous cell interfaces without  porosity jumps become much easier to solve. 
Meanwhile, discontinuous interfacial fluxes commented in Subsection \ref{subsec:stati-contact-trouble} no longer arise and the integrals of the nozzling term over gaseous cells can be evaluated accurately based on the formula \eqref{eq:BN3-eta}. Therefore such a modified Godunov method is well-balanced. We improve the numerical integration approach to be consistent with the governing equations \eqref{eq:BN-Euler}.

In \cite{karni_hybrid_2010}, Karni and Hern\'andez-Due\~nas had exemplified that the non-conservative scheme can suppress spurious oscillations near porosity interfaces.
However, the conservation property is necessary for schemes to capture shocks.
It is difficult to distinguish shocks and porosity jumps, mainly for the two waves coinciding or interacting with each other.
Therefore, it is not easy to choose the conservative or non-conservative scheme in some parts of the computational domain.
The Karni method is to determine locations of porosity jumps by using the gradient of the porosity.
The non-conservative method is applied near porosity jumps.
In this section, we design a projection method automatically switching between conservative and non-conservative formulae.
It maintains a unified formulation over the entire computational domain, no matter  whether the porosity is discontinuous or not.
This method projects solid contacts in Lagrangian coordinates back to Eulerian coordinates.
The mass of each phase,  total momentum, and Riemann invariants for the $\lambda_0$- field keep constant during the projection.
The correct use of  Riemann invariants suppress spurious oscillations. 
The total energy is non-conservative only at porosity jumps, and conservation errors disappear gradually as jumps get smaller.
Thus the conservation of the scheme can be well guaranteed in cells containing weak solid contacts but strong shocks.



\subsection{Algorithmic steps of a staggered-projection scheme}

In this section, a staggered-projection Godunov-based method is proposed with detailed justification.  It consists of four steps that will be specified below. For convenience of numerical  implementation, the algorithm is diagramed at the end of this section. 

\subsubsection{Initial states within gas-solid staggered cells}

\begin{figure}[htb]
\begin{center}
\includegraphics[width=0.65\textwidth]{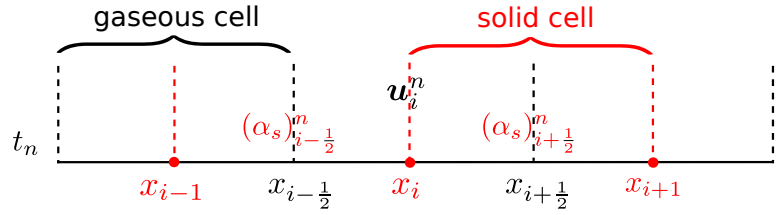}
\end{center}
\caption{ Staggered gas-solid grid cells.}\label{fig:BN-stag-cell}
\end{figure}

As shown in Figure \ref{fig:BN-stag-cell}, a black regular grid,  a {\em gaseous grid},  is fixed.
It is assumed that a porosity jump is  only present  at the center $x=x_{i}$ of each gaseous grid cell $I_i=[x_{i-\frac12},x_{i+\frac12}]$.
Red points of porosity jumps inside gaseous cells can be treated as interfaces of {\em solid grid} cells.
For the Godunov scheme, the initial data are assumed to satisfy a piece-wise constant distribution at time $t_n$. The solid volume fraction $\alpha_s$ takes constants $(\alpha_s)_{i-\frac12}^n$ and $(\alpha_s)_{i+\frac12}^n$ in solid cells $[x_{i-1},x_i]$ and $[x_i,x_{i+1}]$, respectively. The vector $\bm{u}$ is constant $\bm{u}_{i-\frac12,i}^n$ and $\bm{u}_{i,i+\frac12}^n$ in the smaller intervals $[x_{i-\frac12},x_i]$ and $[x_i,x_{i+\frac12}]$, respectively.
Then, the cell average of $\bm{u}$ over  $I_i$
is denoted as $\bm{u}_i^n=\frac12\left(\bm{u}_{i-\frac12,i}^n+\bm{u}_{i,i+\frac12}^n\right)$. We suppose that there is only one wave at the porosity jump, i.e.,  the solid contact emanating at $x=x_i$.
The Riemann invariants $\psi$, $\psi=u_s, \eta_g, Q, P, H$, associated with  $\lambda_0$-field,  are constant across the solid contact.
Then $\psi_{i-\frac12,i}^n=\psi_{i,i+\frac12}^n=\psi_i^n$ in the gaseous cell $I_i$.
In addition, we make another hypothesis that the solid density $\rho_s$ is constant in each gaseous cell.
It means that $(\rho_s)_{i-\frac12,i}^n=(\rho_s)_{i,i+\frac12}^n=(\rho_s)_{i}^n$ and initial discontinuities of the solid density only appear at gaseous cell interfaces.
Based on this hypothesis, solid contact waves  associated with the $\lambda_0$, $\lambda_{2,s}$- fields are separated. The solid contact $\lambda_0$-waves generate at centers of gaseous cells, and the solid contact $\lambda_{2,s}$-waves generate at the boundaries of gaseous cells.
Hence, system \eqref{eq:BN-Euler} is  decoupled  into two Euler equations for the gas and solid phases at gaseous cell interfaces.
The solid phase does not affect the gas phase, and gaseous parameters do not change across solid contacts.
Thus, the local Riemann problem $\textbf{RP}\left(\xi; \bm{u}_{i,i+\frac12}^n,\bm{u}_{i+\frac12,i+1}^n\right)$ at the gaseous cell interface $x_{i+\frac12}$ is easier to be solved than  the Riemann problem in Section \ref{sec:BN-RP}.

\begin{figure}[htb]
\begin{center}
\includegraphics[width=0.5\textwidth]{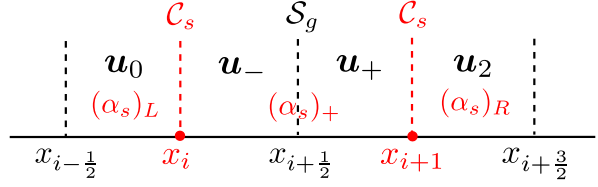}
\end{center}
\caption{Coalescence of gaseous shock and solid contact in staggered cells.}\label{fig:BN-stag-cell1}
\end{figure}

\begin{rem}
For the coalescence of gaseous shock and solid contact in Figure \ref{fig:shock-contact}, the staggered gas-solid grid can decompose the resonance wave.
Through an ideal schematic diagram in Figure \ref{fig:BN-stag-cell1}, the resonant wave is divided into three waves: $\mathcal{C}_s(\bm{u}_0,\bm{u}_-)$ at the solid cell interface $x=x_{i}$, $\mathcal{S}_g(\bm{u}_-,\bm{u}_+)$ at the gaseous cell interface $x=x_{i+\frac12}$ and $\mathcal{C}_s(\bm{u}_+,\bm{u}_2)$ at $x=x_{i+1}$. In this way, these three waves can be solved individually.
\end{rem}

\subsubsection{A Godunov scheme over gaseous cells}

The Godunov scheme \eqref{eq:BN-Godunov} for system \eqref{eq:BN-Euler} is utilized to update the cell average $\bm{u}_i^n$ over the cell $I_i$. The numerical flux $f_{i+\frac12}^n$ is obtained by solving the local Riemann problem $\textbf{RP}\left(\xi;\bm{u}_{i,i+\frac12}^n,\bm{u}_{i+\frac12,i+1}^n\right)$ at the cell interface $x_{i+\frac12}$. The time step $\Delta t$ is restricted by the CFL condition to  avoid wave interactions.

Next  the integral of the nozzling term is evaluated.
Note that solid contacts at centers of gaseous cells propagate along fluid trajectories with the velocity $(u_s)_i^n$.
We set 
\begin{equation*}
(p_g)_{\max}=\max\left\{(p_g)_{i-\frac12,i}^n,(p_g)_{i,i+\frac12}^n\right\},\quad (p_g)_{\min}=\min\left\{(p_g)_{i-\frac12,i}^n,(p_g)_{i,i+\frac12}^n\right\}
\end{equation*}
as the maximum and minimum values of gaseous pressure in $I_i$, respectively.
By the mean-value theorem, there exists a $(\overline{p_g})_i^{n} \in \left[(p_g)_{\min},(p_g)_{\max}\right]$ such that  there holds
\begin{equation}\label{eq:source-stag}
\frac{1}{\Delta t}\int_{t_n}^{t_{n+1}}\int_{x_{i-\frac{1}{2}}}^{x_{i+\frac{1}{2}}}p_g(\alpha_s)_x\mathrm{d}x\mathrm{d}t
=\frac{1}{\Delta t}\int_{t_n}^{t_{n+1}}(\overline{p_g})_i^{n}\int_{\xi_L}^{\xi_R} (\alpha_s)_\xi\mathrm{d}\xi\mathrm{d}t
=(\overline{p_g})_i^{n} \left((\alpha_s)_{i+\frac12}^n-(\alpha_s)_{i-\frac12}^n\right),
\end{equation}
where $\xi=(x-x_i)/(t-t_n)$, $\xi_L=(u_s)_i^n-\epsilon$ and $\xi_R=(u_s)_i^n+\epsilon$.
For smooth solutions, this approximation  is consistent with  first-order accuracy.

On the other hand, recall that local solutions across solid contacts satisfy the system \eqref{eq:BN-eta}. Then  we have
\begin{equation*}
\begin{aligned}
&\frac{1}{\Delta t}\int_{t_n}^{t_{n+1}}\int_{x_{i-\frac{1}{2}}}^{x_{i+\frac{1}{2}}}p_g(\alpha_s)_x\mathrm{d}x\mathrm{d}t
=\frac{1}{\Delta t}\int_{t_n}^{t_{n+1}}\int_{\xi_L}^{\xi_R} p_g(\alpha_s)_\xi \mathrm{d}\xi \mathrm{d}t\\
=&\frac{1}{\Delta t}\int_{t_n}^{t_{n+1}}\int_{\xi_L}^{\xi_R}  (\alpha_s p_s)_\xi \mathrm{d}\xi \mathrm{d}t
=(\alpha_s)_{i+\frac12}^n (p_s)_{i,i+\frac12}^n-(\alpha_s)_{i-\frac12}^n (p_s)_{i-\frac12,i}^n.
\end{aligned}
\end{equation*}
Comparison between these two evaluations yields 
\begin{equation*}
(\overline{p_g})_i^n= \frac{(\alpha_s)_{i+\frac12}^n (p_s)_{i,i+\frac12}^n-(\alpha_s)_{i-\frac12}^n (p_s)_{i-\frac12,i}^n}{(\alpha_s)_{i+\frac12}^n-(\alpha_s)_{i-\frac12}^n}=:\mathcal{P}_g,
\end{equation*}
for $(\alpha_s)_{i+\frac12}^n\neq (\alpha_s)_{i-\frac12}^n$.
Since $(\overline{p_g})_i^n\in \left[(p_g)_{\min},(p_g)_{\max}\right]$ holds,
a  computation method for $\overline{p_g}$ is proposed as 
\begin{equation}\label{eq:cal-pg}
(\overline{p_g})_i^n=\left\{
\begin{aligned}
&\frac12\left[(p_g)_{i-\frac12,i}^n+(p_g)_{i,i+\frac12}^n\right], && \text{ if } \left|(\alpha_s)_{i+\frac12}^n - (\alpha_s)_{i-\frac12}^n\right|<\varepsilon,\\
&(p_g)_{\min}, && \text{ if } \mathcal{P}_g < (p_g)_{\min},\\
&(p_g)_{\max}, && \text{ if } \mathcal{P}_g > (p_g)_{\max},\\
&\mathcal{P}_g, && \text{ otherwise,}
\end{aligned}\right.
\end{equation}
for small positive number $\varepsilon$, typically $\varepsilon=10^{-6}$.
Then  the full set of integrals for non-conservative products are summarized as, in addition to  \eqref{eq:source-stag},
\begin{equation*}
\frac{1}{\Delta t}\int_{t_n}^{t_{n+1}}\int_{x_{i-\frac12}}^{x_{i+\frac{1}{2}}}p_g u_s(\alpha_s)_x\mathrm{d}x\mathrm{d}t = (\overline{p_g})_i^n (u_s)_i^n \left[(\alpha_s)_{i+\frac12}^n-(\alpha_s)_{i-\frac12}^n\right].
\end{equation*}
In addition, 
\begin{equation*}
\frac{1}{\Delta t}\int_{t_n}^{t_{n+1}}\int_{x_{i-\frac{1}{2}}}^{x_{i+\frac12}}-u_s(\alpha_s)_x\mathrm{d}x\mathrm{d}t = -(u_s)_i^n \left[(\alpha_s)_{i+\frac12}^n-(\alpha_s)_{i-\frac12}^n\right].
\end{equation*}
In the case of large porosity jumps, the integral of the nozzling term $\bm{S}_i^n$ is computed almost exactly.
In the absence of porosity jumps, $(\alpha_s)_{i+\frac12}^n=(\alpha_s)_{i-\frac12}^n$, the numerical integral $\bm{S}_i^n$ is identical to $\bm{0}$.
In this case, this Godunov method reduces to the standard Godunov scheme about the Euler equations for the gas or  solid phase.

\begin{figure}[!htb]
\begin{center}
\includegraphics[width=0.7\textwidth]{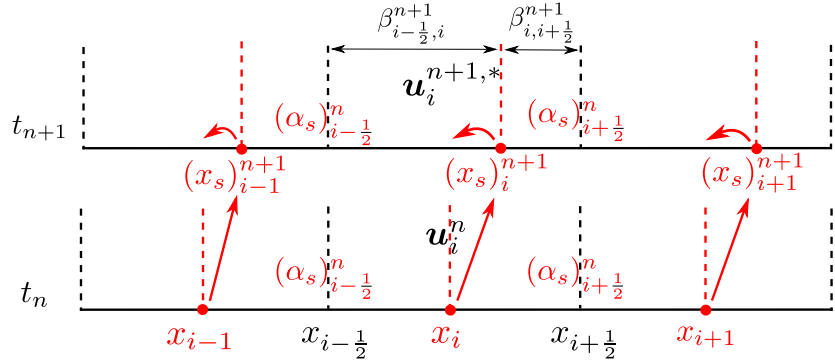}
\caption{Moving solid contacts and solution distribution  over time level $[t_n, t_{n+1}]$.}\label{fig:BN-stag-cell2}
\end{center}
\end{figure}

\subsubsection{Moving solid contacts and limiting states on both sides}
Denote a solid contact inside $I_i$ as 
\begin{equation}\label{eq:cal-xs}
(x_s)_i(t) =x_i+(u_s)_i^n ( t-t_n), \ \ \ \ t_n\leq t \leq t_{n+1}, 
\end{equation}
as shown in Figure \ref{fig:BN-stag-cell2}, in which 
\begin{equation}\label{eq:cal-beta}
\beta_{i-\frac12,i}^{n+1}=\frac{(x_s)_i^{n+1}-x_{i-\frac12}}{x_{i+\frac12}-x_{i-\frac12}},\quad
\beta_{i,i+\frac12}^{n+1}=\frac{x_{i+\frac12}-(x_s)_i^{n+1}}{x_{i+\frac12}-x_{i-\frac12}},
\end{equation}
represent the proportions of the interval $[x_{i-\frac12},(x_s)_i^{n+1}]$ and $[(x_s)_i^{n+1},x_{i+\frac12}]$ relative to the length of  the  whole cell $I_i$, respectively. The  average of $\bm{u}$ over the intervals $[x_{i-\frac12},(x_s)_i^{n+1}]$ and $[(x_s)_i^{n+1},x_{i+\frac12}]$ are denoted as $\bm{u}_{i-\frac12,i}^{n+1,*}$ and $\bm{u}_{i,i+\frac12}^{n+1,*}$, respectively.
Then the cell average of $\bm{u}$ over the cell $I_i$ at time $t_{n+1}$ is
\begin{equation}\label{eq:BN-stag-cell-CON-EQ}
\bm{u}_i^{n+1,*} = \beta_{i-\frac12,i}^{n+1} \bm{u}_{i-\frac12,i}^{n+1,*} + \beta_{i,i+\frac12}^{n+1} \bm{u}_{i,i+\frac12}^{n+1,*},
\end{equation}
which is the same as that obtained directly by the Godunov scheme \eqref{eq:BN-Godunov}, 
\begin{equation}\label{eq:BN-Godunov-tmp}
\bm{u}_i^{n+1,*}=\bm{u}_i^n-\frac{\Delta t}{\Delta x}\left(\bm{f}_{i+\frac{1}{2}}^n-\bm{f}_{i-\frac{1}{2}}^n\right)+\frac{\Delta t}{\Delta x}\bm{S}_i^n.
\end{equation}
 The solid volume fraction $\alpha_s$ in the scheme needs to be solved by a Lagrangian method.
The advection of the solid volume fraction indicates 
\begin{equation}\label{eq:alpha-s-result}
(\alpha_s)_{i-\frac12,i}^{n+1,*}=(\alpha_s)_{i-\frac12}^n,\quad
(\alpha_s)_{i,i+\frac12}^{n+1,*}=(\alpha_s)_{i+\frac12}^n.
\end{equation}
In light of the continuity of  Riemann invariants associated with  the $\lambda_0$- field across the solid contact, we have 
\begin{equation}\label{eq:BN-stag-cell-RI-EQ}
\psi_{i-\frac12,i}^{n+1,*}=\psi_{i,i+\frac12}^{n+1,*},\quad \psi=u_s, \eta_g, Q, P, H.
\end{equation}
It is observed that the density  $\rho_s$ is not among  these Riemann invariants, thus the solid density acts as a free variable.
Then we  assume that the discontinuity of the solid  density only appears at cell interfaces of the cell $I_i$, implying
\begin{equation}\label{eq:rho-s-suppose}
(\rho_s)_{i-\frac12,i}^{n+1,*}=(\rho_s)_{i,i+\frac12}^{n+1,*}=(\rho_s)_i^{n+1,*}=(\alpha_s \rho_s)_i^{n+1,*}/(\alpha_s)_i^{n+1,*}.
\end{equation}
 The Riemann invariant $u_s$ is calculated as
\begin{equation}\label{eq:u-s-result}
(u_s)_{i-\frac12,i}^{n+1,*}=(u_s)_{i,i+\frac12}^{n+1,*}=(u_s)_i^{n+1,*}=(\alpha_s \rho_s u_s)_i^{n+1,*}/(\alpha_s \rho_s)_i^{n+1,*}.
\end{equation}
In view of  \eqref{eq:Q-inv-cal},  the Riemann invariant $Q$ remains constant across the solid contact and is obtained through
\begin{equation}\label{eq:Q-result}
\begin{aligned}
Q_{i-\frac12,i}^{n+1,*}&=Q_{i,i+\frac12}^{n+1,*}=Q_{i}^{n+1,*}\\
&= \mathcal{M}_{i}^{n+1,*} - \left[(\alpha_s \rho_s)_{i}^{n+1,*} + (\alpha_g \rho_g)_{i}^{n+1,*}\right] (u_s)_i^{n+1,*}.
\end{aligned}
\end{equation}
Therefore, the Godunov scheme resolves variables $\alpha_s,\rho_s,u_s,Q$ in intervals $[x_{i-\frac12}, (x_s)_i^{n+1}]$ and $[(x_s)_i^{n+1}, x_{i+\frac12}]$.
Recalling $\alpha_g=1-\alpha_s$ and making a variable substitution $u_g=Q/(\alpha_g \rho_g)+u_s$, it remains to  solve  the equations
\eqref{eq:BN-stag-cell-CON-EQ} and 
\eqref{eq:BN-stag-cell-RI-EQ}. 
Some of them are
\begin{subequations}\label{eq:non-linear-system}
\begin{align}
&\mathcal{U}_i^{n+1,*} = \beta_{i-\frac12,i}^{n+1} \mathcal{U}_{i-\frac12,i}^{n+1,*} + \beta_{i,i+\frac12}^{n+1} \mathcal{U}_{i,i+\frac12}^{n+1,*},&&\  
\mathcal{U}=\alpha_g \rho_g, \alpha_g \rho_g E_g,\\
&\psi_{i-\frac12,i}^{n+1,*}=\psi_{i,i+\frac12}^{n+1,*},&&\ 
\psi=\eta_g, H,
\end{align}
\end{subequations}
with four unknown thermodynamic variables of the gas phase
$(\rho_g)_{i-\frac12,i}^{n+1,*},(p_g)_{i-\frac12,i}^{n+1,*}$
and
$(\rho_g)_{i,i+\frac12}^{n+1,*},(p_g)_{i,i+\frac12}^{n+1,*}$.
The nonlinear four-equation algebraic system \eqref{eq:non-linear-system} may be solved, say, by the Newton-Raphson iterative method.
Sometimes, the Newton-Raphson method does not converge or the nonlinear system has no solution. This situation is closely related to the case that the Riemann problem has no solution.
As described in \cite{deledicque_exact_2007}, the conditions for which there is no solution are not special cases but do arise often.
For the situations that the Newton-Raphson method is not applicable, the least-square method is applied to solve the nonlinear system approximately. See \ref{app-nonlinear} for details.
Next, we solve solid pressures $(p_s)_{i-\frac12,i}^{n+1,*}$ and $(p_s)_{i,i+\frac12}^{n+1,*}$ by the two equations
\begin{subequations}\label{eq:non-linear-system2}
\begin{align}
&(\alpha_s \rho_s E_s)_i^{n+1,*} = \beta_{i-\frac12,i}^{n+1} (\alpha_s \rho_s E_s)_{i-\frac12,i}^{n+1,*} + \beta_{i,i+\frac12}^{n+1} (\alpha_s \rho_s E_s)_{i,i+\frac12}^{n+1,*},\\
&P_{i-\frac12,i}^{n+1,*}=P_{i,i+\frac12}^{n+1,*}.
\end{align}
\end{subequations}
For the special case where both phases meet the EOS for polytropic gases, these two equations are linear and easy to solve.
Through the above processes, the Riemann invariants $P_{i-\frac12,i}^{n+1,*}$, $H_{i-\frac12,i}^{n+1,*}$, $(\eta_g)_{i-\frac12,i}^{n+1,*}$ are evaluated.

Across an individual solid contact, all Riemann invariants for the $\lambda_0$-field are computed exactly, suppressing  spurious oscillations.
This step  not only relies on the conservative formulation but also legitimately determines the distribution of Riemann invariants.

\subsubsection{Projection of Riemann invariants and computation of volume fractions}

\begin{figure}[htb]
\begin{center}
\includegraphics[width=0.7\textwidth]{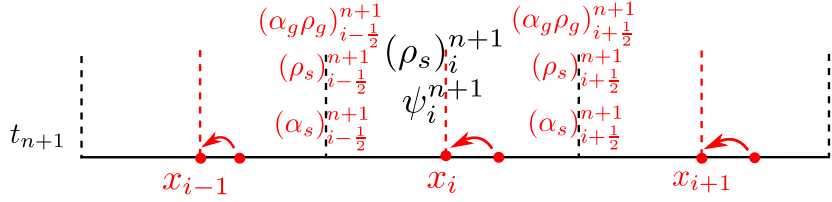}
\caption{Projection of the solution and variable distribution in solid cells at time $t_n$.}\label{fig:BN-stag-cell3}
\end{center}
\end{figure}

At this step, a basic requirement is that  the solid contact  does not cross gaseous cell interfaces. For this purpose,  solid contacts are fixed in gaseous cell centers and the solid grid is an Eulerian grid, inspiring a projection procedure on the gas-solid Eulerian grid.
In order to suppress spurious oscillations in the vicinity of porosity interfaces, all Riemann invariants associated with  the $\lambda_0$-field remain unchanged after the projection.
In other words, the projection is based on a target that all Riemann invariants $\psi=u_s, \eta_g, Q, P, H$ keep constant $\psi_i^{n+1}=\psi_{i-\frac12,i}^{n+1,*}$ in the cell $I_i$. 
According to the  assumption \eqref{eq:rho-s-suppose} on the solid density $\rho_s$, 
the projected solid density  is constant $(\rho_s)_i^{n+1}=(\rho_s)_i^{n+1,*}$.
In fact, the projected solution can be regarded as the  authentic solution at time $t_{n+1}$, as shown in Figure \ref{fig:BN-stag-cell3}.
We have obtained six independent variables $(\rho_s)_i^{n+1}$ and $\psi_i^{n+1}$.
It remains to calculate the seventh independent variable $\alpha_g$ or $\alpha_s$, which is constant in the solid cell $[x_i,x_{i+1}]$.

There is a convenient method to evaluate $\alpha_s$. Recall in system  \eqref{eq:BN-Euler} that the solid density satisfies a conservation law
\begin{equation}\label{eq:rho-s}
(\rho_s)_t+(\rho_s u_s)_x=0.
\end{equation}
Application of  the Godunov scheme over  the solid cell $[x_i,x_{i+1}]$ yields
\begin{equation}\label{eq:God-stag-rho-s}
(\rho_s)_{i+\frac12}^{n+1}=(\rho_s)_{i+\frac12}^n-\frac{\Delta t}{\Delta x}\left[(\rho_s u_s)_{i+1}^n-(\rho_s u_s)_i^n\right],
\end{equation}
where $(\rho_s)_{i+\frac12}^n=\frac12(\rho_s)_i^n+\frac12(\rho_s)_{i+1}^n$ is the integral average of the solid density in $[x_i,x_{i+1}]$ at time $t_n$.
The Godunov scheme for the solid phase by recalling system \eqref{eq:BN-Euler} again takes 
\begin{equation}\label{eq:God-stag-alpha-rho-s}
(\alpha_s\rho_s)_{i+\frac12}^{n+1}=(\alpha_s\rho_s)_{i+\frac12}^n-\frac{\Delta t}{\Delta x}\left[(\alpha_s\rho_s u_s)_{i+1}^n-(\alpha_s\rho_s u_s)_i^n\right]
\end{equation}
over the cell $[x_i, x_{i+1}]$, where the integral average $(\alpha_s\rho_s)_{i+\frac12}^n=(\alpha_s)_{i+\frac12}^n(\rho_s)_{i+\frac12}^n$.
In the numerical flux along the solid cell interface $x_i$, 
\begin{equation*}
(\alpha_s)_i^n=\left\{
\begin{aligned}
&(\alpha_s)_{i-\frac12}^n, & \text{ if } (u_s)_i^n>0,\\
&(\alpha_s)_{i+\frac12}^n, & \text{ if } (u_s)_i^n\leq 0.
\end{aligned}\right.
\end{equation*}
Thus the solid volume fraction at time $t_{n+1}$ is updated as 
\begin{equation}\label{eq:alpha-cal}
(\alpha_s)_{i+\frac12}^{n+1}=(\alpha_s\rho_s)_{i+\frac12}^{n+1} \big/ (\rho_s)_{i+\frac12}^{n+1}.
\end{equation}
With $(\alpha_s)_{i\pm\frac12}^{n+1}$, $\psi_i^{n+1}$ and $(\rho_s)_{i}^{n+1}$ available, the distribution $\bm{u}_{i-\frac12,i}^{n+1}$ and $\bm{u}_{i,i+\frac12}^{n+1}$ at time $t_{n+1}$ are constructed.
 Here the computation of the vector $\bm{u}$ involves a root-finding process of an equation for $\rho_g$ as indicated in \cite{karni_hybrid_2010}. The specific manipulation  is described  detailedly in \ref{app-nonlinear}.

The whole procedure  is  summarized  in \textbf{Algorithm \ref{alo:1-o}} below.
In the computational region where the porosity is constant, the staggered-projection Godunov-type scheme completely degenerates to the standard Godunov scheme.
In the region containing porosity jumps, this scheme is conservative before the projection step, and during the projection process, six independent variables remain unchanged except for the porosity.
Therefore, the projection rarely impairs the conservation of the Godunov-type scheme.
In the later section, numerical  experiments corroborate that this scheme significantly suppresses spurious oscillations near porosity interfaces and accurately captures shock waves of each phase.

\begin{algorithm}[ht]
\caption{First-order staggered-projection Godunov-type scheme}\label{alo:1-o}
\begin{algorithmic}[1]
\Require Cell averages $(\rho_s)_i^n$ and $\psi_i^n$ in $[x_{i-\frac12},x_{i+\frac12}]$ for $\psi=u_s, \eta_g, Q, P, H$; $(\alpha_s)_{i+\frac12}^n$ in $[x_i,x_{i+1}]$.
\Ensure $(\rho_s)_i^{n+1}$ and $\psi_i^{n+1}$ in $[x_{i-\frac12},x_{i+\frac12}]$; $(\alpha_s)_{i+\frac12}^{n+1}$ in $[x_i,x_{i+1}]$.

\hspace*{-0.51in}{\textbf{Procedure:}}

\State {Solve the Riemann problem $\textbf{RP}\left(\xi; \bm{u}_{i,i+\frac12}^n,\bm{u}_{i+\frac12,i+1}^n\right)$ at $x_{i+\frac12}$ to obtain the Riemann solution $\bm{u}_{i+\frac12}^n=\textbf{RP}\left(\xi=0; \bm{u}_{i,i+\frac12}^n,\bm{u}_{i+\frac12,i+1}^n\right)$}.
\State {Evaluate the nozzling term $\bm{S}_i^n=\bm{h}\left((\overline{p_g})_i^n,(u_s)_i^n\right)\left[(\alpha_s)_{i+\frac12}^n-(\alpha_s)_{i-\frac12}^n\right]$ with  $(\overline{p_g})_i^n$ in \eqref{eq:cal-pg}}.
\State {Utilize the Godunov scheme \eqref{eq:BN-Godunov-tmp} to get the updated cell average $\bm{u}_i^{n+1,*}$}.
\State {Calculate the position of the solid contact $(x_s)_i^{n+1}$ according to \eqref{eq:cal-xs} and obtain the proportion $\beta_{i-\frac12,i}^{n+1}=1-\beta_{i,i+\frac12}^{n+1}$ in \eqref{eq:cal-beta}}.
\State {Determine $\upsilon_{i-\frac12,i}^{n+1,*}$ and $\upsilon_{i,i+\frac12}^{n+1,*}$ for $\upsilon=\alpha_s, \rho_s, u_s, Q$ by \eqref{eq:alpha-s-result}, \eqref{eq:rho-s-suppose}, \eqref{eq:u-s-result} and \eqref{eq:Q-result} and for $\upsilon=\rho_g, p_g, p_s$ by solving the algebraic systems \eqref{eq:non-linear-system} and \eqref{eq:non-linear-system2}. Then determine $(\rho_s)_i^{n+1}=(\rho_s)_{i-\frac12,i}^{n+1,*}$ and $\psi_i^{n+1}=\psi_{i-\frac12,i}^{n+1,*}$ for $\psi=u_s,\eta_g,Q,P,H$}.
\State {By the projection $(x_s)_i^{n+1}$ onto  $x_i$, compute $(\alpha_s)_{i+\frac12}^{n+1}$ represented in \eqref{eq:alpha-cal}} with  $(\rho_s)_{i+\frac12}^{n+1}$ in \eqref{eq:God-stag-rho-s} and $(\alpha_s \rho_s)_{i+\frac12}^{n+1}$ in \eqref{eq:God-stag-alpha-rho-s}.
\end{algorithmic}
\end{algorithm}

\begin{rem}
It's worth noting that the computation of the volume fraction \eqref{eq:alpha-cal} causes slight conservation errors of the individual-phase mass across porosity jumps, for which, however, there is a more sophisticated evaluation method of the volume fraction in conform with  the conservation of mass.
We project the conservative variables $(\alpha_g \rho_g)$ onto solid cells and then evaluate the integral average $(\alpha_g \rho_g)_{i+\frac12}^{n+1}$ in $[x_i,x_{i+1}]$. In fact, by assuming $u_s>0$ (similarly if $u_s<0$), we have
\begin{equation*}
(\alpha_g \rho_g)_{i+\frac12}^{n+1} = \left(\frac12 -\beta_{i,i+\frac12}^{n+1}\right)(\alpha_g \rho_g)_{i-\frac12,i}^{n+1,*} +\beta_{i,i+\frac12}^{n+1}(\alpha_g \rho_g)_{i,i+\frac12}^{n+1,*}
+ \frac12 (\alpha_g \rho_g)_{i+\frac12,i+1}^{n+1,*}.
\end{equation*}
After the projection, the integral average is denoted as
\begin{equation*}
(\alpha_g \rho_g)_{i+\frac12}^{n+1} = \frac12 (\alpha_g \rho_g)_{i,i+\frac12}^{n+1} + \frac12 (\alpha_g \rho_g)_{i+\frac12,i+1}^{n+1}.
\end{equation*}
Together with the available  Riemann invariants $\psi_i^{n+1}$ in $[x_i,x_{i+\frac12}]$ and $\psi_{i+1}^{n+1}$ in $[x_{i+\frac12},x_{i+1}]$, $\psi=Q,H,\eta_g$,  the porosity in the solid cell $[x_i,x_{i+1}]$ 
\begin{equation*}
(\alpha_g)_{i+\frac12}^{n+1} = (\alpha_g)_{i,i+\frac12}^{n+1} = (\alpha_g)_{i+\frac12,i+1}^{n+1}
\end{equation*}
is obtained  by using the Newton-Raphson iteration.
Then, the  solid density in the cell $I_i$ is computed by the conservation of $\alpha_s\rho_s$
\begin{equation*}
(\rho_s)_{i-\frac12,i}^{n+1}=(\alpha_s \rho_s)_i^{n+1,*}/(\alpha_s)_i^{n+1},
\end{equation*}
where $(\alpha_s)_i^{n+1}=1-(\alpha_g)_i^{n+1}=1-\frac12\left[(\alpha_g)_{i-\frac12}^{n+1}+(\alpha_g)_{i+\frac12}^{n+1}\right]$.
Considering constant $u_s$ across porosity jumps, we may also notice that the total momentum $\mathcal{M}=(\alpha_g\rho_g+\alpha_s\rho_s)u_s+Q$ is conservative since $\alpha_g\rho_g,\alpha_s\rho_s$ and $Q$ are conservative. 
\end{rem}

\subsection{Extension to second-order accuracy}
The generalized Riemann problem (GRP) approach is adopted to extend the above projection method to second order accuracy \cite{ben-artzi_second-order_1984,ben-artzi_generalized_2003}. The version we will use is that for general hyperbolic laws \cite{Li-2}.  The resulting scheme, when combined with the above staggered projection,  is a temporal-spatial coupling staggered second order method \cite{Li-3}. 
In the vicinity of the porosity jump,  algorithmic steps that are different from the first-order version are listed below.

In the first-order staggered-projection  scheme, the Riemann invariants $\psi=u_s,\eta_g,Q,P,H$ and solid density $\rho_s$ are constant in the gaseous cell $I_i$, while the solid volume fraction $\alpha_s$ is constant in the solid cell $[x_i,x_{i+1}]$.
In order to design a second-order version, we assume that $\phi$ and $\rho_s$ are piece-wise linear in gaseous cells, and $\alpha_s$ are piece-wise linear in solid cells.
Since the Riemann invariants are continuous across solid contacts, rather than the primitive variables $\textbf{v}$ or $\bm{u}$, the local reconstruction reasonably applies for the Riemann invariants,   $\bm{\omega}:=[\rho_s,u_s,P,Q,H,\eta_g]$, 
\begin{equation}\label{eq:omega-linear}
\widetilde{\bm{\omega}}_i(x)=\bm{\omega}_i^n+(\bm{\omega}_x)_i^n(x-x_i),
\end{equation}
where $\bm{\omega}_i^n$ is calculated by the cell average $\bm{u}_i^n$ over $I_i$, $(\bm{\omega}_x)_i^n$ is the slope of the vector $\bm{\omega}$.
In the solid cell $[x_i,x_{i+1}]$, $\alpha_s$ has a piece-wise linear representation 
\begin{equation}\label{eq:alpha-s-linear}
(\widetilde{\alpha_s})_{i+\frac12}(x)=(\alpha_s)_{i+\frac12}^n+((\alpha_s)_x)_{i+\frac12}^n(x-x_{i+\frac12}),
\end{equation}
where $(\alpha_s)_{i+\frac12}^n$ is the cell average and $((\alpha_s)_x)_{i+\frac12}^n$ is the slope.

With such ``initial" data at time level $t=t_n$, the GRP needs high order numerical fluxes.  
Denote $\bm{w}=[\alpha_s,\bm{\omega}]$, $\widetilde{\bm{w}}_{i,i+\frac12}(x)=\left[(\widetilde{\alpha_s})_{i+\frac12}(x),\widetilde{\bm{\omega}}_i(x)\right]^\top$ and $\widetilde{\bm{w}}_{i+\frac12,i+1}(x)=\left[(\widetilde{\alpha_s})_{i+\frac12}(x),\widetilde{\bm{\omega}}_{i+1}(x)\right]^\top$ with limiting states
\begin{equation*}
\bm{w}_{i+\frac12,-}^n=\lim\limits_{x\rightarrow x_{i+\frac12,-}}\widetilde{\bm{w}}_{i,i+\frac12}(x),\quad
\bm{w}_{i+\frac12,+}^n=\lim\limits_{x\rightarrow x_{i+\frac12,+}}\widetilde{\bm{w}}_{i+\frac12,i+1}(x).
\end{equation*}
This GRP is written as $\textbf{GRP}\left(\widetilde{\bm{w}}_{i,i+\frac12}(x),\widetilde{\bm{w}}_{i+\frac12,i+1}(x)\right)$.
In this paper, an acoustic GRP solver  \cite{ben-artzi_generalized_2003,toro_riemann_1997} is utilized to approximately solve the GRP, which depends on  the associated  Riemann problem 
with the initial data
\begin{equation*}
\bm{u}(x,t_n)=\left\{
\begin{aligned}
&\bm{u}\left(\bm{w}_{i+\frac12,-}^n\right), & x<x_{i+\frac{1}{2}},\\
&\bm{u}\left(\bm{w}_{i+\frac12,+}^n\right), & x>x_{i+\frac{1}{2}}. 
\end{aligned}
\right.
\end{equation*}
Denote again  the associated Riemann solution as $\bm{w}_{i+\frac12}^n=\bm{w}\left(\textbf{RP}(\xi = 0)\right)$.
Here the recovery of the vector $\bm{u}$ from the Riemann invariants is described in detail in \ref{app-nonlinear}.
Using the acoustic approximation $\bm{w}_{i+\frac12}^n\approx \bm{w}_{i+\frac12,-}^n \approx \bm{w}_{i+\frac12,+}^n$, 
the instantaneous temporal derivative is determined according to
\begin{equation}\label{eq:omega-t}
(\bm{w}_t)_{i+\frac{1}{2}}^n=
\bm{R}(\bm{z}_t)_{i+\frac{1}{2}}^n=
-[\bm{R}\bm{\Lambda}_+\bm{R}^{-1} (\bm{w}_x)_{i}^n
+ \bm{R}\bm{\Lambda}_-\bm{R}^{-1} (\bm{w}_x)_{i+1}^n],
\end{equation}
where $\bm{R}=\bm{R}(\bm{w}_{i+\frac12}^n)$, $\bm{z}=\bm{R}^{-1}\bm{\omega}$, $\bm{\Lambda}=\bm{\Lambda}(\bm{w}_{i+\frac12}^n)$, $\bm{\Lambda}_+ = \text{diag}(\max(\lambda_i,0))$ and $\bm{\Lambda}_- = \text{diag}(\min(\lambda_i, 0))$. Relavant characteristic decompositions are put in \ref{app-RI} with notations $\bm{R}$, $\bm{\Lambda}$.

Then the GRP scheme updates the solution of the homogeneous BN model \eqref{eq:BN-Euler}  in the following formula, 
\begin{equation}\label{eq:BN-GRP}
\bm{u}_i^{n+1,*}=\bm{u}_i^n-\frac{\Delta t}{\Delta x}\left(\bm{f}_{i+\frac{1}{2}}^{n+\frac12}-\bm{f}_{i-\frac{1}{2}}^{n+\frac12}\right)+\frac{\Delta t}{\Delta x}\bm{S}_i^{n+\frac12},
\end{equation}
where the numerical flux is taken  as
\begin{equation}\label{eq:omega-mid-point}
\bm{f}_{i+\frac{1}{2}}^{n+\frac12}=\bm{f}\left(\bm{w}_{i+\frac{1}{2}}^{n+\frac12}\right), \ \ \ 
\bm{w}_{i+\frac{1}{2}}^{n+\frac12}=\bm{w}_{i+\frac{1}{2}}^n+\frac{\Delta t}{2}(\bm{w}_t)_{i+\frac12}^{n}.
\end{equation}
 The  nozzling term  is approximated as
\begin{equation}\label{eq:S_i_cal}
\bm{S}_i^{n+\frac12}=\bm{h}\left((\overline{p_g})_i^{n+\frac12},(u_s)_i^{n+\frac12}\right)\left[(\alpha_s)_{i+\frac12}^n-(\alpha_s)_{i-\frac12}^n\right],
\end{equation}
and
\begin{equation}\label{eq:cal-pg-2o}
(\overline{p_g})_i^{n+\frac12}=\left\{
\begin{aligned}
&\frac12\left[(p_g)_{i,-}^{n+\frac12}+(p_g)_{i,+}^{n+\frac12}\right], && \text{ if } \left|(\alpha_s)_{i+\frac12}^{n} - (\alpha_s)_{i-\frac12}^{n}\right|<\varepsilon,\\
&(p_g)_{\min}, && \text{ if } \mathcal{P}_g < (p_g)_{\min},\\
&(p_g)_{\max}, && \text{ if } \mathcal{P}_g > (p_g)_{\max},\\
&\mathcal{P}_g, && \text{ otherwise,}
\end{aligned}\right.
\end{equation}
with
\begin{equation*}
(p_g)_{\max}=\max\left\{(p_g)_{i,-}^{n+\frac12},(p_g)_{i,+}^{n+\frac12}\right\},\quad
(p_g)_{\min}=\min\left\{(p_g)_{i,-}^{n+\frac12},(p_g)_{i,+}^{n+\frac12}\right\},
\end{equation*}
\begin{equation*}
\mathcal{P}_g = \frac{(\alpha_s)_{i+\frac12}^{n+\frac12} (p_s)_{i+\frac12,+}^{n+\frac12}-(\alpha_s)_{i-\frac12}^{n+\frac12} (p_s)_{i-\frac12,-}^{n+\frac12}}{(\alpha_s)_{i+\frac12}^{n}-(\alpha_s)_{i-\frac12}^{n}}.
\end{equation*}
The mid-point values $(p_g)_{i,-}^{n+\frac12}$, $(p_g)_{i,+}^{n+\frac12}$ at the solid cell interface $x_i$ are given by
\begin{equation*}
\bm{w}_{i,-}^{n+\frac12}=\bm{w}_{i,-}^n+\frac{\Delta t}{2} (\bm{w}_t)_{i,-}^n,\quad
\bm{w}_{i,+}^{n+\frac12}=\bm{w}_{i,+}^n+\frac{\Delta t}{2}(\bm{w}_t)_{i,+}^n,
\end{equation*}
which achieves  second-order accuracy of the numerical integration.
Here, the limiting states are
\begin{equation*}
\begin{aligned}
&\bm{w}_{i,-}^n=\lim\limits_{x\rightarrow x_i-}\widetilde{\bm{w}}_{i-\frac12,i}(x),
&&\bm{w}_{i,+}^n=\lim\limits_{x\rightarrow x_i+}\widetilde{\bm{w}}_{i,i+\frac12}(x),\\
&(\bm{w}_t)_{i,-}^n=-\bm{B}(\bm{w}_{i,-}^n)(\bm{w}_x)_{i-\frac12}^n,\quad
&&(\bm{w}_t)_{i,+}^n=-\bm{B}(\bm{w}_{i,+}^n)(\bm{w}_x)_{i+\frac12}^n.
\end{aligned}
\end{equation*}
Then the states on both sides of solid contacts are evaluated by using the algebraic systems \eqref{eq:non-linear-system} and \eqref{eq:non-linear-system2}. 
So we obtain the Riemann invariants for the $\lambda_0$-field and the solid density. Then the projection step is the same as that for the first-order version.

In order to compute the solid fractions, we first take the mid-point values on the solid interface, 
\begin{equation}\label{w-i-mid}
\bm{w}_{i}^{n+\frac12}=\left\{
\begin{aligned}
&\bm{w}_{i,-}^{n+\frac12},& \text{ if }(u_s)_i^n > 0,\\
&\bm{w}_{i,+}^{n+\frac12},& \text{ if }(u_s)_i^n\leq 0.
\end{aligned}\right.
\end{equation}
The position of the solid contact  at next time level $t=t_{n+1}$ is given as 
\begin{equation}\label{eq:cal-xs-2o}
(x_s)_i^{n+1}= x_i + (u_s)_i^{n+\frac12} \Delta t.
\end{equation}
The solid volume fractions in intervals $[x_{i-\frac12}, (x_s)_i^{n+1}]$ and $[(x_s)_i^{n+1}, x_{i+\frac12}]$ are given in a Lagrangian step
\begin{equation}\label{eq:alpha-s-result-2o}
\begin{aligned}
&(\alpha_s)_{i-\frac12,i}^{n+1,*}=(\alpha_s)_{i-\frac12}^n-\frac{(u_s)_{i-\frac12}^{n+\frac12}\Delta t}{(x_s)_i^{n+1}-x_{i-\frac12}}\left[(\alpha_s)_{i-\frac12}^n-(\alpha_s)_{i-\frac12}^{n+\frac12}\right],\\
&(\alpha_s)_{i,i+\frac12}^{n+1,*}=(\alpha_s)_{i+\frac12}^n-
\frac{(u_s)_{i+\frac12}^{n+\frac12}\Delta t}{x_{i+\frac12}-(x_s)_i^{n+1}}
\left[(\alpha_s)_{i+\frac12}^{n+\frac12}-(\alpha_s)_{i+\frac12}^n\right],
\end{aligned}
\end{equation}
respectively.   The density of solid phase  over the solid cell $[x_i,x_{i+1}]$ is approximated as 
\begin{equation}\label{eq:God-stag-rho-s-2o}
(\rho_s)_{i+\frac12}^{n+1}=(\rho_s)_{i+\frac12}^n-\frac{\Delta t}{\Delta x}\left[(\rho_s u_s)_{i+1}^{n+\frac12}-(\rho_s u_s)_i^{n+\frac12}\right],
\end{equation}
and
\begin{equation}\label{eq:God-stag-alpha-rho-s-2o}
(\alpha_s\rho_s)_{i+\frac12}^{n+1}=(\alpha_s\rho_s)_{i+\frac12}^n-\frac{\Delta t}{\Delta x}\left[(\alpha_s\rho_s u_s)_{i+1}^{n+\frac12}-(\alpha_s\rho_s u_s)_i^{n+\frac12}\right],
\end{equation}
where $(\rho_s u_s)_{i}^{n+\frac12}$ and $(\alpha_s\rho_s u_s)_{i}^{n+\frac12}$ are obtained by the mid-point value at $x=x_i$

Finally, the data need  updating at next time level $t=t_{n+1}$ by  using the minmod limiter to suppress oscillations,  particularly
the  slope $(\bm{\omega}_x)_i^{n+1}$ \cite{van_leer_towards_1979,Li-2}, 
\begin{equation}\label{eq:cal-slope-omega}
(\bm{\omega}_x)_i^{n+1}=\text{minmod}\left(\phi\frac{\bm{\omega}_{i+1}^{n+1}-\bm{\omega}_i^{n+1}}{\Delta x},\frac{\bm{\omega}_{i+\frac12}^{n+1,-}-\bm{\omega}_{i-\frac12}^{n+1,-}}{\Delta x},\phi\frac{\bm{\omega}_i^{n+1}-\bm{\omega}_{i-1}^{n+1}}{\Delta x}\right),
\end{equation}
where $\bm{\omega}_{i+\frac12}^{n+1,-}$ is obtained by $\bm{w}_{i+\frac12}^{n+1,-}=\bm{w}_{i+\frac12}^n+\Delta t (\bm{w}_t)_{i+\frac12}^n$ and 
\begin{equation*}
\text{minmod}(a,b,c)=\left\{
\begin{aligned}
&\min(|a|,|b|,|c|), && \text{ if } a,b,c>0,\\
&-\min(|a|,|b|,|c|), && \text{ if } a,b,c<0,\\
&0 && \text{ otherwise,}
\end{aligned}
\right.
\end{equation*}
with  $\phi\in[0,2)$. 
 The limited slope of $\alpha_s$ in the solid cell $[x_i,x_{i+1}]$ is constructed as
\begin{equation}\label{eq:cal-slope-alpha-s}
((\alpha_s)_x)_{i+\frac12}^{n+1}=\text{minmod}\left(\phi\frac{(\alpha_s)_{i+\frac32}^{n+1}-(\alpha_s)_{i+\frac12}^{n+1}}{\Delta x},\frac{(\alpha_s)_{i+1}^{n+1,-}-(\alpha_s)_{i}^{n+1,-}}{\Delta x},\phi\frac{(\alpha_s)_{i+\frac12}^{n+1}-(\alpha_s)_{i-\frac12}^{n+1}}{\Delta x}\right),
\end{equation}
where $(\alpha_s)_i^{n+1,-}$ is chosen as 
\begin{equation*}
(\alpha_s)_{i}^{n+1,-}=
\left\{
\begin{aligned}
&(\alpha_s)_{i,-}^n + \Delta t ((\alpha_s)_t)_{i,-}^n,
&\text{ if } (u_s)_i^n>0,\\
&(\alpha_s)_{i,+}^n + \Delta t ((\alpha_s)_t)_{i,+}^n,
&\text{ if } (u_s)_i^n\leq 0.
\end{aligned}
\right.
\end{equation*}

The algorithm for the  second-order GRP scheme is summarized in \textbf{Algorithm \ref{alo:2-o}}.

\begin{algorithm}[ht]
\caption{ A second-order staggered-projection GRP scheme}\label{alo:2-o}
\begin{algorithmic}[1]
\Require Piece-wise linear data $\widetilde{\bm{\omega}}_i(x)$ in \eqref{eq:omega-linear} in $[x_{i-\frac12},x_{i+\frac12}]$ for $\bm{\omega}=[\rho_s,u_s,Q,P,H,\eta_g]$;
$(\widetilde{\alpha_s})_{i+\frac12}(x)$ in \eqref{eq:alpha-s-linear} in $[x_i,x_{i+1}]$.
\Ensure $\bm{\omega}_i^{n+1}$ and $(\bm{\omega}_x)_i^{n+1}$ in $[x_{i-\frac12},x_{i+\frac12}]$; $(\alpha_s)_{i+\frac12}^{n+1}$ and $((\alpha_s)_x)_{i+\frac12}^{n+1}$ in $[x_i,x_{i+1}]$.

\hspace*{-0.51in}{\textbf{Procedure:}}

\State {Solve the Riemann problem $\textbf{RP}\left(\xi;\bm{u}\left(\bm{w}_{i+\frac12,-}^n\right),\bm{u}\left(\bm{w}_{i+\frac12,+}^n\right)\right)$ at $x_{i+\frac12}$ to obtain the Riemann solution $\bm{w}_{i+\frac12}^n=\bm{w}\left(\textbf{RP}(\xi=0)\right)$}.
\State {Determine $(\bm{w}_t)_{i+\frac{1}{2}}^n$ by \eqref{eq:omega-t}} using  the acoustic approximation. The numerical fluxes  are given in \eqref{eq:omega-mid-point}.
\State {Evaluate the nozzling term $\bm{S}_i^{n+\frac12}$ in \eqref{eq:S_i_cal}
by $(\overline{p_g})_i^n$ in \eqref{eq:cal-pg-2o}}.
\State {Utilize the GRP scheme \eqref{eq:BN-GRP} to get the updated cell average $\bm{u}_i^{n+1,*}$}.
\State {Calculate $(x_s)_i^{n+1}$ is  according to \eqref{eq:cal-xs-2o} and obtain $\beta_{i-\frac12,i}^{n+1}=1-\beta_{i,i+\frac12}^{n+1}$ in \eqref{eq:cal-beta}}.
\State {Implement Procedure 5 in \textbf{Algorithm \ref{alo:1-o}}. In this algorithm, $(\alpha_s)_{i-\frac12,i}^{n+1,*}$ and $(\alpha_s)_{i,i+\frac12}^{n+1,*}$ are determined by \eqref{eq:alpha-s-result-2o}}.
\State {With  the projection from $(x_s)_i^{n+1}$ onto $ x_i$, compute $(\alpha_s)_{i+\frac12}^{n+1}$ in \eqref{eq:alpha-cal}}, $(\rho_s)_{i+\frac12}^{n+1}$ in \eqref{eq:God-stag-rho-s-2o} and $(\alpha_s \rho_s)_{i+\frac12}^{n+1}$ in \eqref{eq:God-stag-alpha-rho-s-2o}.
\State {Update the slopes $(\bm{\omega}_x)_i^{n+1}$ and $((\alpha_s)_x)_{i+\frac12}^{n+1}$ by \eqref{eq:cal-slope-omega} and \eqref{eq:cal-slope-alpha-s}}, respectively.
\end{algorithmic}
\end{algorithm}

\subsection{Extension to two dimensions}

 The staggered-projection Godunov-type scheme can be extended  over structural meshes in  two dimensions. In this section we adopt the dimensional splitting  just to show this method works well. 
We write the two-dimensional homogeneous BN model as
\begin{equation}\label{eq:BN-Euler2}
\bm{u}_t+\bm{f}(\bm{u})_x+\bm{g}(\bm{u})_y=\bm{h}(\bm{u})\cdot\left((\alpha_s)_x,(\alpha_s)_y\right)^\top,
\end{equation}
with
\begin{equation*}
\bm{u}=
\begin{bmatrix}
\alpha_s\\
\alpha_s \rho_s\\
\alpha_s \rho_s u_s\\
\alpha_s \rho_s v_s\\
\alpha_s \rho_s E_s\\
\alpha_g \rho_g\\
\alpha_g \rho_g u_g\\
\alpha_g \rho_g v_g\\
\alpha_g \rho_g E_g
\end{bmatrix},\
\bm{f}=
\begin{bmatrix}
0\\
\alpha_s \rho_s u_s\\
\alpha_s \rho_s u_s^2+\alpha_s p_s\\
\alpha_s \rho_s u_s v_s\\
\alpha_s u_s (\rho_s E_s + p_s)\\
\alpha_g \rho_g u_g\\
\alpha_g \rho_g u_g^2+\alpha_g p_g\\
\alpha_g \rho_g u_g v_g\\
\alpha_g u_g (\rho_g E_g  + p_g)
\end{bmatrix},\
\bm{g}=
\begin{bmatrix}
0\\
\alpha_s \rho_s v_s\\
\alpha_s \rho_s v_s u_s\\
\alpha_s \rho_s v_s^2+\alpha_s p_s\\
\alpha_s v_s (\rho_s E_s + p_s)\\
\alpha_g \rho_g v_g\\
\alpha_g \rho_g v_g u_g\\
\alpha_g \rho_g v_g^2+\alpha_g p_g\\
\alpha_g v_g (\rho_g E_g  + p_g)
\end{bmatrix},\ 
\bm{h}=
\begin{bmatrix}
-u_s & -u_s\\
0 & 0\\
p_g & 0\\
0 & p_g\\
p_g u_s & p_g v_s\\
0 & 0\\
-p_g & 0\\
0 & -p_g \\
-p_g u_s & -p_g v_s
\end{bmatrix},
\end{equation*}
where $(u_k,v_k)$ is the velocity for the phase $k$ and $E_k=e_k+\frac{1}{2}(u_k^2+v_k^2)$. 
We split the system \eqref{eq:BN-Euler2} into two  subsystems%
\begin{subequations}\label{eq:BN-Euler2-sub}
\begin{align}
&\bm{u}_t+\bm{f}(\bm{u})_x=\bm{h}_1(\bm{u})(\alpha_s)_x,\\
&\bm{u}_t+\bm{g}(\bm{u})_y=\bm{h}_2(\bm{u})(\alpha_s)_y,
\end{align}
\end{subequations}
with $[\bm{h}_1,\bm{h}_2]=\bm{h}$,  and denote $\mathcal{L}^{(\Delta t)}$, $\mathcal{L}_x^{(\Delta t)}$ and $\mathcal{L}_y^{(\Delta t)}$ as approximate solution operators for \eqref{eq:BN-Euler2}, (\ref{eq:BN-Euler2-sub}a) and (\ref{eq:BN-Euler2-sub}b) with a time step increment, respectively.
Assume that the solution operators $\mathcal{L}_x^{(\Delta t)}$ and $\mathcal{L}_y^{(\Delta t)}$ are space-time second-order accurate. Then the Strang splitting algorithm \cite{strang_construction_1968} provides a second order approximation to $\mathcal{L}_{\Delta t}$,
\begin{equation}
\mathcal{L}^{(\Delta t)}=\mathcal{L}_x^{\left(\frac{\Delta t}{2}\right)}\mathcal{L}_y^{(\Delta t)}\mathcal{L}_x^{\left(\frac{\Delta t}{2}\right)}.
\end{equation}

\begin{figure}[htb]
\begin{center}
\includegraphics[width=0.5\textwidth]{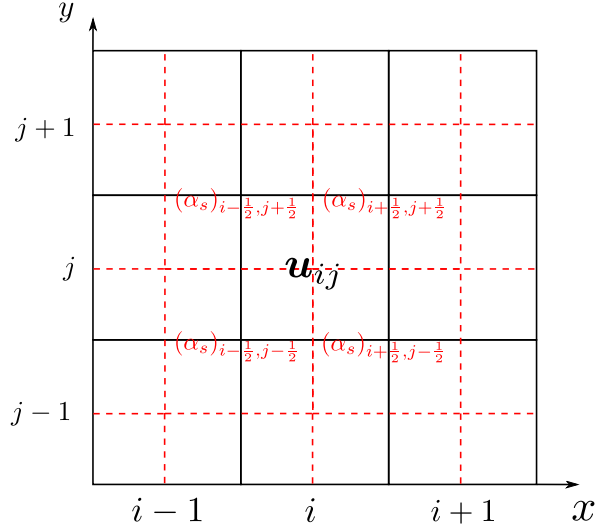}
\end{center}
\caption{2-D gas-solid staggered regular grid}\label{fig:BN-stag-2D}
\end{figure}

The solution operators $\mathcal{L}_x^{(\Delta t)}$ and $\mathcal{L}_y^{(\Delta t)}$ can be specified as the staggered-projection GRP scheme in the $x$- and $y$-directions,
over a 2-D gas-solid staggered grid shown in Figure \ref{fig:BN-stag-2D}.
The Riemann invariants $\psi$ and solid density $\rho_s$ are piece-wise linear in the black-line  gaseous cells $\Omega_{ij}$, while the volume fraction $\alpha_s$ satisfies piece-wise linear distribution in the red-dashed solid cells $\Omega_{i+\frac12,j+\frac12}$.
Take the solution operator $\mathcal{L}_x^{(\Delta t)}$ as an example.
Like the staggered-projection Godunov-type scheme for the 1-D case, 
within  one time step, the Riemann invariants for the $\lambda_0$- field and the solid density in the gaseous cell $\Omega_{ij}$ are obtained.
The cell average of $(\rho_s)_{i+\frac12,j+\frac12}^{n+1}$ and $(\alpha_s\rho_s)_{i+\frac12,j+\frac12}^{n+1}$ is updated by the GRP scheme
in the solid cell $\Omega_{i+\frac12,j+\frac12}$.
Then, the cell average of $\alpha_s$ in $\Omega_{i+\frac12,j+\frac12}$ at time $t_{n+1}$ is evaluated by
\begin{equation*}
(\alpha_s)_{i+\frac12,j+\frac12}^{n+1}
=\mathcal{L}_x^{(\Delta t)}(\alpha_s)_{i+\frac12,j+\frac12}^n
=(\alpha_s\rho_s)_{i+\frac12,j+\frac12}^{n+1} \big/ (\rho_s)_{i+\frac12,j+\frac12}^{n+1}.
\end{equation*}
Other  steps are identical to the 1-D staggered-projection GRP scheme.
Based on these Riemann invariants and the solid density in the gaseous cell, as well as the porosity in the solid cell, we can obtain the distribution of $\bm{u}$ on the gas-solid staggered grid $\bm{u}^{n+1}=\mathcal{L}_x^{(\Delta t)}\bm{u}^n$. Thus we have
\begin{equation*}
\bm{u}^{n+1}=\mathcal{L}^{(\Delta t)}\bm{u}^n=\mathcal{L}_x^{\left(\frac{\Delta t}{2}\right)}\mathcal{L}_y^{(\Delta t)}\mathcal{L}_x^{\left(\frac{\Delta t}{2}\right)}\bm{u}^n. 
\end{equation*}

\section{Numerical examples}\label{sec:ex-BN}

In this section, several numerical examples including porosity jumps are tested, separately for the quasi-1-D duct flow model, the one- and two-dimensional homogeneous BN models.
In order to avoid difficulties of thermodynamic modeling for the two phases, we assume that both phases meet the EOS for polytropic gases
\begin{equation*}
e_k=e_k(\rho_k,p_k) = \frac{p_k}{(\gamma_k -1)\rho_k},
\end{equation*}
where constant $\gamma_k=\Gamma_k+1$ is the specific heat ratio for the phase $k$.
The sound speed and entropy for the phase $k$ are represented as
\begin{equation*}
c_k=\sqrt{\frac{\gamma_k p_k}{\rho_k}},\quad
\eta_k=\frac{p_k}{\rho_k^{\gamma_k}},
\end{equation*}
respectively. 

For these  examples, numerical simulations are based on regular grids with staggered gas-solid grid cells.
The initial data are given by cell averages $(\rho_s)_i^{0}$ and exact $\psi_i^{0}$ in gaseous cells and cell averages $(\alpha_s)_{i+\frac12}^{0}$ in solid cells, and exact solutions are found by using  the routine package CONSTRUCT \cite{andrianov_CONSTRUCT}.
The numerical results of staggered-projection Godunov-type schemes are displayed, for which numerical fluxes are evaluated by the exact Riemann solver or the acoustic GRP solver, and the time step $\Delta t$ is restricted by the CFL condition
\begin{equation*}
\Delta t=\frac{\text{CFL}\cdot \Delta x/2}{\max\limits_{\bm{u}}\{|\lambda_{i,k}(\bm{u})|+|u_s|;\, l=1,2,3,\, k=s,g\}}
\end{equation*}
with the number $\text{CFL}=0.9$.

For all 1-D examples, we present the numerical results with first- and second-order accuracy, respectively.
In the graphs of numerical results,
the red  `{\color{red}$\bm{\times}$}s' represent the numerical solution  by the   first-order   scheme, the black  `{\color{black}$\bm{+}$}s' represent the numerical solution by the second GRP scheme, and the solid blue line represents the exact solution for each case.
It turns out that each numerical solution presents  essentially no spurious oscillation near porosity interfaces and totally accord with the corresponding exact solution.

\subsection{Quasi-1-D duct flow model} 

The system of Euler equations in a duct of variable cross-section \eqref{eq:deter-var2} embodies the situation of the stationary solid phase.
It is regarded as a reduced system of the 1-D homogeneous BN model, which effectually reflects gas dynamics in the two-phase flow.
At first, a numerical experiment with two cases  is carried out for this simplified model. The staggered Godunov-type scheme, when applied to this model,  does not need to consider the motion of contacts and the projection, only involving an Eulerian step actually.
The computational domain  $[0,0.06]$ is composed of $M=111$ regular grid cells, and the initial position of the discontinuity is $x = 0.02$.
The states on both sides of the discontinuity are denoted as $\bm{u}_L$ and $\bm{u}_R$.
The specific heat ratio of the gas is taken as $\gamma = 1.23$.

\begin{table}[htb]
\begin{center}
\caption{Initial data for  duct flows}\label{tab:duct-init}\vspace{0.1cm}

\begin{tabular}{c|cccccccc}
\hline
Case & $A_{L}$ &$\rho_{L}$ & $u_{L}$& $p_{L}$ & $A_{R}$ &$\rho_{R}$ & $u_{R}$& $p_{R}$ \\
\hline
\uppercase\expandafter{\romannumeral1} & 1 & 151.13 & 212.31 & $2.4836\times 10^8$ & 0.25 & 95.199 & 1348.2 & $1.4067\times 10^8$ \\
\uppercase\expandafter{\romannumeral2} & 1 & 169.34 & 0 & $2.96\times 10^8$ & 0.25 & 0.76278 & 0 & $1\times 10^5$\\
\hline
\end{tabular}
\end{center}
\end{table}

\begin{example}[Shock-tube problems]

Two  cases for the duct flow with a discontinuous cross-section are simulated. The initial data are given in Table \ref{tab:duct-init}.
The first case is picked  up from \cite{karni_hybrid_2010}, corresponding to an isolated $0$-contact.  The numerical solution  and the exact solution are shown in Figure \ref{fig:duct-ex1}(a), which shows that the current staggered  scheme preserves the constant entropy across the $0$-contact, approaching accurate simulation. In contrast,   a conservative scheme based on a split-step algorithm in \cite{karni_hybrid_2010} fails, leading to  spurious oscillations.

\begin{figure}[tp]
\centering
\begin{minipage}{0.8\linewidth}
\begin{center}
\includegraphics[width=\textwidth]{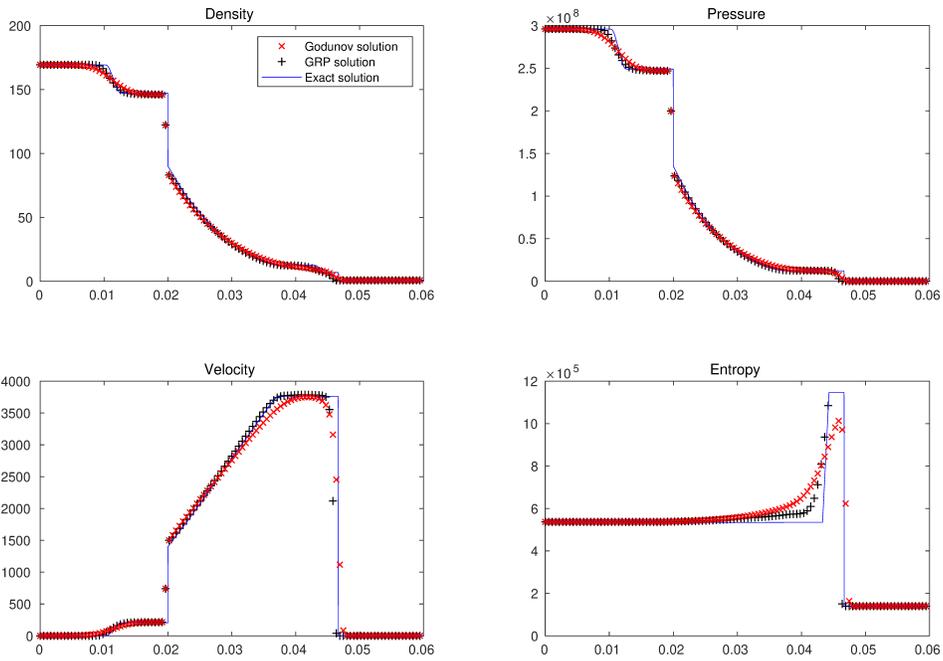}

(a) Isolated $0$-contact
\end{center}
\end{minipage}
\begin{minipage}{0.8\linewidth}
\begin{center}
\includegraphics[width=\textwidth]{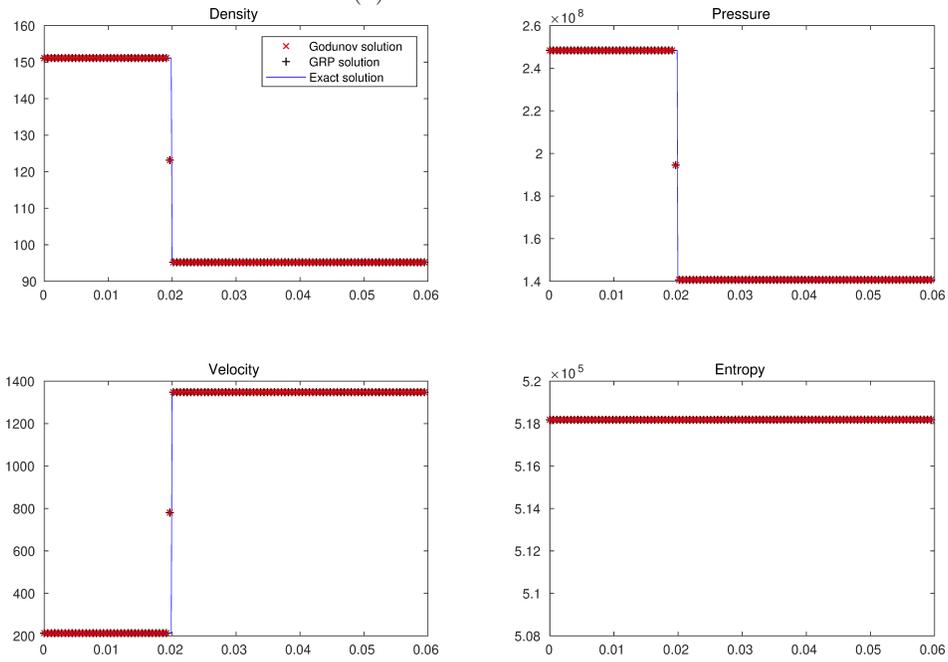}

(b) Riemann problem with a cross-section jump
\end{center}
\end{minipage}
\caption{Numerical results of the staggered Godunov-type scheme and the exact solution at
$t=6.3\times 10^{-6}$.}\label{fig:duct-ex1}
\end{figure}

The second case was ever considered in \cite{lowe_two-phase_2005} with a finer grid, which is a Riemann problem containing  a cross-section jump within a rarefaction wave, a shock and a contact discontinuity propagating to the right.
The coincidence of the $0$-contact and the rarefaction generates a resonance phenomenon.
The numerical solution by the current scheme  and the exact solution are shown in Figure \ref{fig:duct-ex1}(b).
Since the entropy cannot remain constant across the $0$-contact, the numerical solution computed by the conservative scheme based on operator splitting in \cite{karni_hybrid_2010} seriously deviated from the exact solution, probably due to the resonance.
In contrast, the numerical solutions by the current  scheme are in good agreement with the exact solution. 

\end{example}

\subsection{1-D homogeneous BN model} 

In the first three examples for the 1-D  homogeneous BN model below, the computational domain $[0,1]$ is divided into $M=300$ regular grid cells, and the initial position of the discontinuity is located at $x=0.5$.
Initial data for all the examples are listed in Table \ref{tab:BN-init}.
The specific heat ratios for the two phases are taken as $\gamma_s=\gamma_g=1.4$.

\begin{table}[htb]
\begin{center}
\caption{Initial data for the 1-D homogeneous BN model}\label{tab:BN-init}
\begin{tabular}{c|ccccccccc}
\hline
Case & Phase $k$ & $\alpha_{kL}$ & $\rho_{kL}$& $u_{kL}$& $p_{kL}$ & $\alpha_{kR}$ & $\rho_{kR}$& $u_{kR}$& $p_{kR}$\\
\hline
\multirow{2}{*}{\uppercase\expandafter{\romannumeral1}}
& $s$(solid) & 0.8 & 2 & 0.3 & 5 & 0.3 & 2 & 0.3 & 12.8567\\
& $g$(gas)   & 0.2 & 1 & 2 & 1 & 0.7 & 0.1941 & 2.8011 & 0.1\\
\hline
\multirow{2}{*}{\uppercase\expandafter{\romannumeral2}}
& $s$ & 0.1 & 0.2068 & 1.4166 & 0.0416 & 0.2 & 2.2263 & 0.9366 & 6\\
& $g$ & 0.9 & 0.5806 & 1.5833 &1.375 & 0.8 & 0.4890 & -0.70138 & 0.986\\
\hline
\multirow{2}{*}{\uppercase\expandafter{\romannumeral3}}
& $s$ & 0.5 & 2.1917 & -0.995 & 3 & 0.1 & 0.6333 & -1.1421 & 2.5011\\
& $g$ & 0.5 & 6.3311 & -0.789 & 1 & 0.9 & 0.4141 & -0.6741 & 0.0291\\
\hline
\multirow{5}{*}{\uppercase\expandafter{\romannumeral4}}
& $s$ & 0.3 & 0.5476 & 0      & 0.328 & 0.7 & 1.04   & 0      &  1.22\\
& $g$ & 0.7 & 2.933  & 0.4136 & 2.5   & 0.3 & 2.0462 & 0.7114 & 1.5096\\
\cline{3-10}
&\multirow{2}{*}{$s$, $g$}
 & $\alpha_{sM}$ & $\rho_{sM}$& $u_{sM}$& $p_{sM}$ & $\alpha_{gM}$ & $\rho_{gM}$& $u_{gM}$& $p_{gM}$\\
\cline{3-10}
 & & 0.3 & 0.5476 & 0      & 0.328  & 0.7 & 2.5154 & 0.248  & 2.0155\\
\hline
\end{tabular}
\end{center}
\end{table}

\begin{example}[Case \uppercase\expandafter{\romannumeral1}. A single solid contact]\label{ex:test-1}

This example is \textit{Test~1} in \cite{andrianov_riemann_2004}. 
The solution to this problem consists of a single solid contact propagating to the right with the velocity  $0.3$.
\begin{figure}[htbp]
\begin{center}
\begin{minipage}{0.8\linewidth}
\begin{center}
\includegraphics[width=\textwidth]{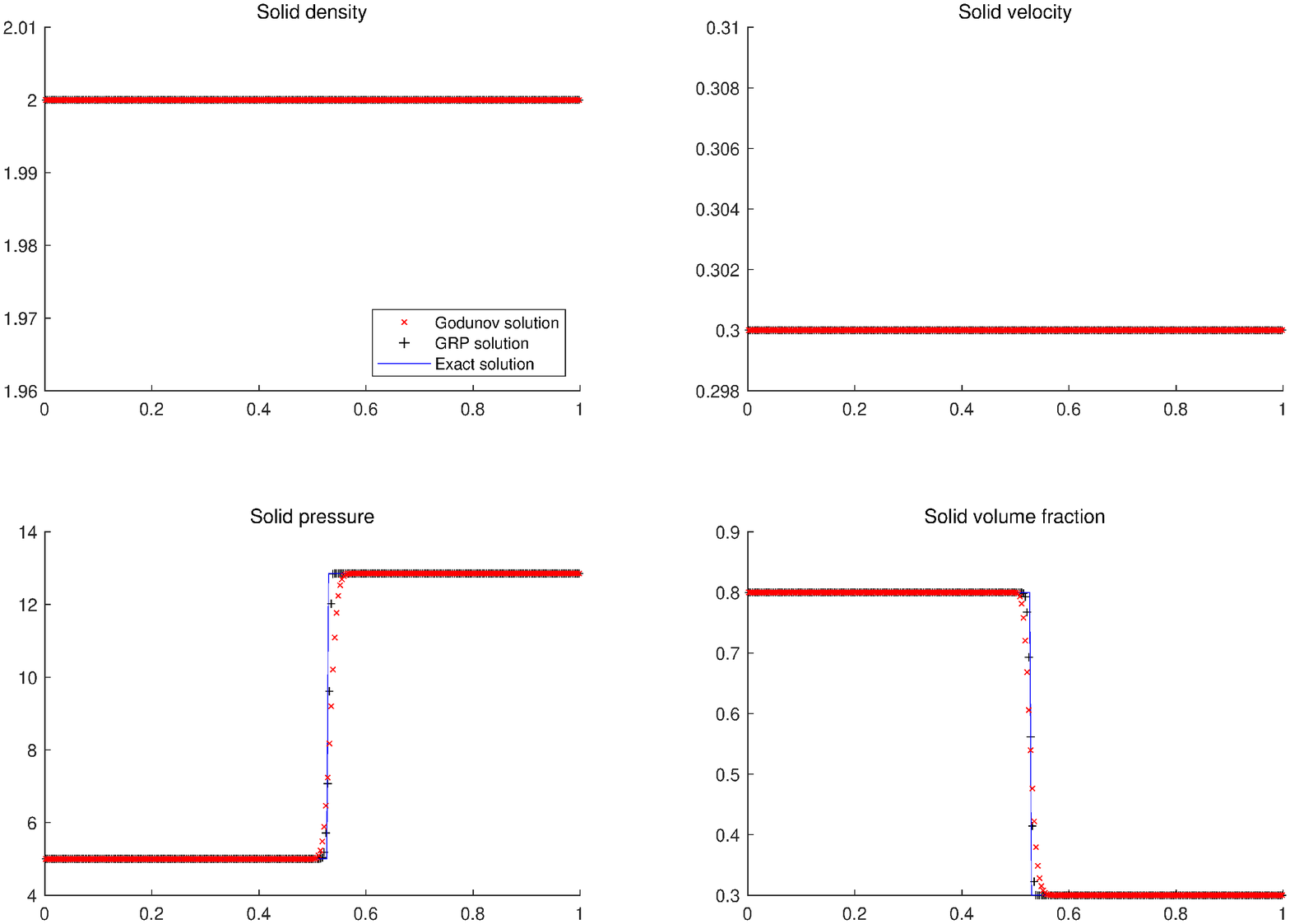}
\end{center}
\end{minipage}
\vfill
\ \\

\ \\
\begin{minipage}{0.8\linewidth}
\begin{center}
\includegraphics[width=\textwidth]{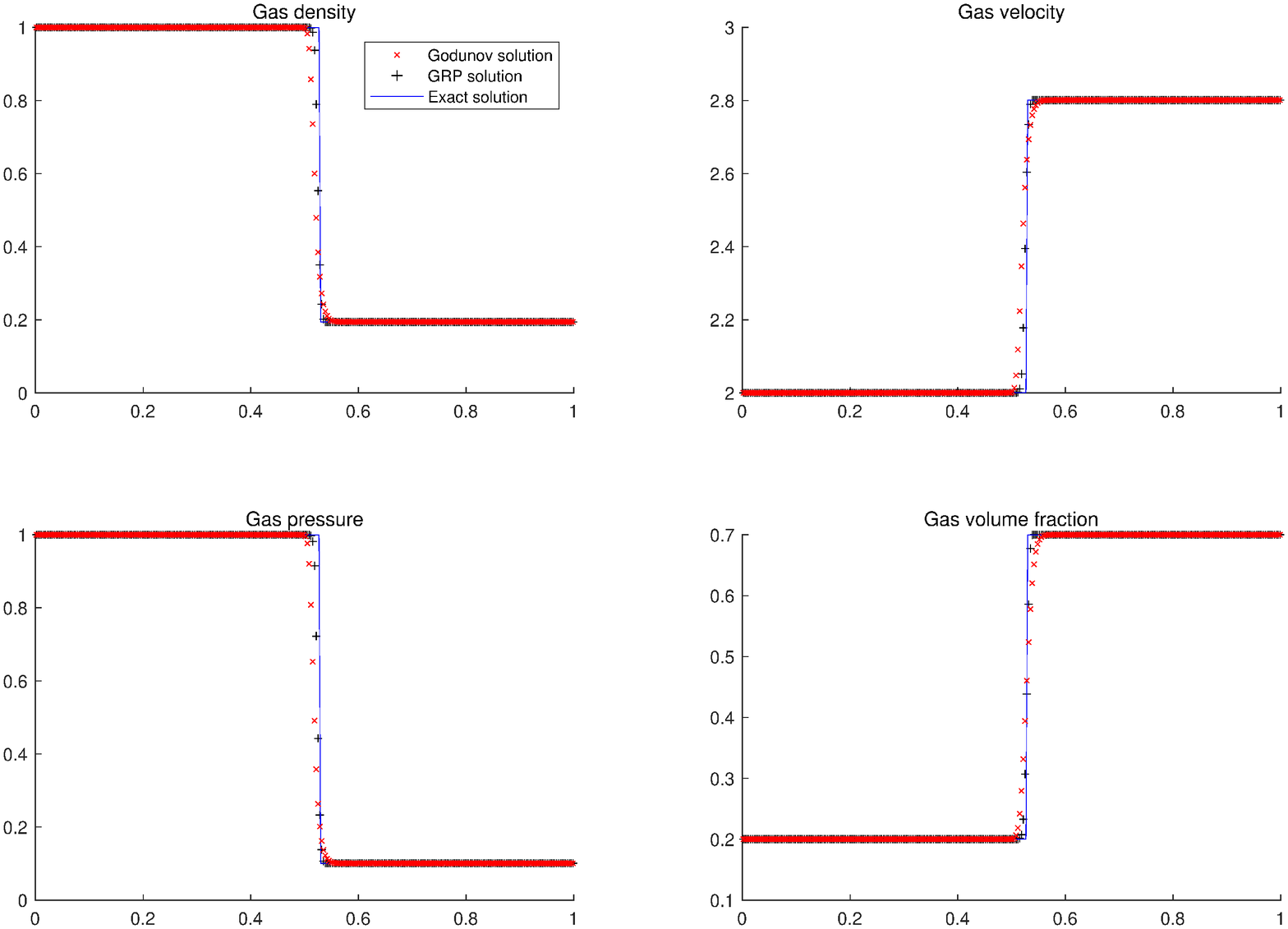}
\end{center}
\end{minipage}
\end{center}
\caption{A single solid contact. Numerical results  by the current staggered-projection scheme and the exact solution are shown at $t=0.1$.}\label{fig:isolate-solid}
\end{figure}
This example was also simulated in \cite{deledicque_exact_2007} by the standard Godunov
scheme with an exact Riemann solver, where   oscillations are present in the vicinity of the solid contact.
The reason was explained in Subsection \ref{subsec:solid-contact-oscillation}.
The numerical results by the current schemes  are shown in Figure \ref{fig:isolate-solid} in nice agreement with the exact solution at $t=0.1$. 

\end{example}

\begin{example}[Case \uppercase\expandafter{\romannumeral2}. 
Coinciding shocks and rarefactions]

This example is \textit{Test~2} taken in \cite{andrianov_riemann_2004}, and the   solution consists of
two coinciding left-going shocks for the gas and solid phase and a right-going gaseous shock within a right-going solid rarefaction wave.
\begin{figure}[htbp]
\begin{center}
\begin{minipage}{0.8\linewidth}
\begin{center}
\includegraphics[width=\textwidth]{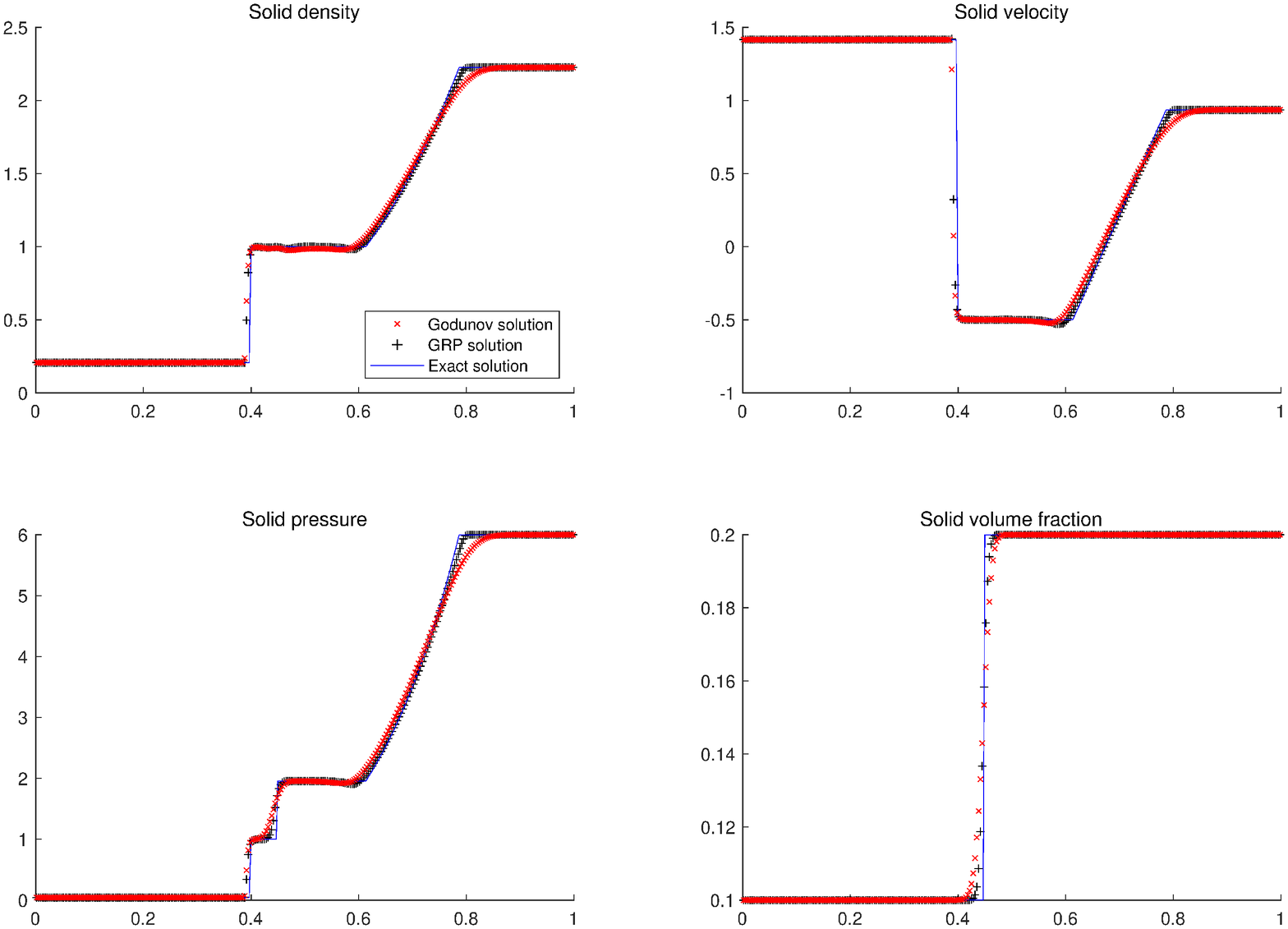}
\end{center}
\end{minipage}
\vfill
\ \\

\ \\
\begin{minipage}{0.8\linewidth}
\begin{center}
\includegraphics[width=\textwidth]{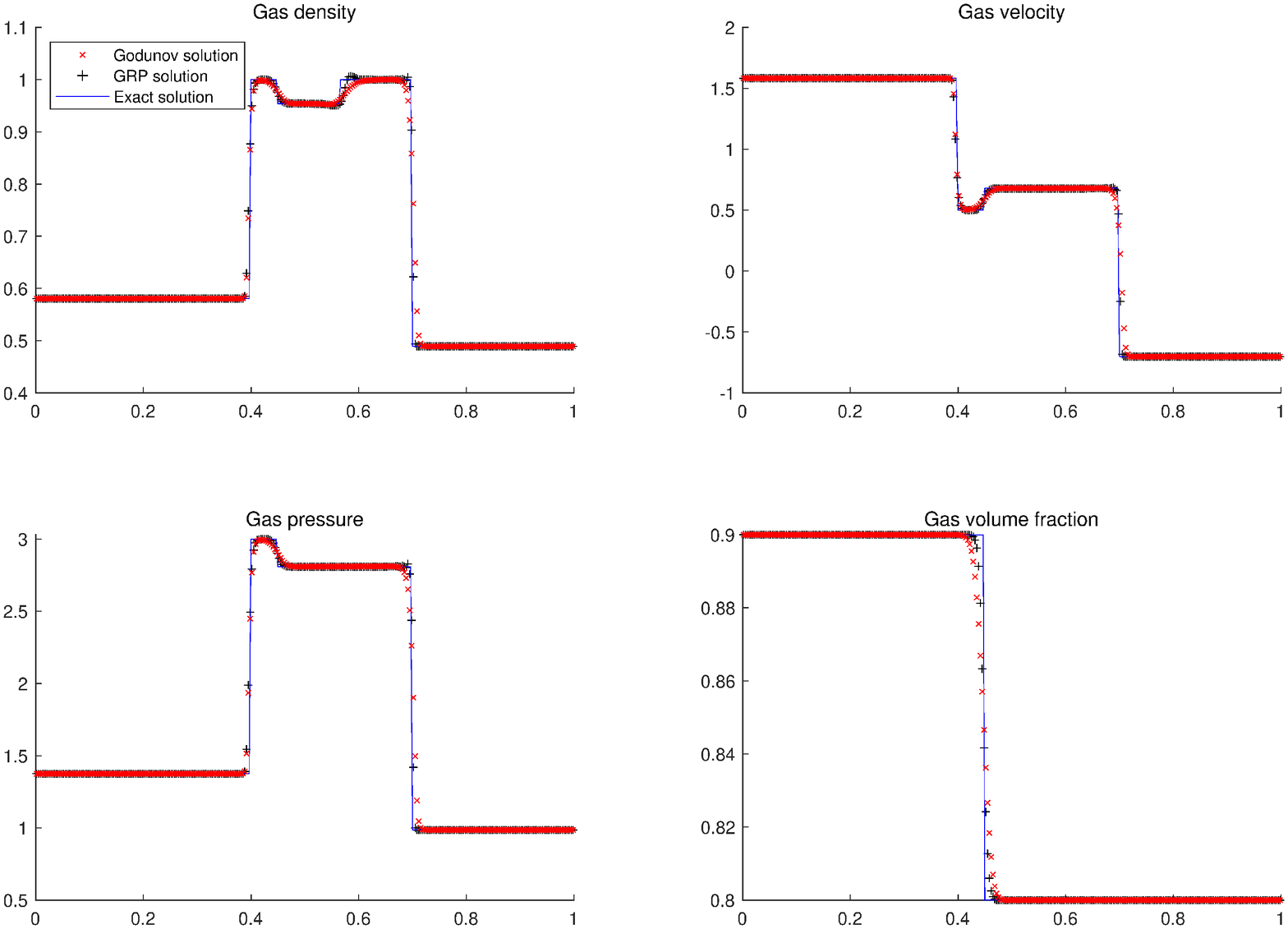}
\end{center}
\end{minipage}
\end{center}
\caption{ Coinciding shocks and rarefactions. Numerical results  by the current staggered-projection scheme and the exact solution are shown  at $t=0.1$.}\label{fig:rare-shock}
\end{figure}
Our numerical results  are presented in Figure \ref{fig:rare-shock}.
Compared with the numerical solution by  a finite-volume Roe method presented in \cite{andrianov_riemann_2004}, our schemes perform better in capturing shocks and resolving the gaseous rarefaction waves.

\end{example}

\begin{example}[Case \uppercase\expandafter{\romannumeral3}. A gas shock approaching a solid contact]\label{ex:test-3}

This example is \textit{Test~4} in \cite{andrianov_riemann_2004}.
In the solution  the solid phase contains a left-going rarefaction wave, a contact and a right-moving shock, the same as the gas phase.
This example involves the Riemann problem  demonstrated in 
Subsection \ref{subsec:coalescence}, i.e., the gaseous shock coincides with the solid contact, leading to  a resonance phenomenon that causes  difficulties to solve such a problem numerically.
The Riemann invariants are no longer constant across the resonant wave containing a solid contact, and the conservation property  is a main factor  to reduce noticeable conservation errors near the shock.
\begin{figure}[htbp]
\begin{center}
\begin{minipage}{0.8\linewidth}
\begin{center}
\includegraphics[width=\textwidth]{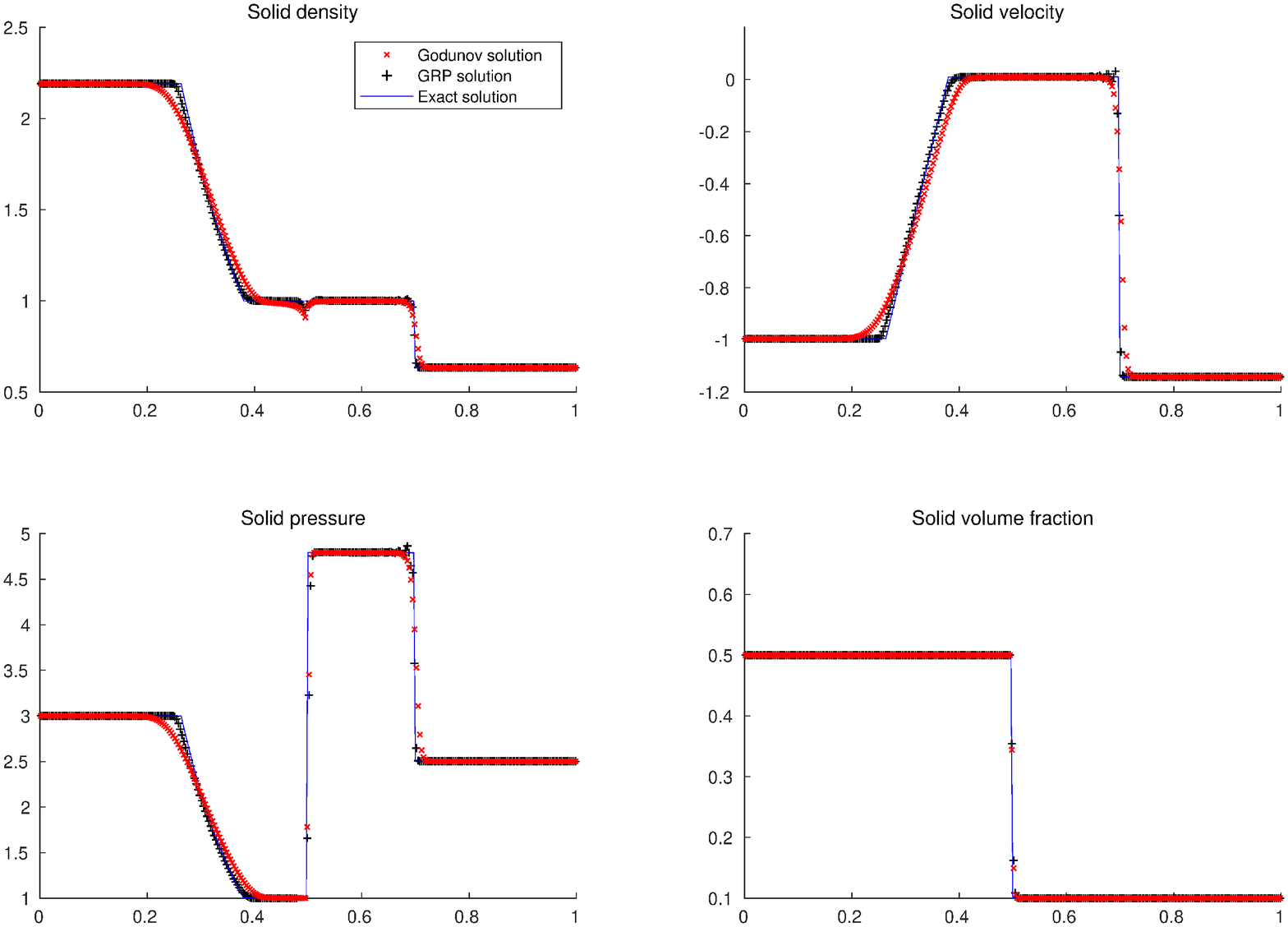}
\end{center}
\end{minipage}
\vfill
\ \\

\ \\
\begin{minipage}{0.8\linewidth}
\begin{center}
\includegraphics[width=\textwidth]{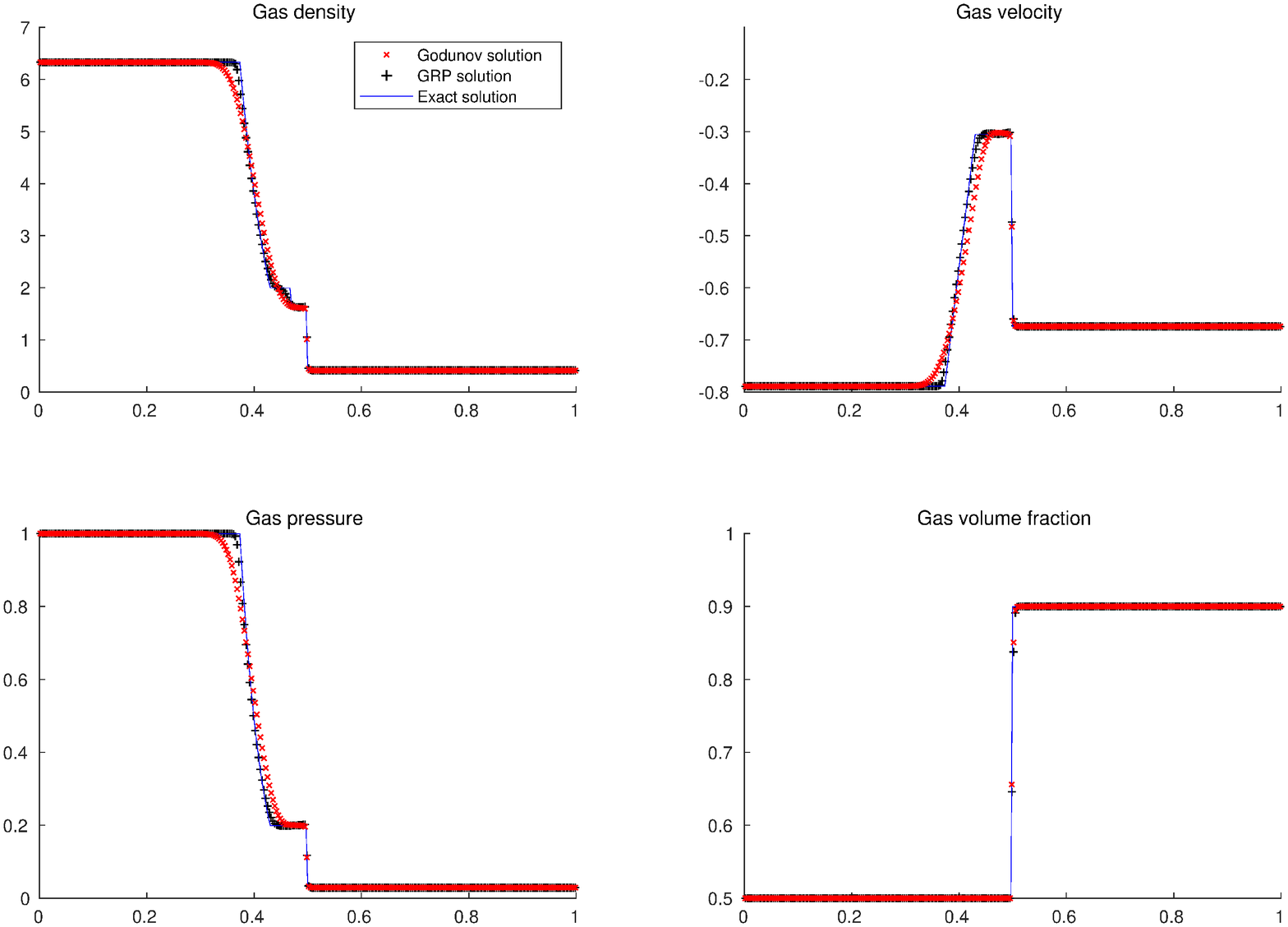}
\end{center}
\end{minipage}
\end{center}
\caption{A gas shock approaching a solid contact. Numerical results  by the current staggered-projection scheme and the exact solution are shown at $t=0.1$.}\label{fig:supersonic}
\end{figure}
For this problem, the numerical results by the current staggered-projection scheme are shown in Figure \ref{fig:supersonic}.
Compared with the numerical solution in \cite{andrianov_riemann_2004} with drastic spurious oscillations, errors in the current solutions are reduced significantly.
\end{example}

\begin{example}[Case \uppercase\expandafter{\romannumeral4}. A shock refraction at a porosity interface]

This example is taken from \cite{karni_hybrid_2010}. A gaseous shock propagates to the right and interacts with a stationary porosity interface.
Initial data are given by $\bm{u}_L$, $\bm{u}_M$ and $\bm{u}_R$ in Case \uppercase\expandafter{\romannumeral4} of Table \ref{tab:BN-init}, in which $\bm{u}_M$ is the middle state between the gaseous shock and the right porosity interface.
This example is simulated in the domain $[0,0.06]$ composed of $M=400$ regular grid cells, and the initial position of the porosity interface is $x=0.03$.
\begin{figure}[htbp]
\begin{center}
\begin{minipage}{0.8\linewidth}
\begin{center}
\includegraphics[width=\textwidth]{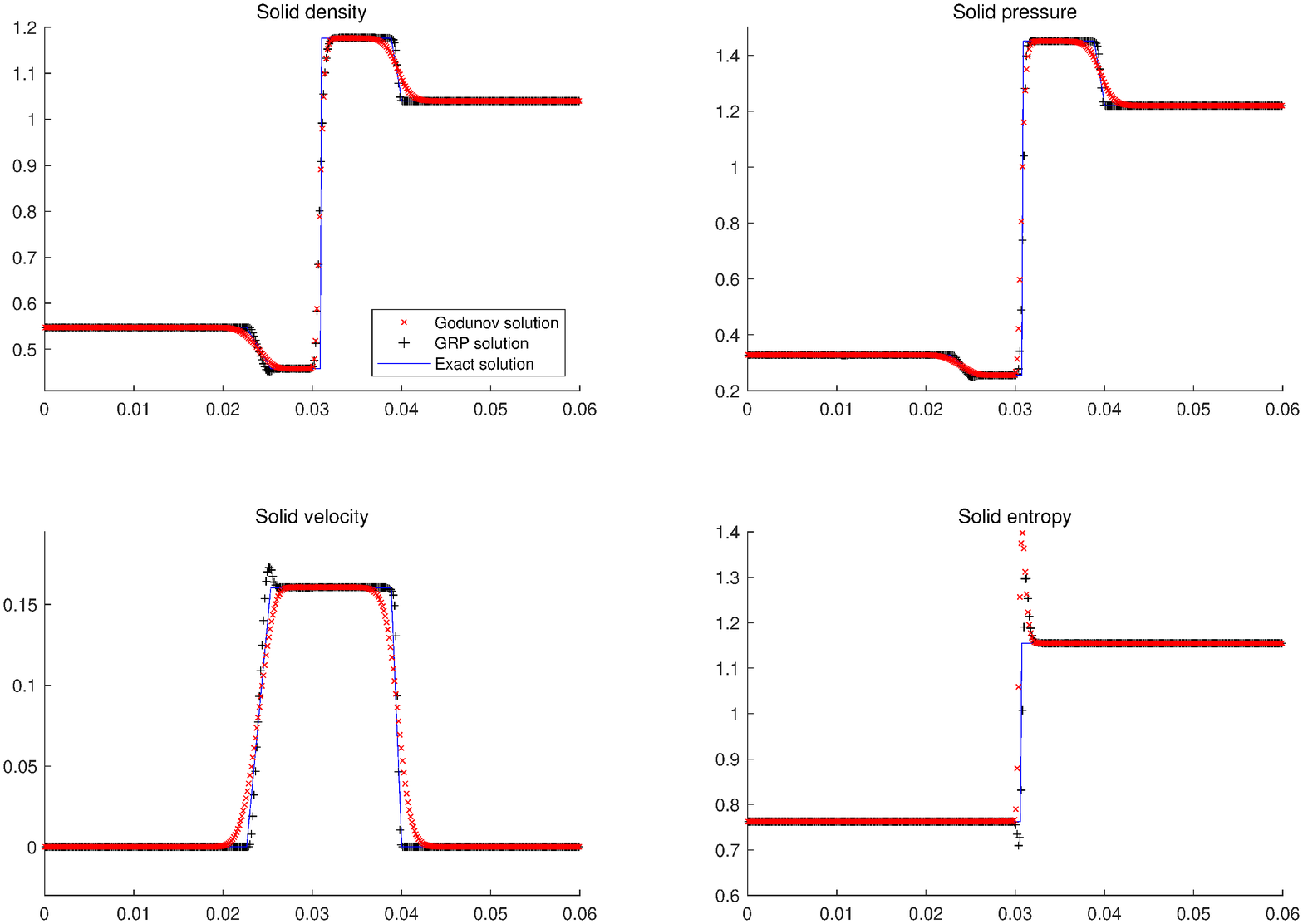}
\end{center}
\end{minipage}
\vfill
\ \\

\ \\
\begin{minipage}{0.8\linewidth}
\begin{center}
\includegraphics[width=\textwidth]{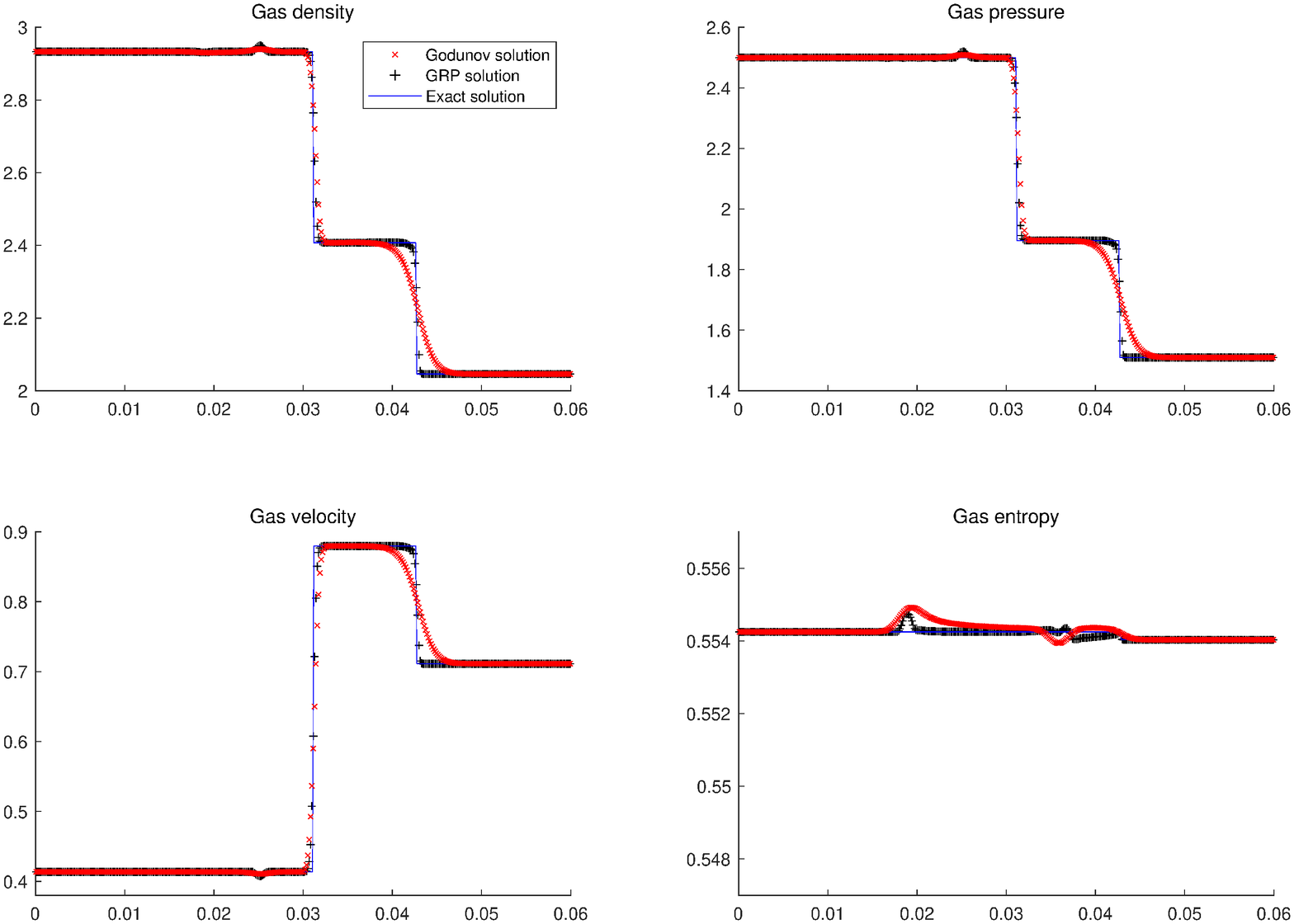}
\end{center}
\end{minipage}
\end{center}
\caption{Shock refraction at a porosity interface. Numerical results  by the current staggered-projection scheme and the exact solution are shown at $t=0.007$ after the refraction.}\label{fig:refraction}
\end{figure}
This example demands  the conservation of numerical schemes and compatible simulation of porosity interfaces.
The numerical solution of an unsplit conservative wave-propagation scheme in \cite{karni_hybrid_2010} exhibited visible errors in the vicinity of the porosity interface.
The numerical solutions by the current staggered-projection scheme  are shown  in Figure \ref{fig:refraction} in  much better agreement with the exact solution at time $t=0.007$ after the interaction of two waves. It is observed  that the current schemes provides a better approximation.  \end{example}

\begin{example}[Accuracy test]\label{ex:acc-test}

This example is an initial value problem in \cite{schwendeman_riemann_2006}.
We take the initial data with a smooth transition of $\alpha_s$ from $0.1$ to $0.9$ and a smooth variation of $v_s$ from $0$ to $1$, specifically,
\begin{equation*}
\alpha_s(x, 0) = 0.5+0.4\tanh(20x-8),\quad
v_s(x, 0) = 0.5+0.5\tanh(20x-10),
\end{equation*} 
and
\begin{equation*}
\rho_s(x, 0) = \rho_g(x, 0) = 1,\quad
p_s(x, 0) = p_g(x, 0) = 1,\quad
v_g(x, 0) = 0
\end{equation*} 
The computational domain $[0,1]$ is divided into $M$ cells for $M=100,200,400,800$, and the left and right numerical boundaries are free boundaries.
In this example, the exact solution is approximated by the staggered-projection GRP scheme with very large number of cells $M=12,800$.

The $L_1$ errors and convergence orders of numerical results for $\bm{u}$ are displayed in Table \ref{tab:Acc-test}. This table shows clearly that the staggered-projection Godunov-type scheme reaches first-order accuracy. In the absence of the limiter, the second-order accuracy of the staggered-projection GRP scheme is achieved. At the local extrema, the minmod limiter reduces the accuracy.

\begin{table}[htb]
\begin{center}
\caption{$L_1$ errors and convergence orders of the vector $\bm{u}$ for Example \ref{ex:acc-test} at $t = 0.1$. The numerical methods are the first-order and second-order staggered-projection schemes with $M$ cells.
}\label{tab:Acc-test}
\begin{tabular}{c|cc|cc|cc}
\hline
 & \multicolumn{2}{c|}{Godunov solution} & \multicolumn{2}{c|}{GRP solution} & \multicolumn{2}{c}{GRP solution (no limiter)}\\
\hline
$M$ & $L_1$ error & order & $L_1$ error & order & $L_1$ error & order \\
\hline
$100$& $1.06\times 10^{-2}$ &      & $1.10\times 10^{-3}$ &      & $3.14\times 10^{-4}$ &     \\
$200$& $5.54\times 10^{-3}$ & 0.94 & $3.17\times 10^{-4}$ & 1.79 & $6.52\times 10^{-5}$ & 2.27 \\
$400$& $2.84\times 10^{-3}$ & 0.96 & $8.50\times 10^{-5}$ & 1.90 & $1.48\times 10^{-5}$  & 2.13 \\
$800$& $1.44\times 10^{-3}$ & 0.98 & $2.20\times 10^{-5}$ & 1.95 & $3.42\times 10^{-6}$ & 2.11 \\
\hline
\end{tabular}
\end{center}
\end{table}

\end{example}

\subsection{2-D homogeneous BN model}

We provide two kinds of examples to show the performance of the current scheme for 2-D problems. The first are the 2-D Riemann problem mimicking those in the context of gas dynamics \cite{Zhang-Zheng, Li-Zhang-1998, Li-4}; and the second is mimicking the shock-bubble interaction problem \cite{haas_interaction_1987}.

\begin{example}[2-D Riemann problems]
The computational domain $[-0.5,0.5]\times[-0.5,0.5]$ is composed of $M\times M$ square cells, and the initial data are composed of piece-wise constant states in the four quadrants, as shown in Table \ref{tab:RP-2D}.
\begin{table}[htb]
\begin{center}
\caption{Initial data for 2-D Riemann problems}\label{tab:RP-2D}
\begin{tabular}{c|cccccccccc}
\hline
Case & Region  & $\alpha_{s}$ & $\rho_{s}$ & $u_{s}$ & $v_{s}$ & $p_{s}$ & $\rho_{g}$ & $u_{g}$ & $v_{g}$ & $p_{g}$\\
\hline
\multirow{4}{*}{\shortstack{\uppercase\expandafter{\romannumeral1}\\
\ \\
$\gamma_s=1.4$\\
$\gamma_g=1.67$}}
&$(x>0,y>0)$& 0.8 & 2 & 0 & 0 & 2 & 1.5 & 0 & 0 & 2\\
&$(x<0,y>0)$& 0.4 & 1 & 0 & 0 & 1 & 0.5 & 0 & 0 & 1\\
&$(x<0,y<0)$& 0.8 & 2 & 0 & 0 & 2 & 1.5 & 0 & 0 & 2\\
&$(x>0,y<0)$& 0.4 & 1 & 0 & 0 & 1 & 0.5 & 0 & 0 & 1\\
\hline
\multirow{4}{*}{\shortstack{\uppercase\expandafter{\romannumeral2}\\
\ \\
$\gamma_s=1.6$\\
$\gamma_g=1.4$}}
&$(x>0,y>0)$& 0.6 & 0.5 & 0 & 0 & 0.6  & 3   & 0 & 0 & 0.3 \\
&$(x<0,y>0)$& 0.4 & 1.2 & 0 & 0 & 0.12 & 1.2 & 0 & 0 & 0.12\\
&$(x<0,y<0)$& 0.6 & 0.5 & 0 & 0 & 0.6  & 3   & 0 & 0 & 0.3 \\
&$(x>0,y<0)$& 0.4 & 1.2 & 0 & 0 & 0.12 & 1.0 & 0 & 0 & 0.12\\
\hline
\end{tabular}
\end{center}
\end{table}
Case \uppercase\expandafter{\romannumeral1} is taken from  \cite{dumbser_high-order_2013,fraysse_upwind_2016} for which the 1-D  initial data  is picked up in \cite{deledicque_exact_2007}. Case \uppercase\expandafter{\romannumeral2} is a 2-D Riemann problem in \cite{daude_computation_2016} containing a supersonic configuration,

The numerical results of the staggered-projection GRP scheme are presented with  $M=200$ in Figure \ref{fig:BN-2D-RP1}, \ref{fig:BN-2D-RP2}.
The corresponding reference solutions are computed by the first-order staggered-projection scheme in a finer regular grid with $M=1000$.
The staggered-projection GRP scheme can capture various waves well and displays high resolution for  Cases \uppercase\expandafter{\romannumeral1},  \uppercase\expandafter{\romannumeral2}.
\begin{figure}[htbp]
\begin{minipage}{.04\linewidth}
$\alpha_s$
\end{minipage}
\begin{minipage}{.48\linewidth}
\begin{center}
\includegraphics[width=0.8\textwidth]{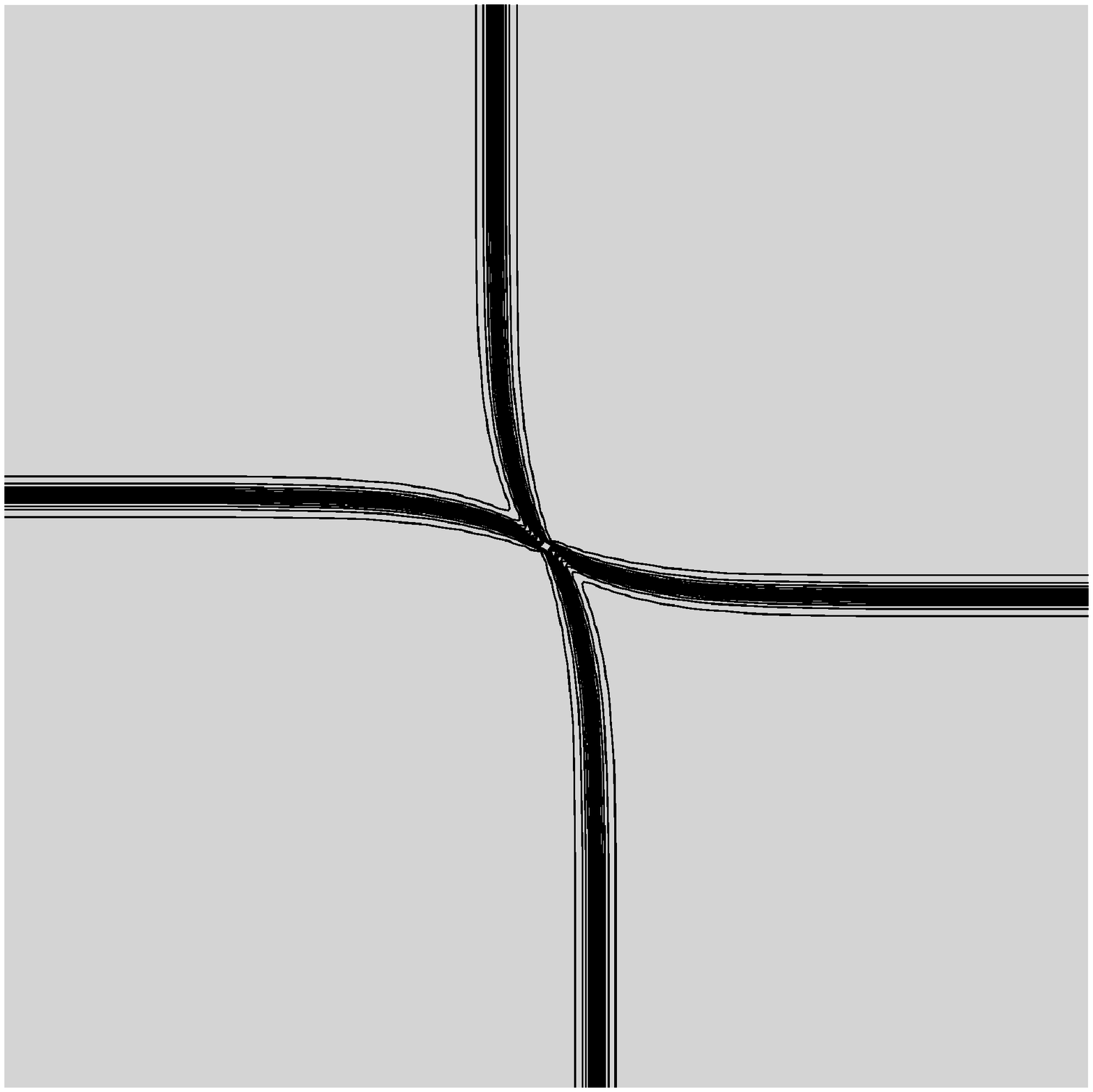}
\end{center}
\end{minipage}
\begin{minipage}{.48\linewidth}
\begin{center}
\includegraphics[width=0.8\textwidth]{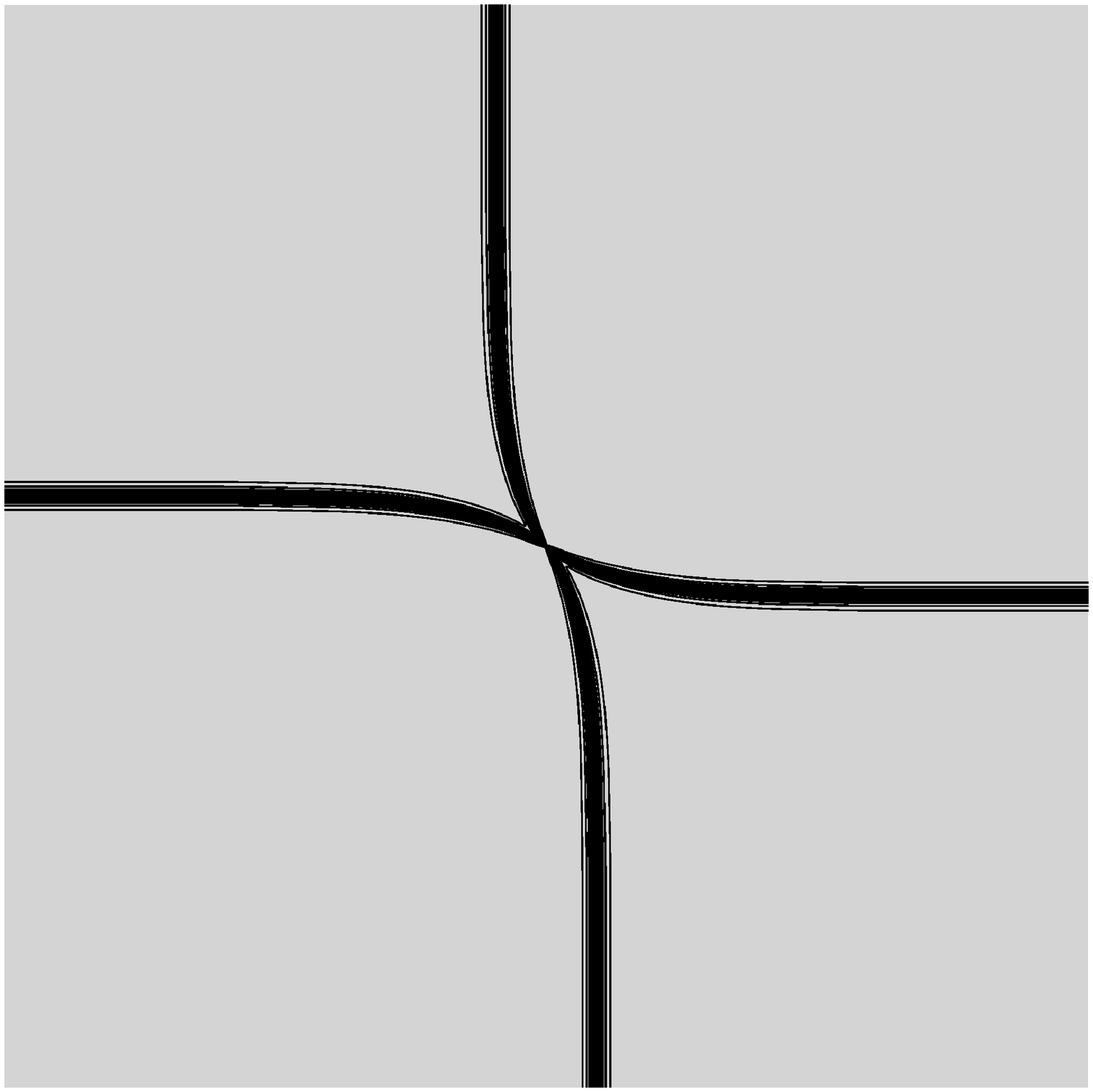}
\end{center}
\end{minipage}
\vspace{0.4em}

\begin{minipage}{.04\linewidth}
$\rho_s$
\end{minipage}
\begin{minipage}{.48\linewidth}
\begin{center}
\includegraphics[width=0.8\textwidth]{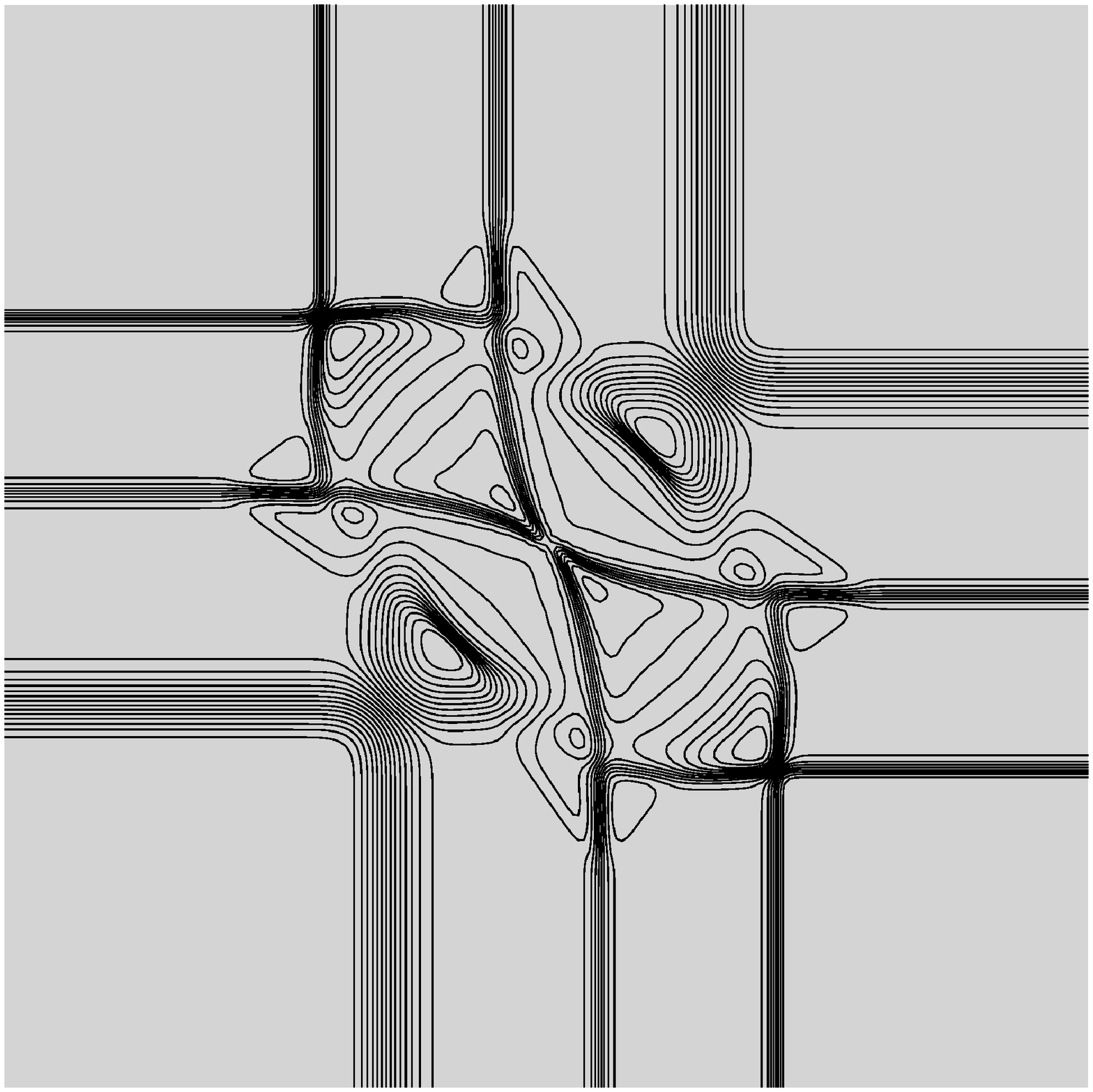}
\end{center}
\end{minipage}
\begin{minipage}{.48\linewidth}
\begin{center}
\includegraphics[width=0.8\textwidth]{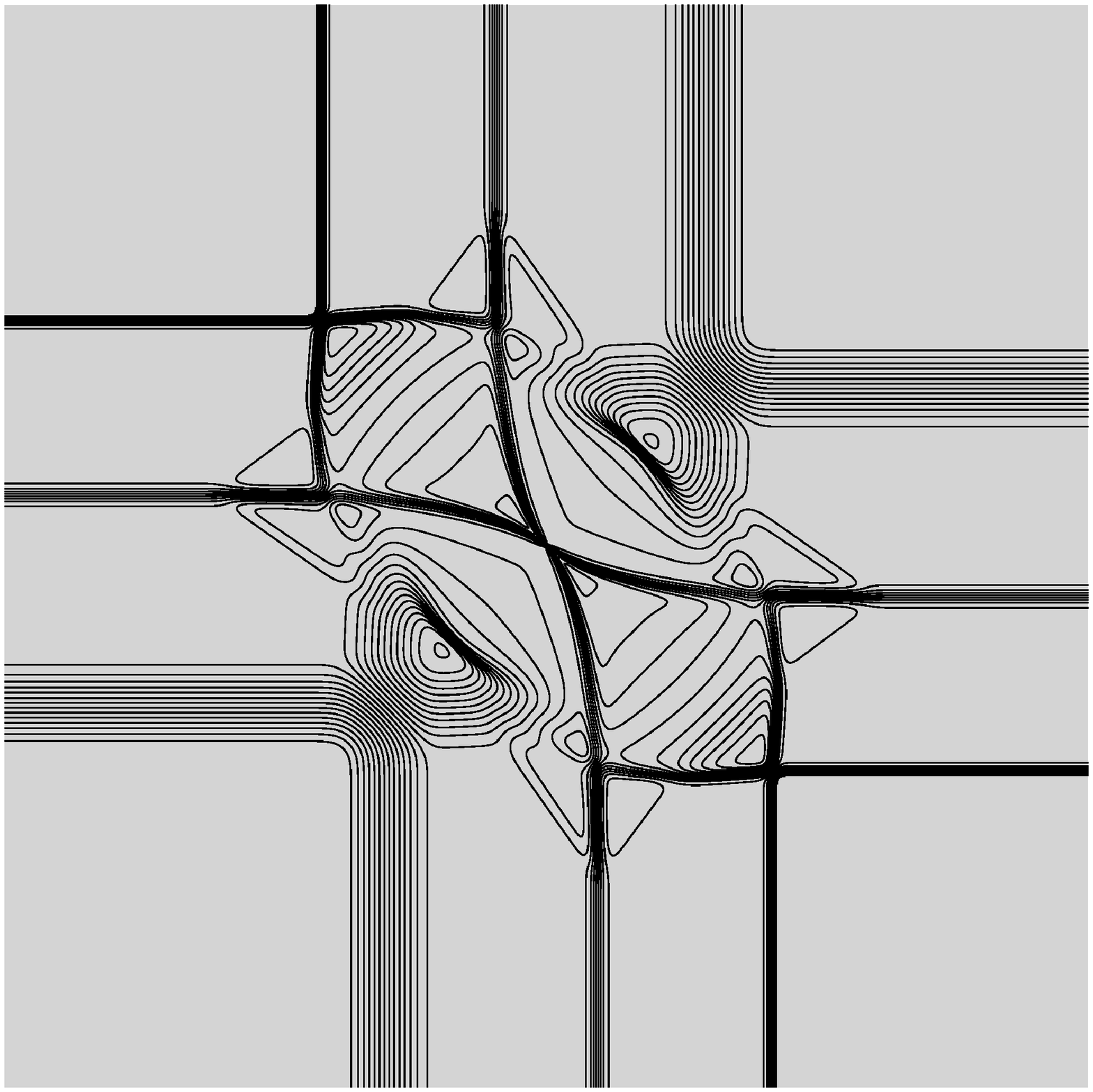}
\end{center}
\end{minipage}
\vspace{0.4em}

\begin{minipage}{.04\linewidth}
$\rho_g$
\end{minipage}
\begin{minipage}{.48\linewidth}
\begin{center}
\includegraphics[width=0.8\textwidth]{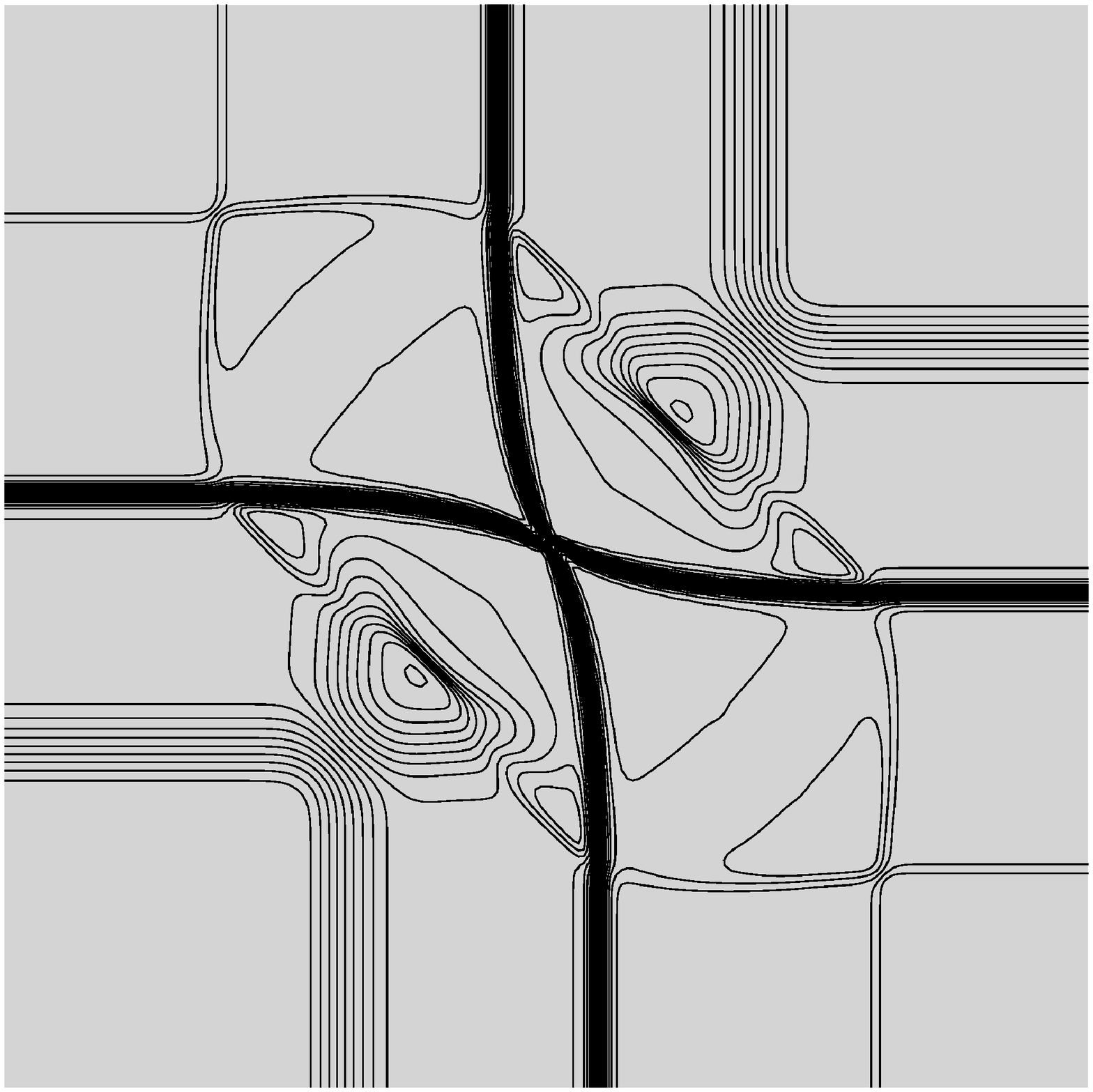}
\end{center}
\end{minipage}
\begin{minipage}{.48\linewidth}
\begin{center}
\includegraphics[width=0.8\textwidth]{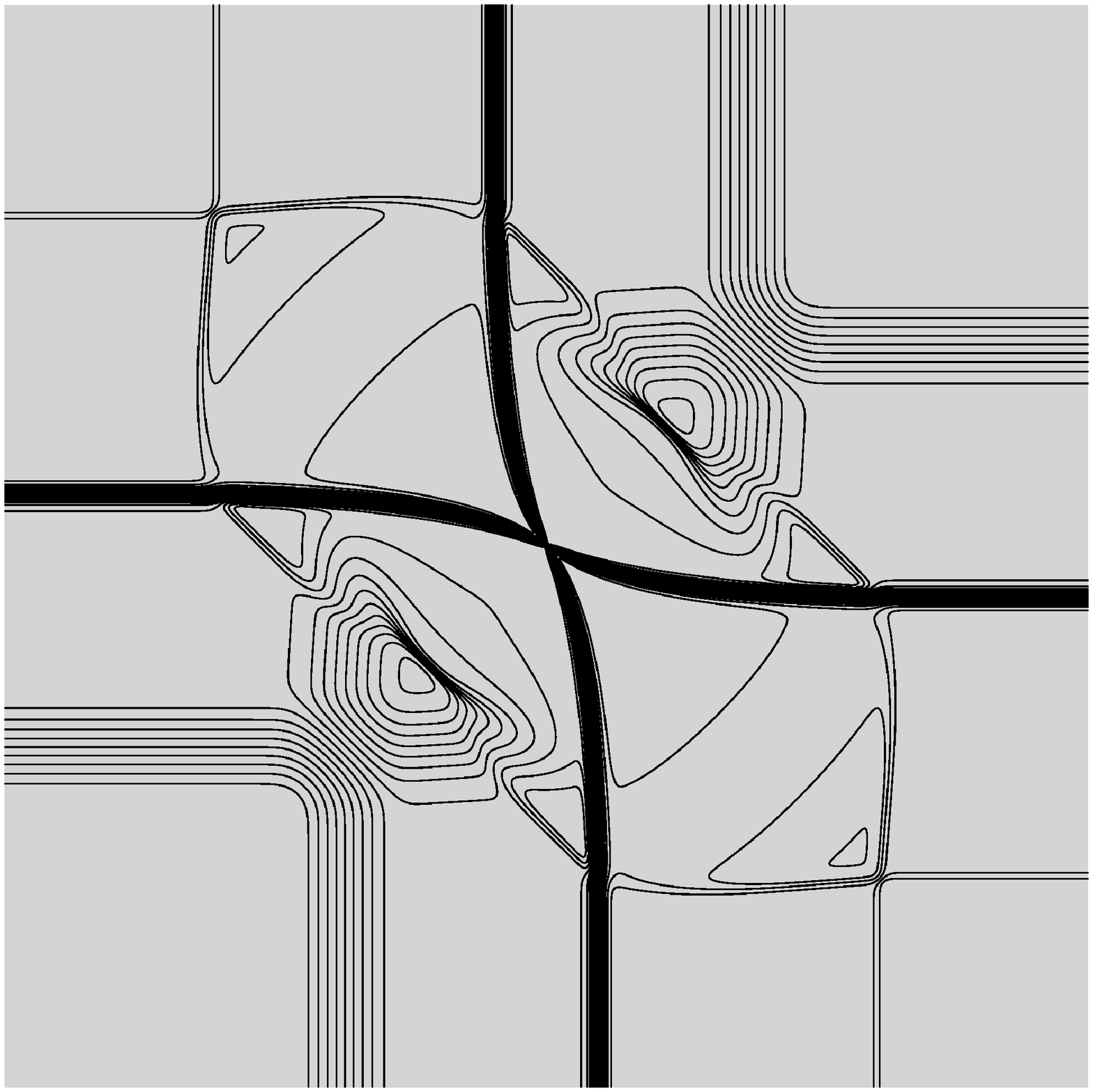}
\end{center}
\end{minipage}
\vspace{0.4em}

\begin{minipage}{.04\linewidth}
\ 
\end{minipage}
\begin{minipage}{.48\linewidth}
\begin{center}
Staggered-projection GRP scheme
\end{center}
\end{minipage}
\begin{minipage}{.48\linewidth}
\begin{center}
Reference solution
\end{center}
\end{minipage}
\caption{2-D Riemann problem \uppercase\expandafter{\romannumeral1}: Numerical results of the staggered-projection GRP scheme ($M=200$) and the reference solution ($M=1000$) at $t=0.15$.}\label{fig:BN-2D-RP1}
\end{figure}

\begin{figure}[htbp]
\begin{minipage}{.04\linewidth}
$\alpha_s$
\end{minipage}
\begin{minipage}{.48\linewidth}
\begin{center}
\includegraphics[width=0.8\textwidth]{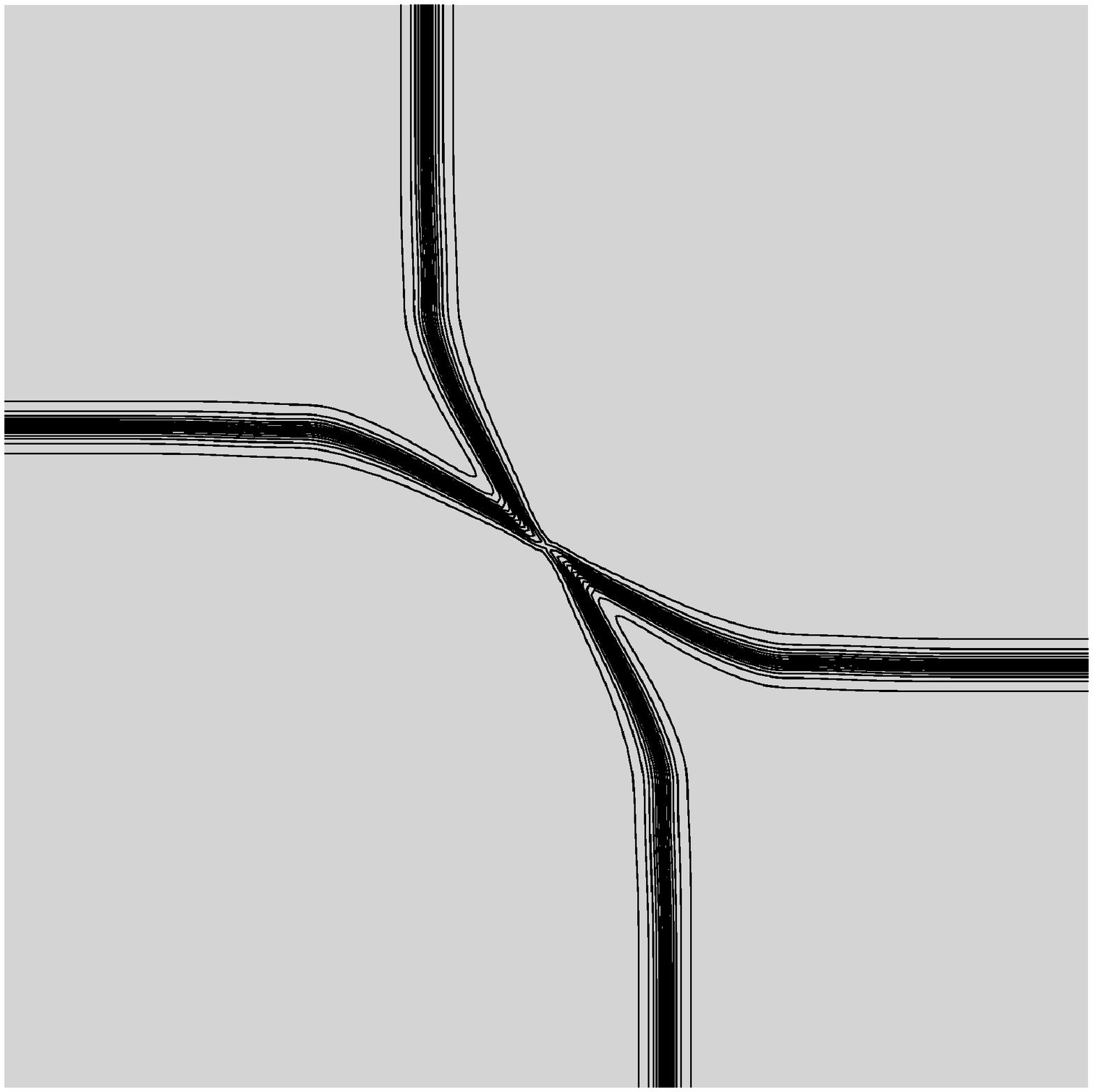}
\end{center}
\end{minipage}
\begin{minipage}{.48\linewidth}
\begin{center}
\includegraphics[width=0.8\textwidth]{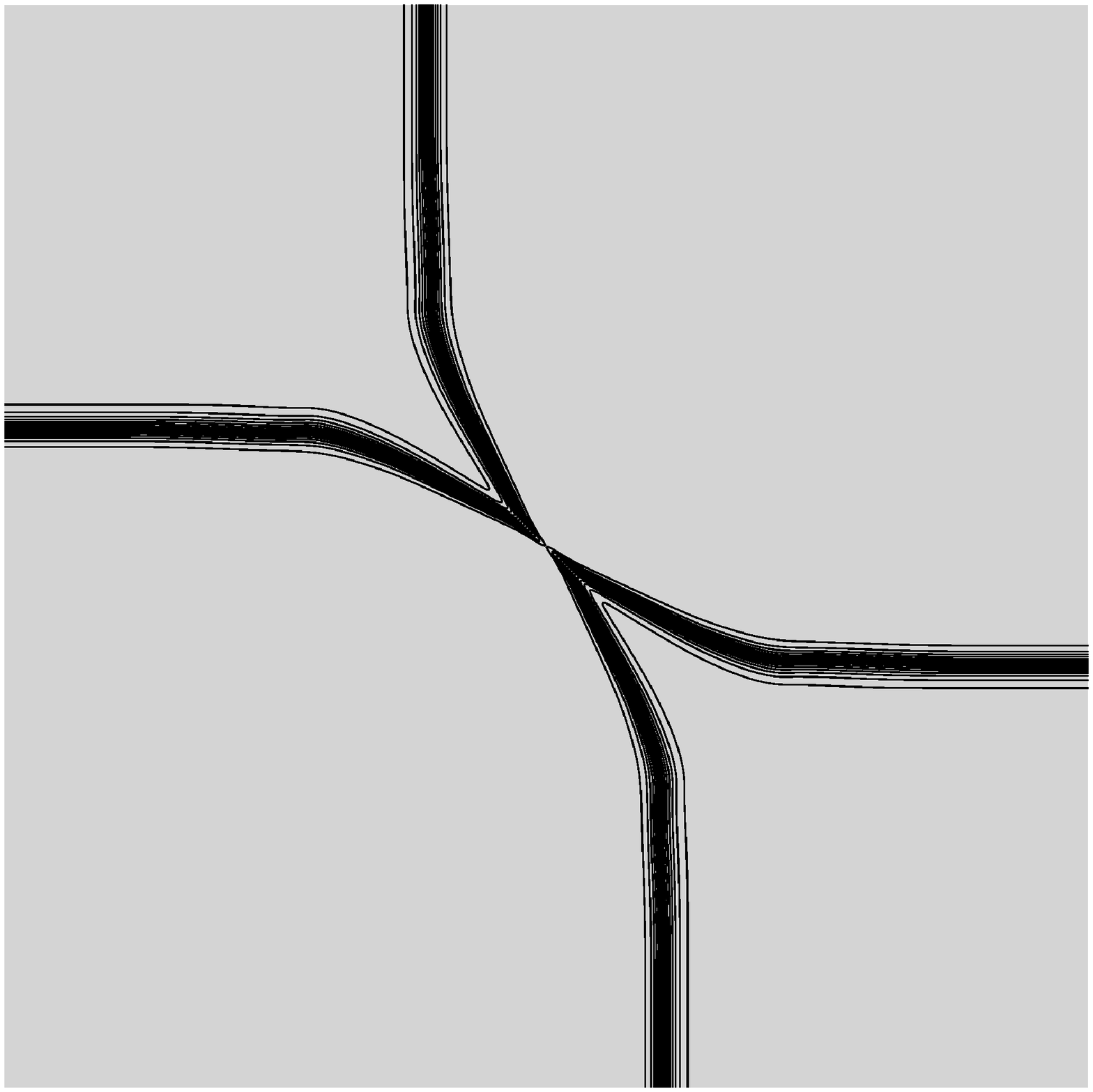}
\end{center}
\end{minipage}
\vspace{0.4em}

\begin{minipage}{.04\linewidth}
$\rho_s$
\end{minipage}
\begin{minipage}{.48\linewidth}
\begin{center}
\includegraphics[width=0.8\textwidth]{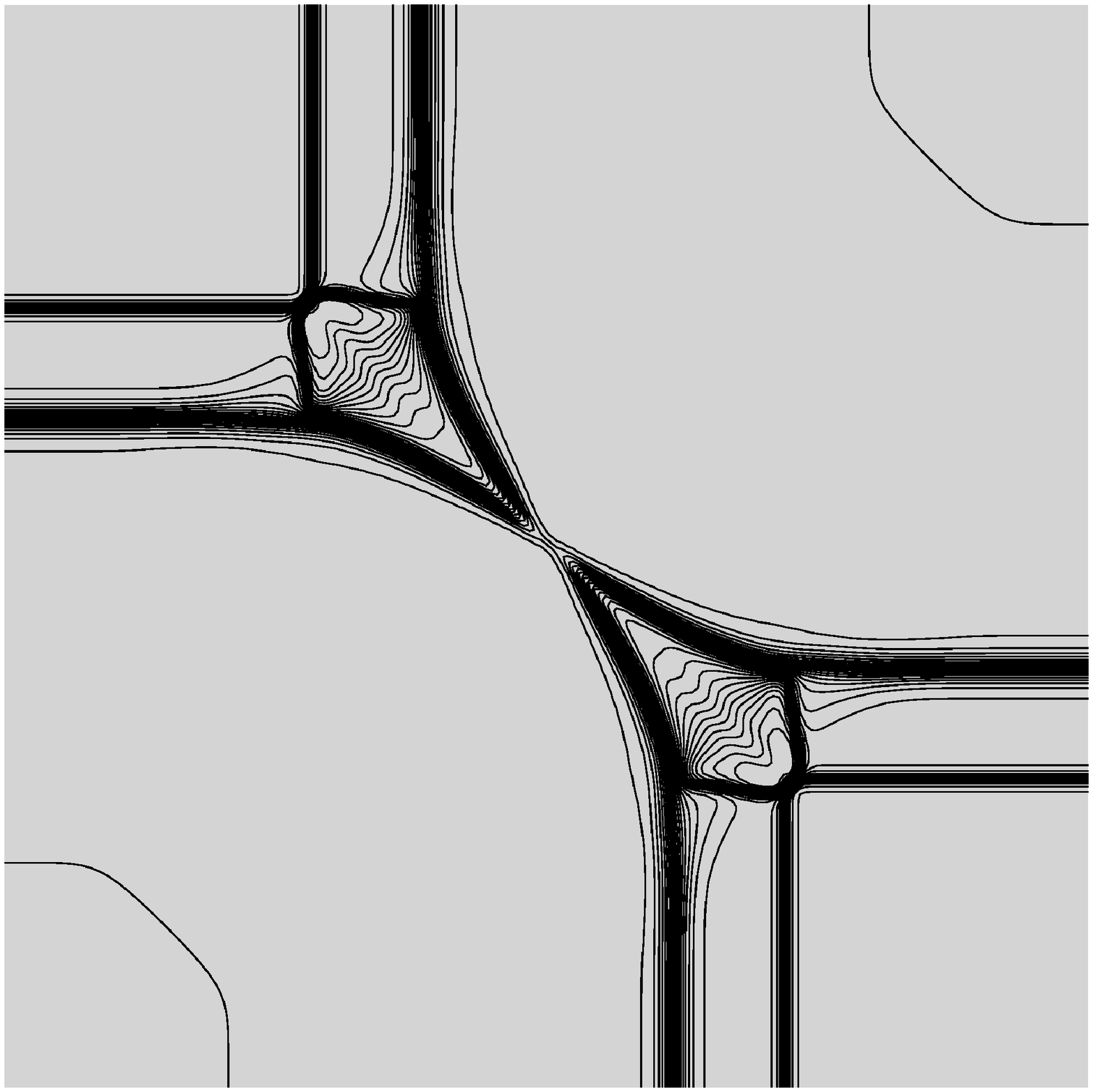}

\end{center}
\end{minipage}
\begin{minipage}{.48\linewidth}
\begin{center}
\includegraphics[width=0.8\textwidth]{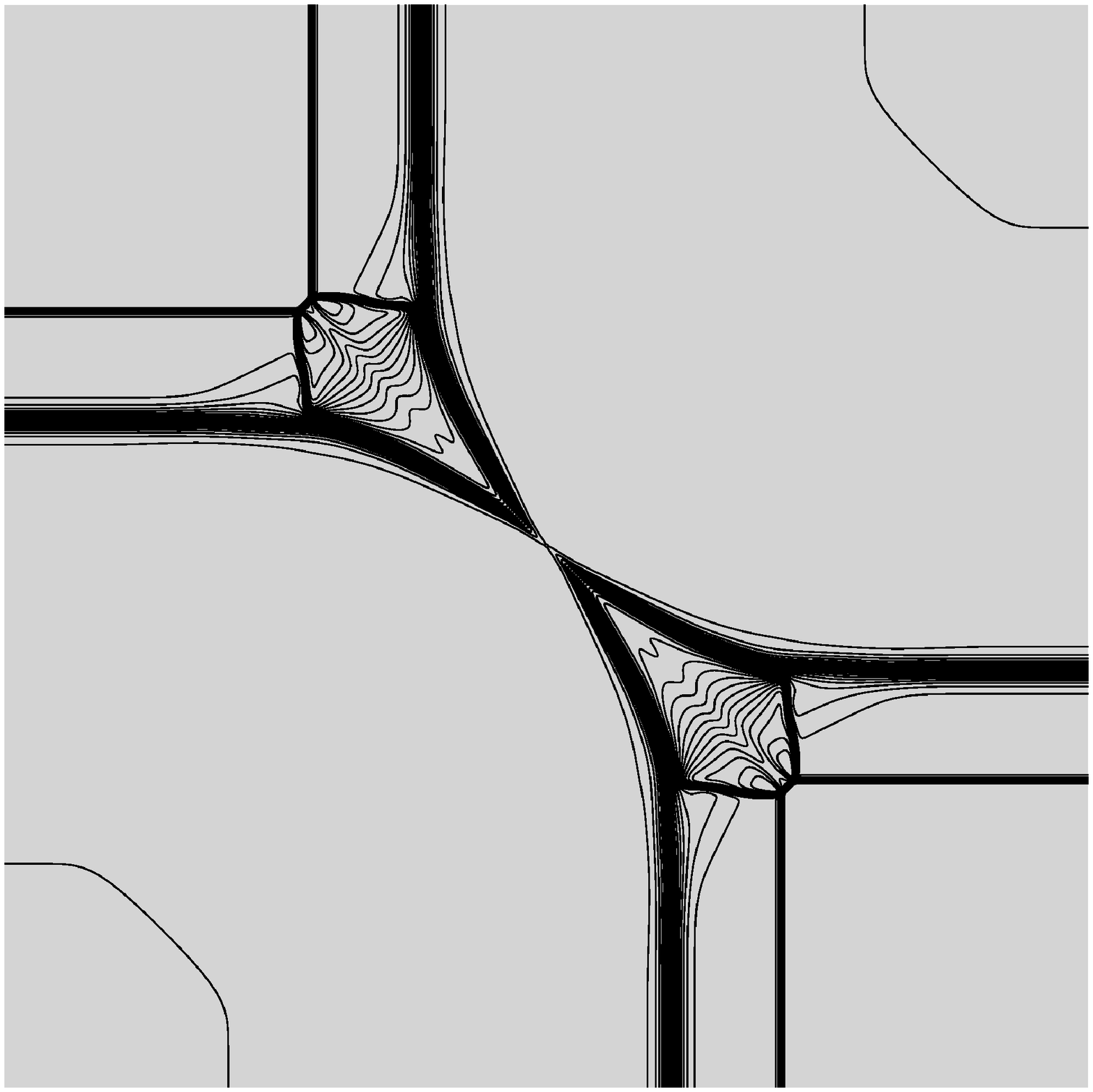}
\end{center}
\end{minipage}
\vspace{0.4em}

\begin{minipage}{.04\linewidth}
$\rho_g$
\end{minipage}
\begin{minipage}{.48\linewidth}
\begin{center}
\includegraphics[width=0.8\textwidth]{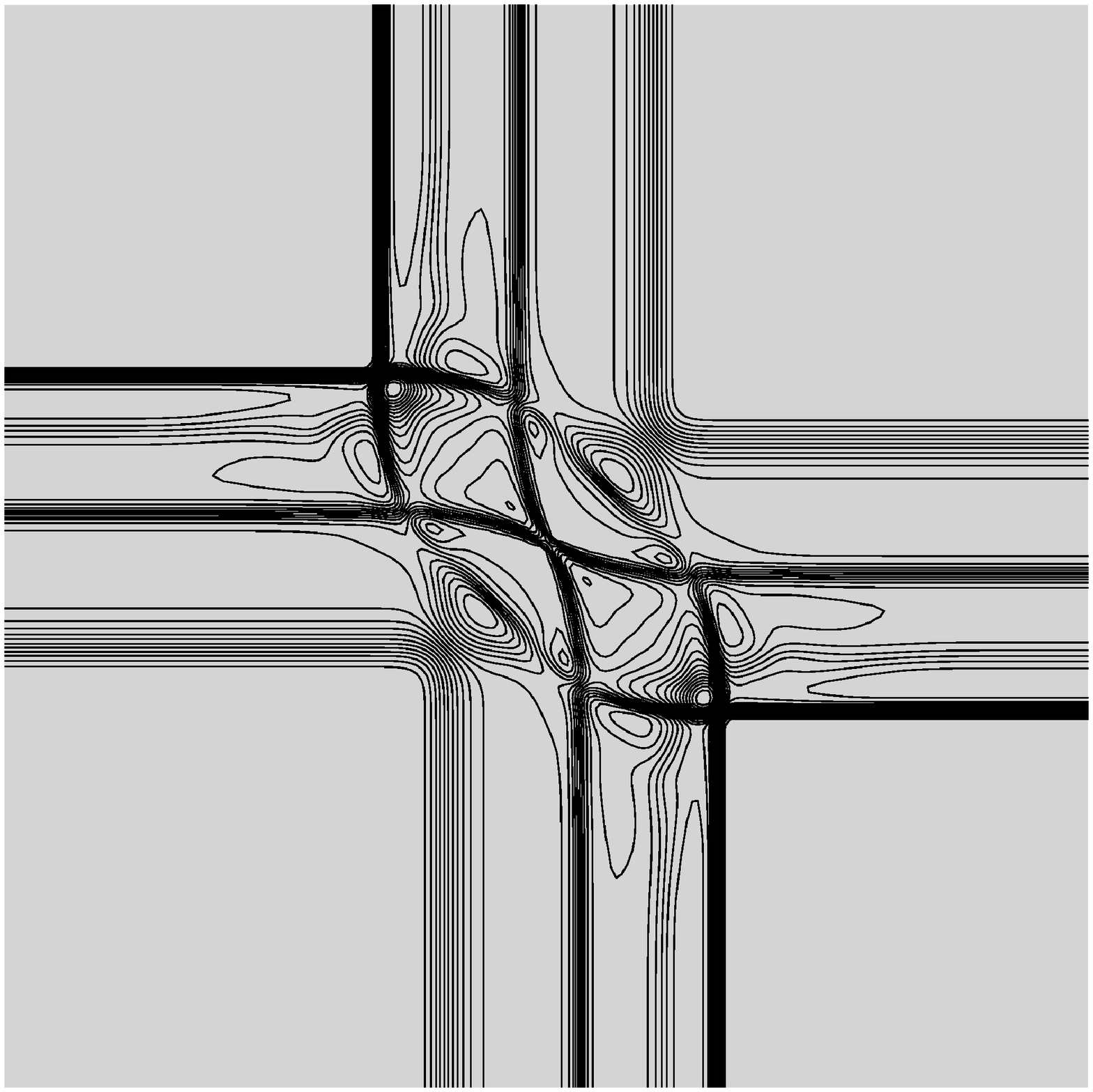}
\end{center}
\end{minipage}
\begin{minipage}{.48\linewidth}
\begin{center}
\includegraphics[width=0.8\textwidth]{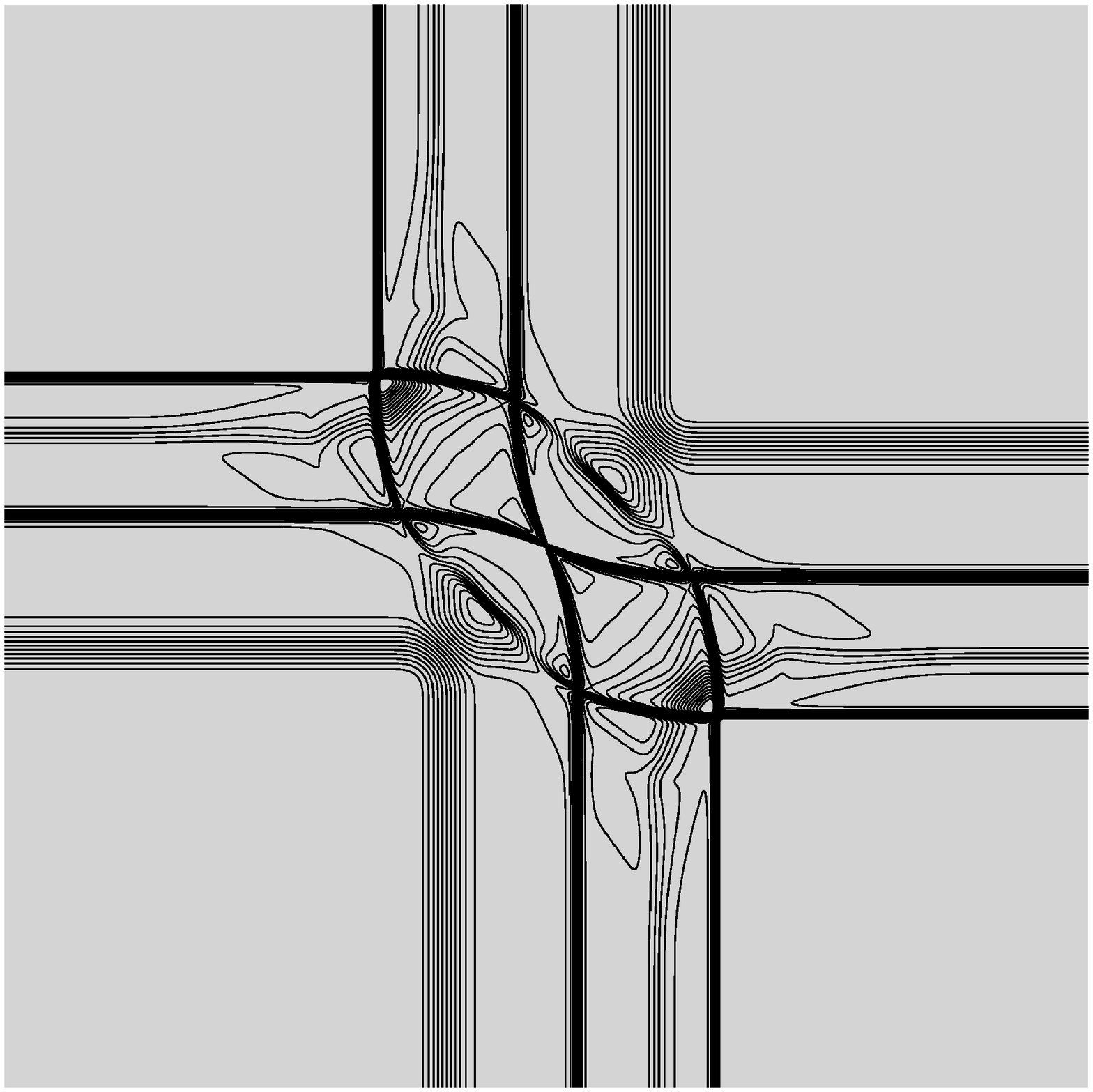}
\end{center}
\end{minipage}
\vspace{0.4em}

\begin{minipage}{.04\linewidth}
\ 
\end{minipage}
\begin{minipage}{.48\linewidth}
\begin{center}
Staggered-projection GRP scheme
\end{center}
\end{minipage}
\begin{minipage}{.48\linewidth}
\begin{center}
Reference solution
\end{center}
\end{minipage}
\caption{2-D Riemann problem \uppercase\expandafter{\romannumeral2}: Numerical results of the staggered-projection GRP scheme ($M=200$) and the reference solution ($M=1000$) at $t=0.1$.}\label{fig:BN-2D-RP2}
\end{figure}

%
%
%
%

\end{example}

\begin{example}[Shock-cylinder interaction]

This example carries out  the numerical simulation of  a physical experiment  in \cite{haas_interaction_1987} that is  used as a benchmark test  evaluating numerical methods of multiphase flows. Some existing numerical results can be found in \cite{daude_computation_2016,lochon_hllc-type_2016}. 
In this example, a weak shock with the shock Mach number $M_s=1.22$ propagates in atmospheric air and collides with a stationary helium cylinder. The computational domain $[0,2.5]\times[0,0.89]$ consists of $560\times 200$ cells and the initial configuration is set in Figure \ref{cylinder}, where the diameter of the cylinder is $D = 0.5$.
\begin{figure}[htb]
\centering
\includegraphics[width=0.6\textwidth]{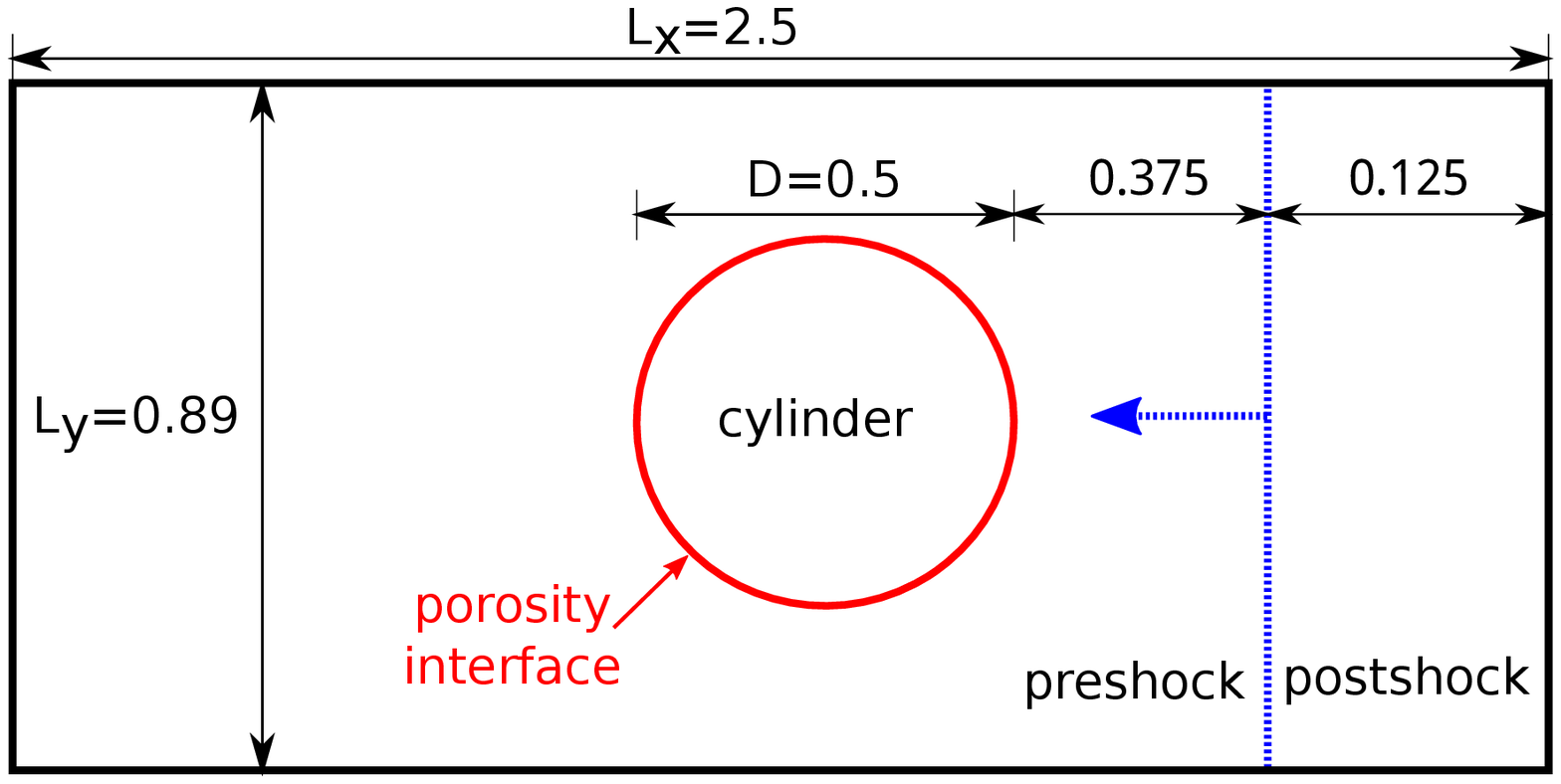}
\caption{Diagram of the shock-cylinder interaction problem}\label{cylinder}
\end{figure}
The upper and lower boundaries are solid walls, whereas the left and right boundaries are non-reflective boundaries.
Two phases inside and outside the cylinder are regarded as polytropic gases, with the specific heat ratio $\gamma_{k1}=1.4$ for air (denoted as the phase $k1$) and $\gamma_{k2}=1.67$ for helium (denoted as the phase $k2$). The initial data are presented in Table \ref{tab:parameters}.

\begin{table}[htbp]
\begin{center}
\caption{Initial data of the shock-cylinder interaction problem}\label{tab:parameters}
\begin{tabular}{c|ccccccc}
\hline
Region  & $\alpha_{k1}$ & $\rho_{k1}$ & $p_{k1}$ & $u_{k1}$ & $\rho_{k2}$ & $p_{k2}$ & $u_{k2}$ \\
\hline
 inside cylinder             & 0.0001 & 1      &     1 &        0 & 0.1821 &1 & 0\\
outside cylinder, pre-shock  & 0.9999 & 1      &     1 &        0 & 0.1821 &1 & 0\\
outside cylinder, post-shock & 0.9999 & 1.3764 &1.5698 &$-0.3947$ & 0.1821 &1 &$-0.3947$\\
\hline 
\end{tabular}
\end{center}
\end{table}

The staggered-projection GRP scheme is applied to simulate this experimental example numerically.
The phase inside the cylinder is regarded as the solid phase or gas phase separately to carry out the simulation, and the differences between the two settings are compared.
Numerical results of the total density $\rho = \alpha_{k1}\rho_{k1}+\alpha_{k2}\rho_{k2}$ in the shock-cylinder interaction problem are displayed in Figure \ref{fig:shock-cylinder}.
It is shown that the staggered-projection GRP scheme can capture interface clearly on a sparse grid.
The instability arising at the interface due to the shock acceleration is known as the Richtmyer-Meshkov instability.
By comparing the numerical results of different settings for the cylindrical phase, it can be found that the interface shapes after the collision are different obviously.
This numerical phenomenon illustrates that when the homogeneous BN two-phase flow model is utilized to simulate two separate phases approximatively, the gas phase and the solid phase have different features because of the nozzling terms.

\begin{figure}[htbp]
\begin{minipage}{.08\linewidth}
$t=0$
\end{minipage}
\begin{minipage}{.46\linewidth}
\begin{center}
\includegraphics[width=0.98\textwidth]{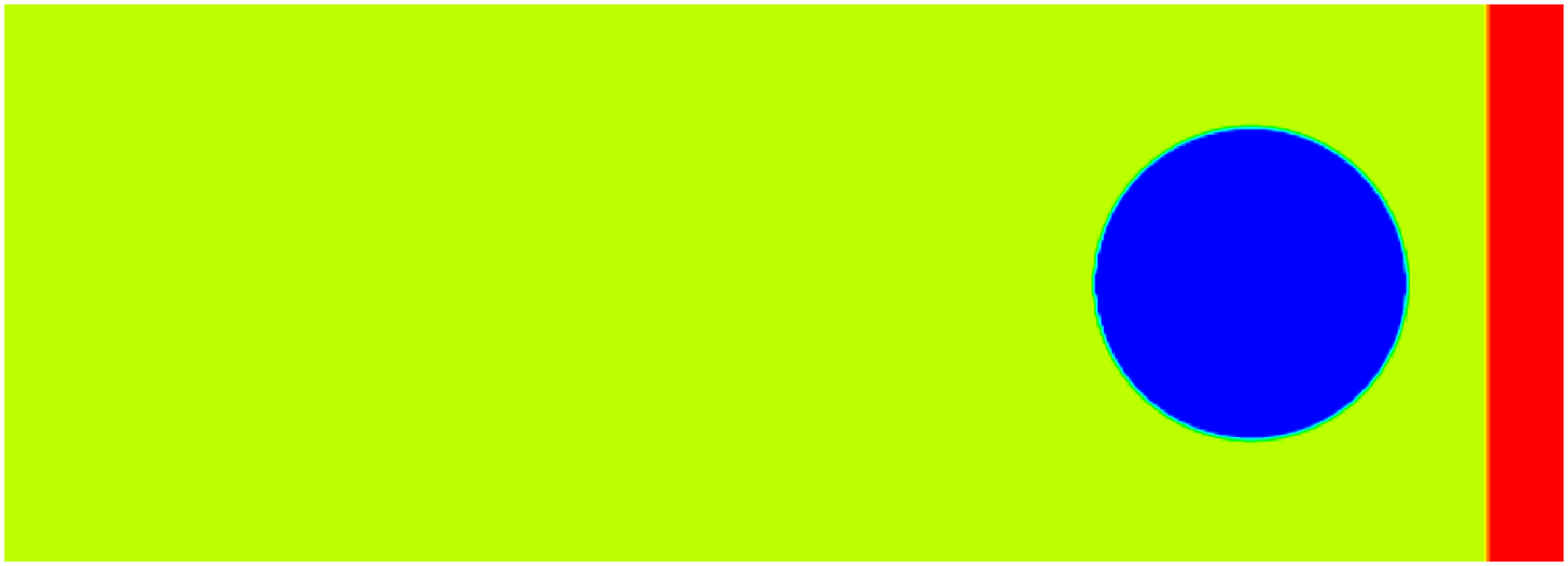}
\end{center}
\end{minipage}
\begin{minipage}{.46\linewidth}
\begin{center}
\includegraphics[width=0.98\textwidth]{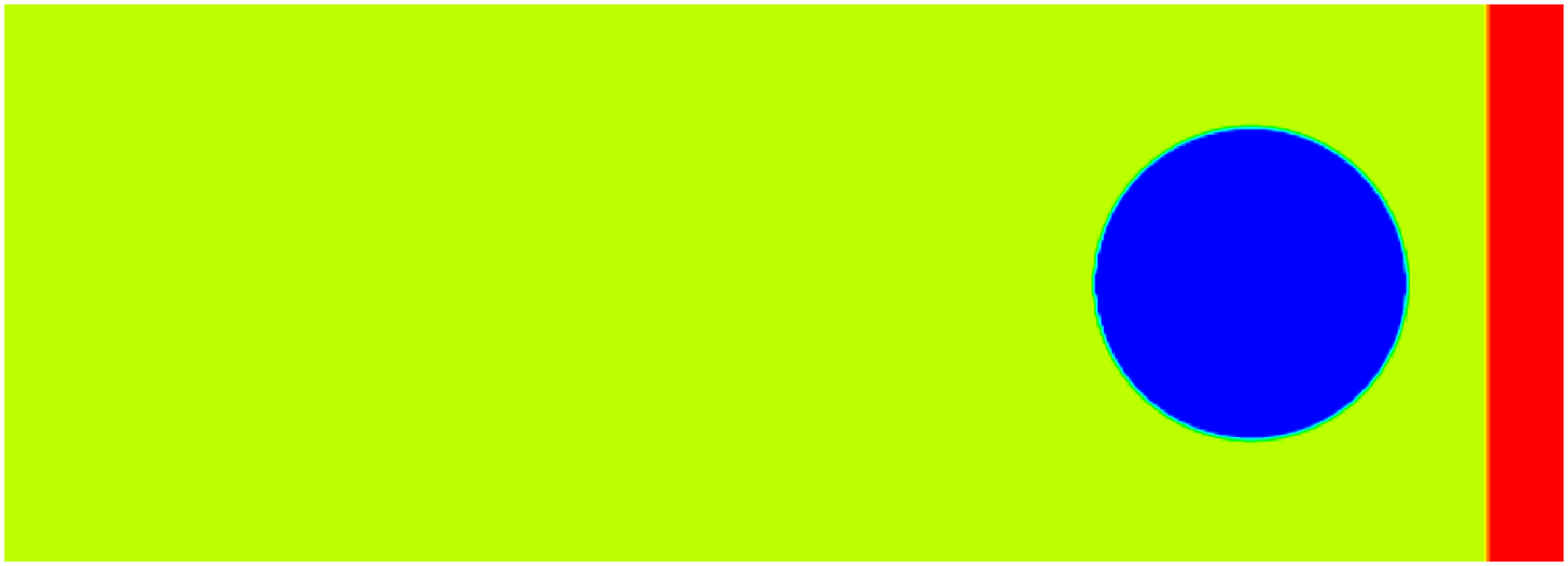}
\end{center}
\end{minipage}
\vspace{0.4em}

\begin{minipage}{.08\linewidth}
$t=0.2$
\end{minipage}
\begin{minipage}{.46\linewidth}
\begin{center}
\includegraphics[width=0.98\textwidth]{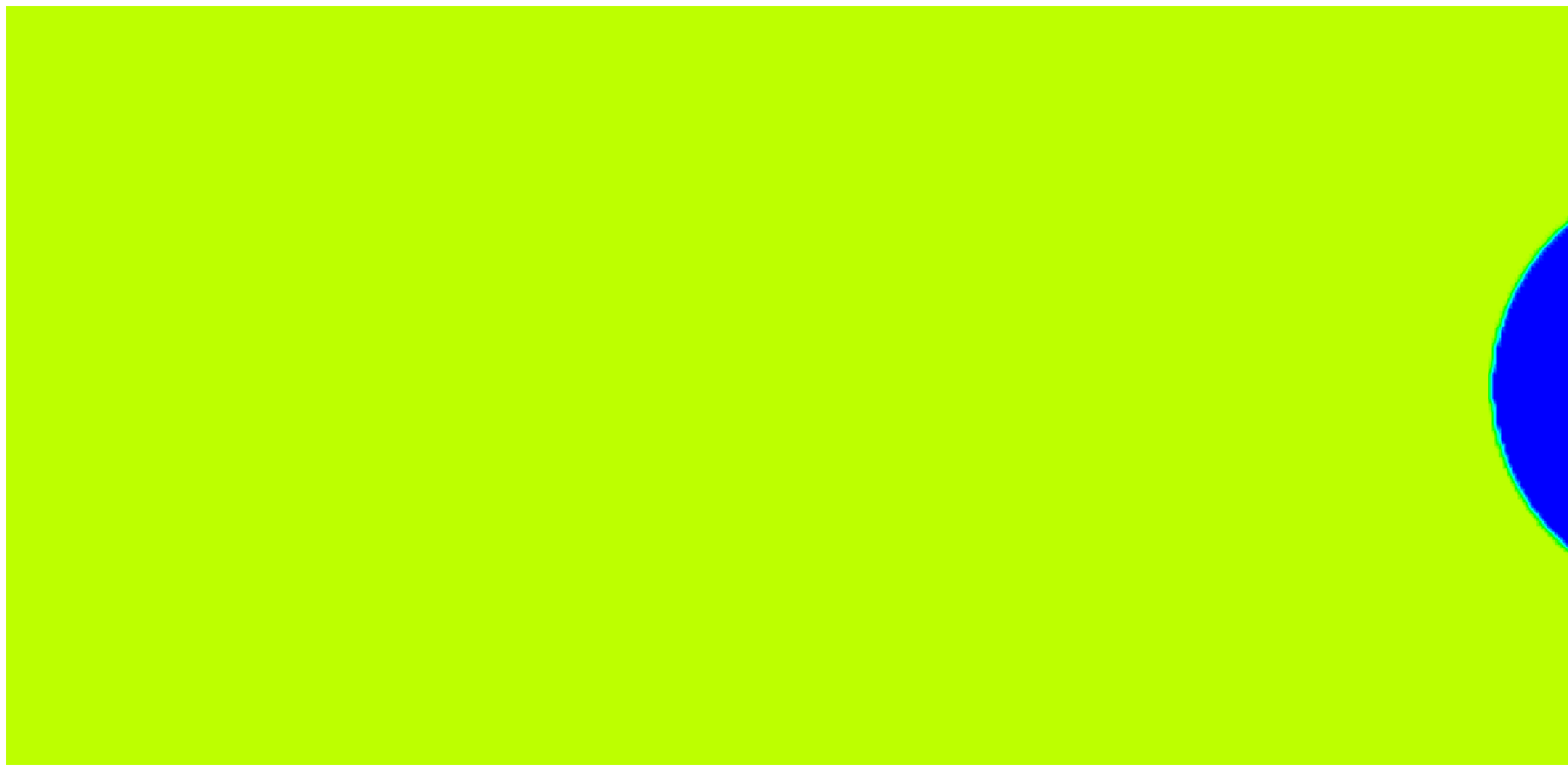}
\end{center}
\end{minipage}
\begin{minipage}{.46\linewidth}
\begin{center}
\includegraphics[width=0.98\textwidth]{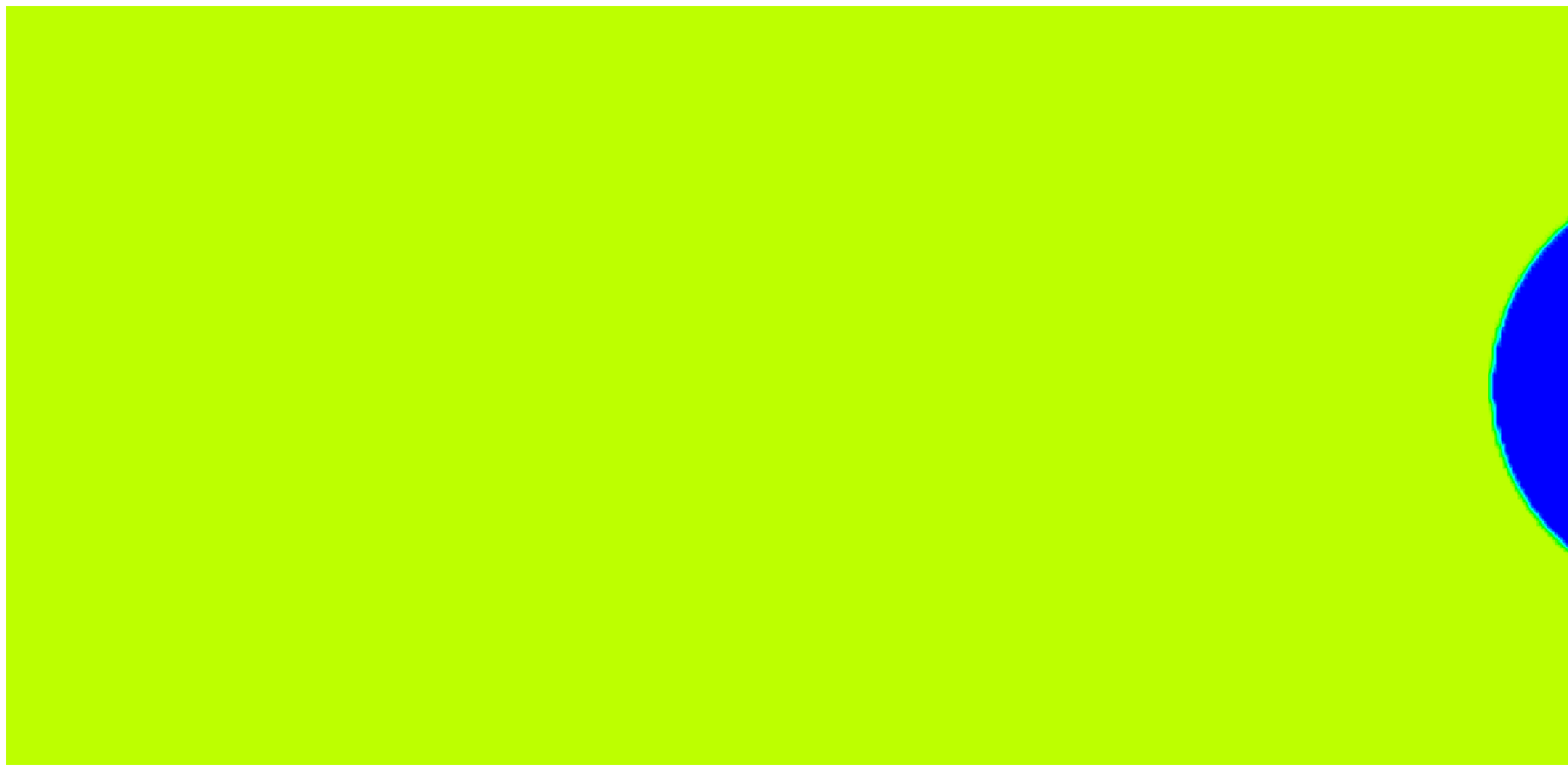}
\end{center}
\end{minipage}
\vspace{0.4em}

\begin{minipage}{.08\linewidth}
$t=0.4$
\end{minipage}
\begin{minipage}{.46\linewidth}
\begin{center}
\includegraphics[width=0.98\textwidth]{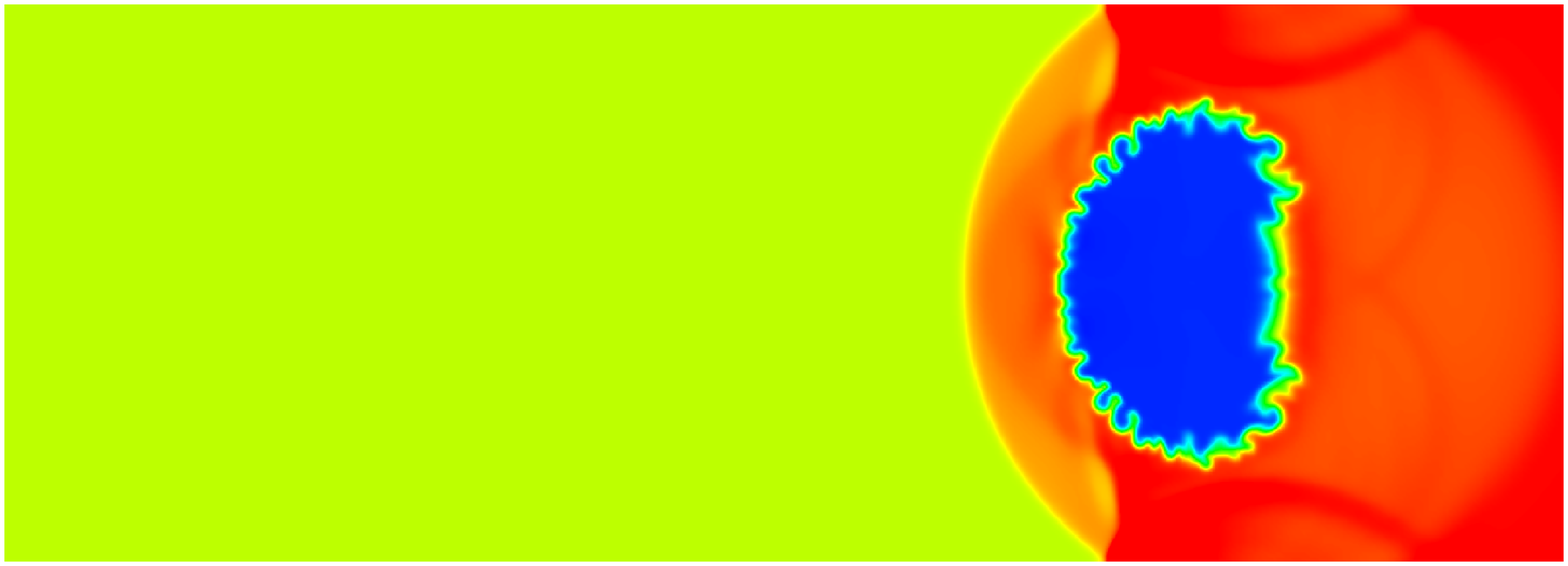}
\end{center}
\end{minipage}
\begin{minipage}{.46\linewidth}
\begin{center}
\includegraphics[width=0.98\textwidth]{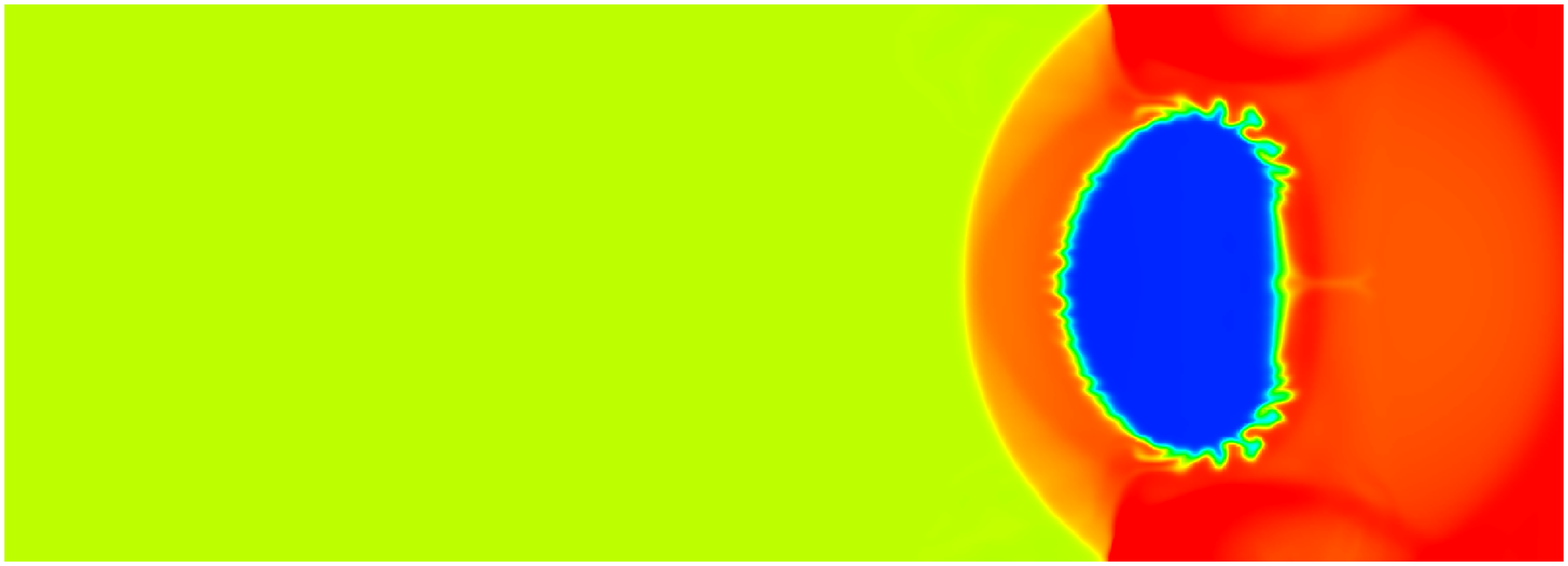}
\end{center}
\end{minipage}
\vspace{0.4em}

\begin{minipage}{.08\linewidth}
$t=0.8$
\end{minipage}
\begin{minipage}{.46\linewidth}
\begin{center}
\includegraphics[width=0.98\textwidth]{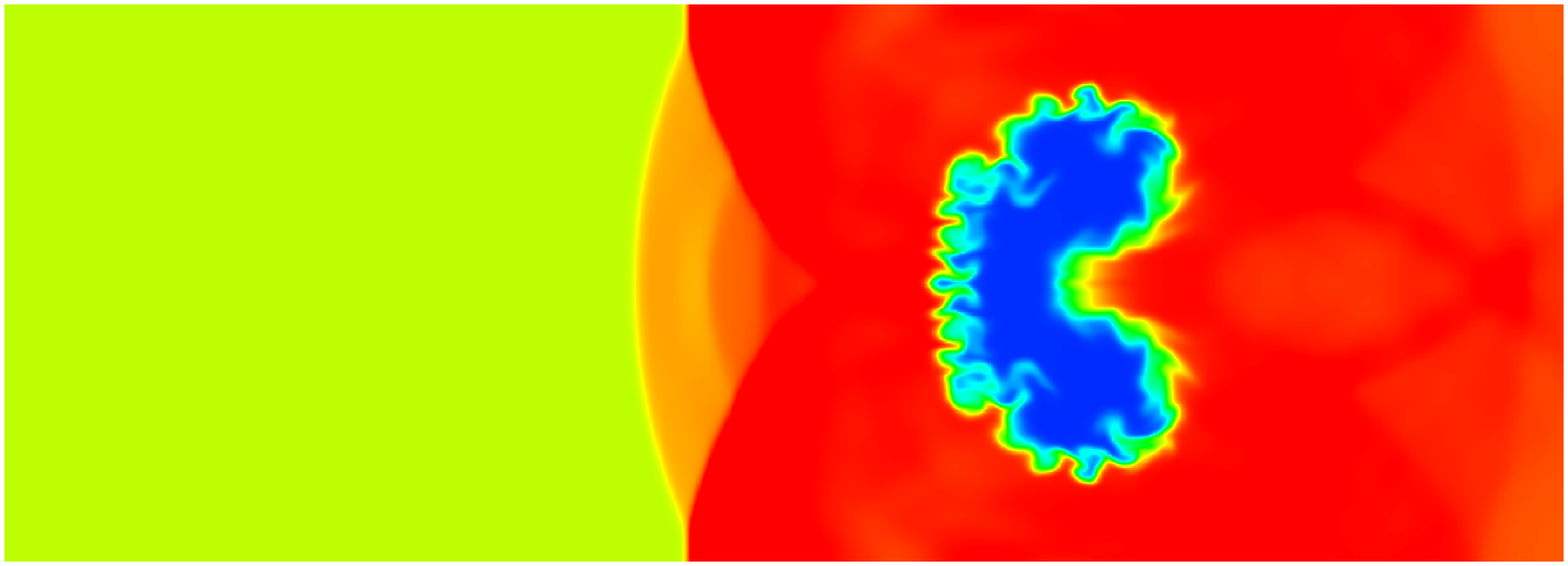}
\end{center}
\end{minipage}
\begin{minipage}{.46\linewidth}
\begin{center}
\includegraphics[width=0.98\textwidth]{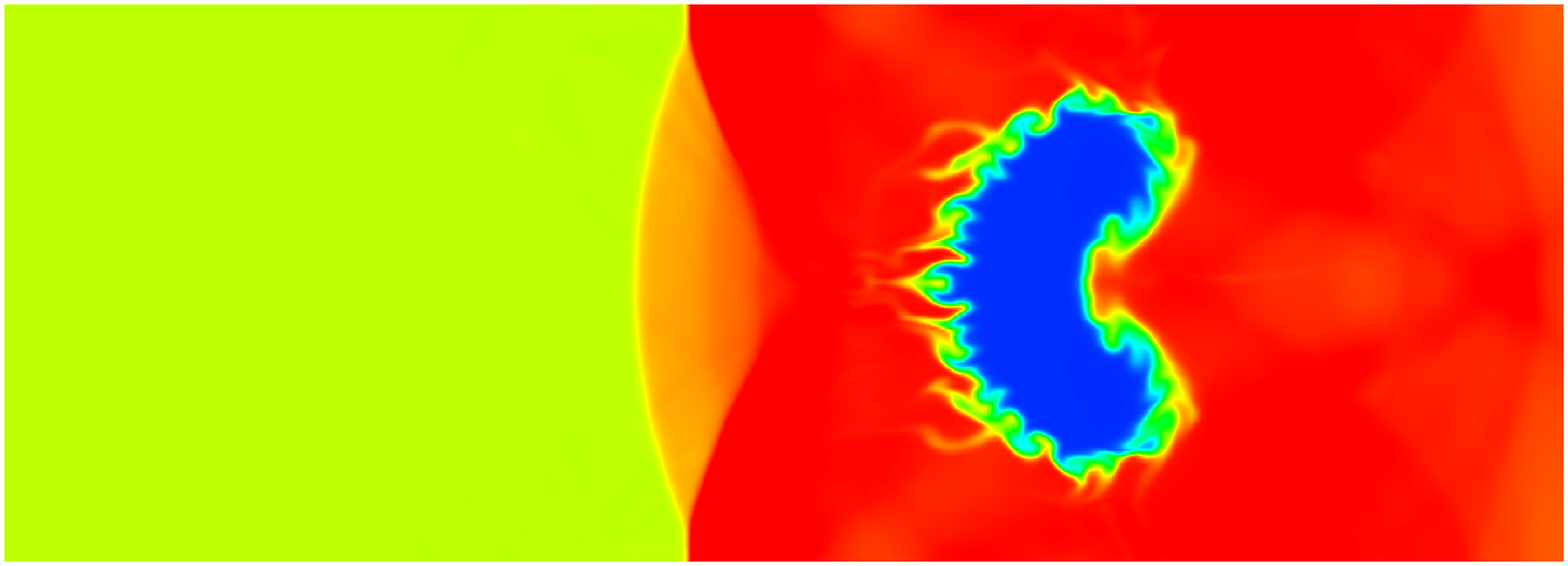}
\end{center}
\end{minipage}
\vspace{0.4em}

\begin{minipage}{.08\linewidth}
$t=1.5$
\end{minipage}
\begin{minipage}{.46\linewidth}
\begin{center}
\includegraphics[width=0.98\textwidth]{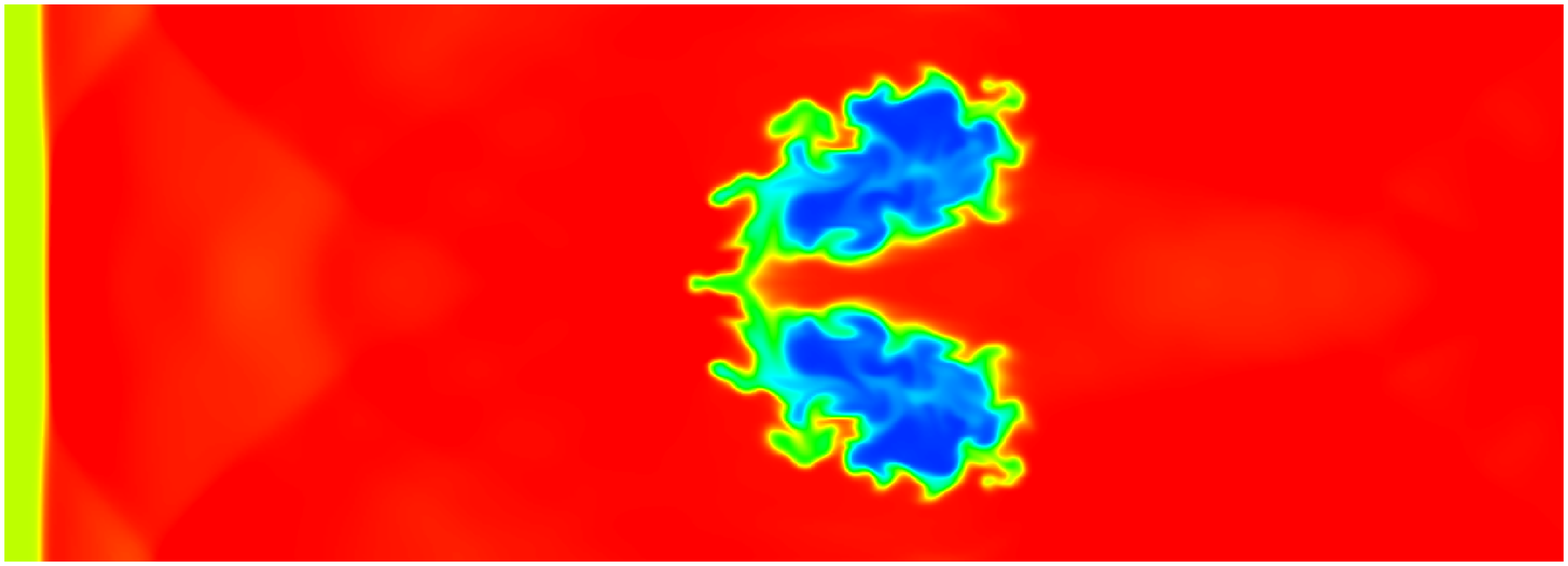}
\end{center}
\end{minipage}
\begin{minipage}{.46\linewidth}
\begin{center}
\includegraphics[width=0.98\textwidth]{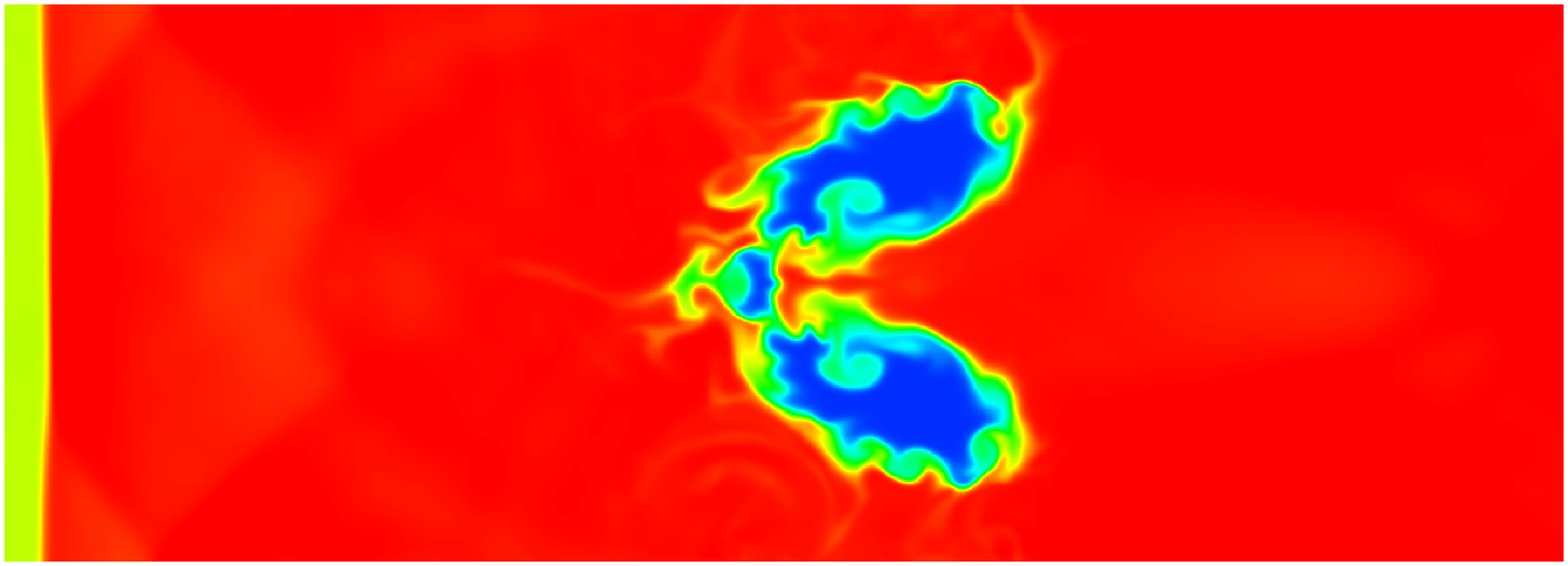}
\end{center}
\end{minipage}
\vspace{0.4em}

\begin{minipage}{.08\linewidth}
$t=3$
\end{minipage}
\begin{minipage}{.46\linewidth}
\begin{center}
\includegraphics[width=0.98\textwidth]{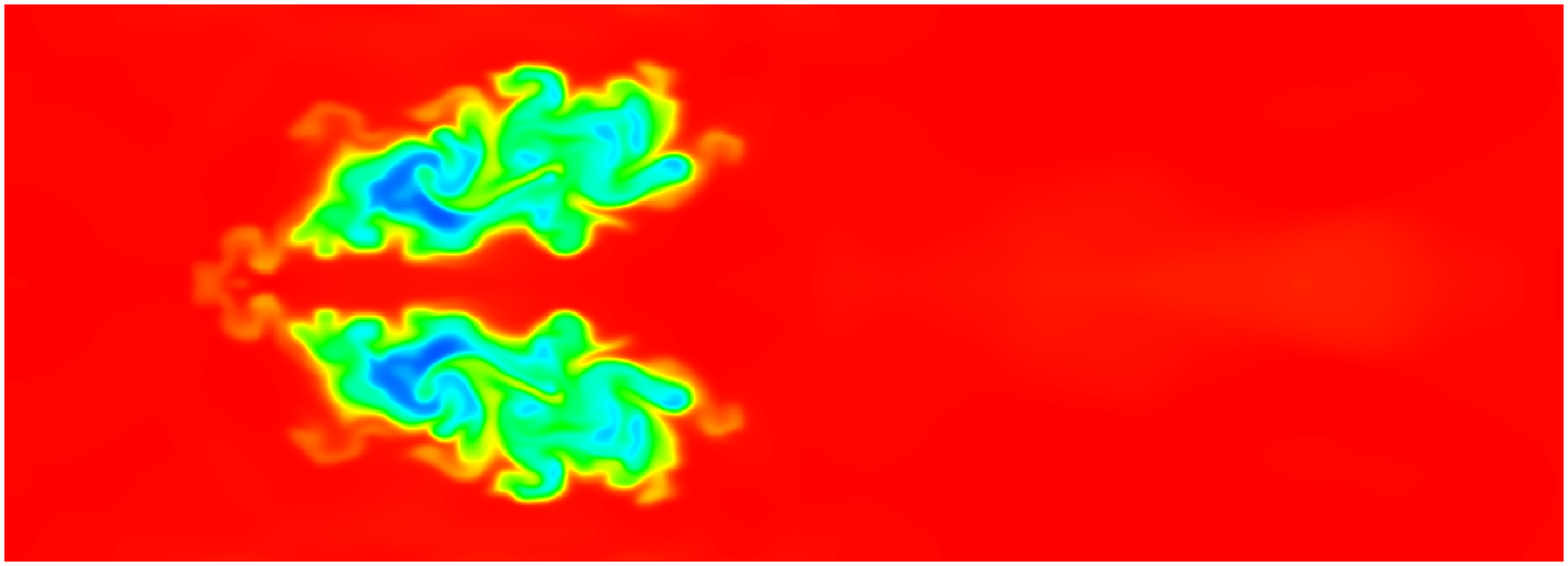}
\end{center}
\end{minipage}
\begin{minipage}{.46\linewidth}
\begin{center}
\includegraphics[width=0.98\textwidth]{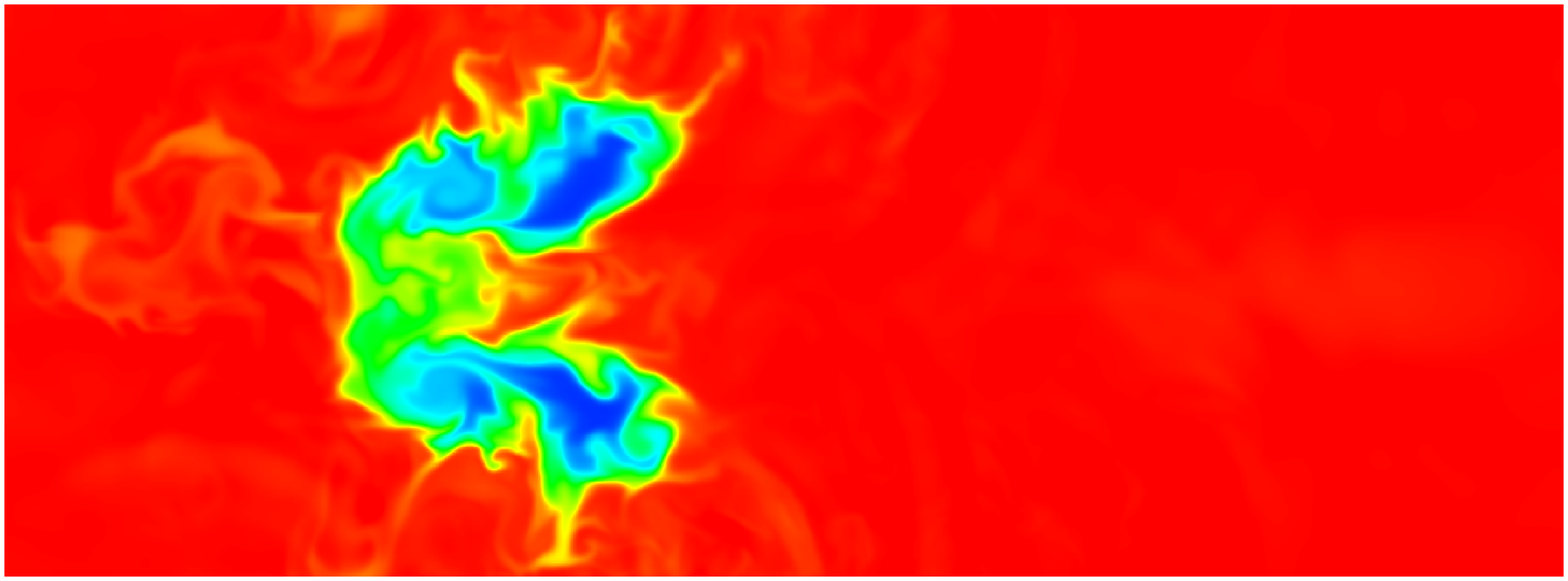}
\end{center}
\end{minipage}
\vspace{0.4em}

\begin{minipage}{.08\linewidth}
\ 
\end{minipage}
\begin{minipage}{.46\linewidth}
\begin{center}
$k1=s$, $k2=g$ (cylindrical gas phase)
\end{center}
\end{minipage}
\begin{minipage}{.46\linewidth}
\begin{center}
$k_1=g$, $k_2=s$ (cylindrical solid phase)
\end{center}
\end{minipage}
\caption{Numerical density plot of the shock-cylinder interaction problem by the staggered-projection GRP scheme for two settings: The cylindrical phase as the gas phase ($k_1=g$) or the solid phase ($k_2=s$).}\label{fig:shock-cylinder}
\end{figure}

\end{example}

\section{Conclusions}\label{sec:conclu}

It is always a challenging problem to simulate  compressible multi-material/phase flows due to the conflict of shock capturing and interface tracking.  Moreover, the interaction  of shocks and interfaces leads to the instability of flow fields, which raises more requirements of numerical methods, even though there were already a lot of  contributions in literature.  To this end,  we make our efforts in this aspect by taking the BN model. 

At first, we realize, through the  detailed analysis of spurious numerical oscillations near material interfaces, that the Riemann invariants play an essential role in the design of numerical methods, they are taken as a kind of key ingredients in practice.  In order to overcome the difficulty resulting from the conflict of conservative and non-conservative requirements,  the staggered strategy is adopted, together with the projection based on the Riemann invariants.  The accuracy is improved through the acoustic  GRP solver.  Several numerical experiments are carried out to  demonstrate  the reasonable performance. 

It is worth noting that the Newton-Raphson iteration method is utilized to solve nonlinear  algebraic equations that result from the transformation between Riemann invariants and primitive variables  and equal Riemann invariants across solid contacts.
Possible non-uniqueness or non-existence of the solutions of the algebraic equations raises a further investigation in the future in the sense that the interaction of nozzling term and flux gradient (resonance) should be more seriously treated and the equations should be more reasonably fitted.


\appendix

\section{Representation of BN model in terms of Riemann invariants}\label{app-RI} 
For smooth solutions, system \eqref{eq:BN-Euler} is  written in terms of these variables
\begin{equation}\label{eq:inv-w}
\bm{w}_t+\bm{B}(\bm{w})\bm{w}_x=\bm{0}, \ \ \ 
\bm{w}=[
\alpha_s,\ 
\rho_s,\ 
u_s,\ 
P,\ 
Q,\ 
H,\ 
\eta_g]^\top,
\end{equation}
with the coefficient $\bm{B}$,
\begin{equation*} 
\bm{B}=
\begin{bmatrix}
u_s & 0 & 0 & 0 & 0 & 0 & 0\\
0 & u_s & \rho_s & 0 & 0 & 0 & 0\\
0 & 0 & u_s & \frac{1}{\rho_s \alpha_s} & -\frac{V}{\rho_s \alpha_s} & -r & r T_g\\
0 & 0 &
\begin{array}{r}
\sum\limits_{k=s}^g\alpha_k\rho_k c_k^2\\
+3 \alpha_g\rho_g V^2
\end{array}
& u_s - 2r V &
\begin{array}{r}
(2r+1) V^2+ c_g^2
\end{array}
& 2\alpha_g\rho_g (r+1) V &
\begin{array}{r}
2 r T_g V \left[-\sum\limits_{k=s}^g\alpha_k\rho_k\right.\\
\left. +\alpha_s\rho_s\frac{\Gamma_g}{2}\right]
\end{array}\\
0 & 0 & 2\alpha_g \rho_g V & -r & 
u_g + r V &
\alpha_g\rho_g(r+1) &
-\alpha_g \rho_g T_g(r+1)\\
0 & 0 & V^2+c_g^2 & -\frac{V}{\rho_s\alpha_s} & \frac{V^2}{\alpha_s \rho_s} + \frac{c_g^2}{\alpha_g \rho_g} & 
u_g + r V &
-T_g (r + \Gamma_g) V\\
0 & 0 & 0 & 0 & 0 & 0 & u_g
\end{bmatrix},
\end{equation*}
where $V=u_g-u_s$ is the relative velocity, $r=\frac{\alpha_g\rho_g}{\alpha_s\rho_s}$ is the ratio of the mass fraction of the two phases and $\Gamma_k=\frac{1}{\rho_k}\frac{\partial p_k(\rho_k,e_k)}{\partial e_k}$ is the Gruneisen coefficient for the phase $k$.
Let $\bm{\Lambda}$ be a diagonal matrix with the eigenvalues $\lambda_i$ of $\bm{B}$,
\begin{equation*}
\bm{\Lambda}(\bm{w})=\text{diag}(\lambda_i)=\text{diag}(u_s,u_s-c_s,u_s,u_s+c_s,u_g-c_g,u_g,u_g+c_g)
\end{equation*}
and $\bm{R}$ be the right (column) eigenvector matrix of $\bm{B}$,
\begin{equation*}
\bm{R}(\bm{w})=\begin{bmatrix}
1&0&0&0&0&0&0\\
0&\frac{1}{c_s}&1&\frac{1}{c_s}&0&0&0\\
0&-{\frac{1}{\rho_s}}&0
&{\frac{1}{\rho_s}}&0&0&0\\
0&\alpha_s c_s+\frac {2 \alpha_g \rho_g V}{\rho_s} &0& 
\alpha_s c_s-\frac {2 \alpha_g \rho_g V}{\rho_s} & u_g-c_g -u_s & -\frac {\alpha_g \rho_g T_g \Gamma_g V^{2}}{c_g^2} & u_g+c_g - u_s\\
0& \frac {\alpha_g \rho_g}{\rho_s} &0
& -\frac {\alpha_g \rho_g}{\rho_s}
&1&-\frac {\alpha_g \rho_g T_g \Gamma_g V}{c_g^2} &1\\
0& \frac{V}{\rho_s} 
&0&- \frac {V}{\rho_s} & -\frac {c_g}{\alpha_g\rho_g} & T_g &{\frac {c_g}{ \alpha_g \rho_g}
}\\
0&0&0&0&0&1&0
\end{bmatrix},
\end{equation*}
so that $\bm{B}\bm{R} = \bm{R}\bm{\Lambda}$. Linearizing the system \eqref{eq:inv-w} around a state $\bm{w}=\bm{w}_*$, it can be diagonalized to obtain
\begin{equation*}
\bm{z}_t+\bm{\Lambda}(\bm{w}_*)\bm{z}_x=\bm{0},
\end{equation*}
where $\bm{z}=\bm{R}(\bm{w}_*)^{-1}\bm{w}$.
 
\section{Fitting nonlinear algebraic systems}\label{app-nonlinear}

As the Newton-Raphson method are not applicable to solve the nonlinear system \eqref{eq:non-linear-system}, the Gauss-Newton method is adopted to solve a least squares problem relevant to the system \eqref{eq:non-linear-system}.
From the equations (\ref{eq:non-linear-system}a), we can obtain symbolic relationships
$(p_g)_{i-\frac12,i}^{n+1,*}=(p_g)_{i-\frac12,i}^{n+1,*}\left((\rho_g)_{i-\frac12,i}^{n+1,*},(\rho_g)_{i,i+\frac12}^{n+1,*}\right)$
and
$(p_g)_{i,i+\frac12}^{n+1,*}=(p_g)_{i,i+\frac12}^{n+1,*}\left((\rho_g)_{i-\frac12,i}^{n+1,*},(\rho_g)_{i,i+\frac12}^{n+1,*}\right)$.
On the basis of the relationships, we come to solve the two equations
\begin{equation*}
\begin{aligned}
& (\eta_g)_{i-\frac12,i}^{n+1,*}-(\eta_g)_{i,i+\frac12}^{n+1,*}=0,\\
& H_{i-\frac12,i}^{n+1,*}-H_{i,i+\frac12}^{n+1,*}=0.
\end{aligned}
\end{equation*}
Then we consider a least squares problem in the form,
\begin{equation*}
\begin{aligned}
&\text{minimize} && \bm{\mathfrak{f}}(\mathfrak{x})=\frac12 \parallel \bm{\mathfrak{g}}(\bm{\mathfrak{x}})\parallel^2,\quad
\bm{\mathfrak{g}}=\left[(\eta_g)_{i-\frac12,i}^{n+1,*}-(\eta_g)_{i,i+\frac12}^{n+1,*}, H_{i-\frac12,i}^{n+1,*}-H_{i,i+\frac12}^{n+1,*}\right]^\top ,\\
&\text{subject to} && \bm{\mathfrak{x}} = \left[(\rho_g)_{i-\frac12,i}^{n+1,*},(\rho_g)_{i,i+\frac12}^{n+1,*}\right]^\top \in (0,\infty)\times(0,\infty).
\end{aligned}
\end{equation*}
The Gauss-Newton method is utilized to minimize the least squares cost $\frac12 \parallel \bm{\mathfrak{g}}(\bm{\mathfrak{x}})\parallel^2$.
 To enhance convergence, a modified form in accordance with the Cholesky factorization scheme is chosen.
As far as the approximate version of the Newton method does not work, we use the Newton-Raphson method.
The detailed procedure of least squares can be found in \cite[Section 1.4.4]{bertsekasnonlinear}.
This least squares solution approximates the  solution of the system \eqref{eq:non-linear-system} by averaging $\eta_g$ and $H$.
This approximation is reasonable since $\eta_g$ and $H$ in essence only reflect the thermodynamic relationship of the gas phase. Hence the error of these two quantities has relatively little influence on the numerical solution.

The above approach can effectively enhance the robustness of the staggered-projection method.
Another part for improving the robustness  is to recover  the vector $\bm{u}$ from the Riemann invariants $\bm{\omega}$, which involves a root-finding process of $\rho_g$ in the equation,
\begin{equation}\label{eq:root-finding}
\mathfrak{h}(\rho_g)=\frac{Q^{2}}{2 \alpha_{g}^{2}} \frac{1}{\rho_{g}^{2}}+\frac{\gamma_{g}}{\gamma_{g}-1} \eta_{g} \rho_{g}^{\gamma_{g}-1}-H=0.
\end{equation}
As pointed out in \cite{karni_hybrid_2010}, this equation may have no root. It may occur when intermediate states are generated when a large initial jump resolves itself into waves. We also solve the equation \eqref{eq:root-finding} approximately based on a nonlinear programming problem
\begin{equation*}
\begin{aligned}
&\text{minimize} && \mathfrak{f}(\rho_g)=\frac12 \mathfrak{h}(\rho_g)^2.
\end{aligned}
\end{equation*}
This problem is also solved using  the Newton method.
Then we recalculate $H$ from $\mathfrak{h}(\rho_g)=0$ in  \eqref{eq:root-finding}.
 It is noted that all above processes satisfy constraints $\rho_g>0$, $p_g>0$ and $p_s>0$.

\bibliography{Bib-BN-stagger}

\end{document}